\catcode`\@=11

\newif\if@fewtab\@fewtabtrue

{\count255=\time\divide\count255 by 60
\xdef\hourmin{\number\count255}
\multiply\count255 by-60\advance\count255 by\time
\xdef\hourmin{\hourmin:\ifnum\count255<10 0\fi\the\count255}}
\def\ps@draft{\let\@mkboth\@gobbletwo
    \def\@oddhead{}
    \def\@oddfoot
       {\hbox to 7 cm{$\scriptstyle Draft\ version:\ \draftdate$
       \hfil}\hskip -7cm\hfil\rm\thepage \hfil}
    \def\@evenhead{}\let\@evenfoot\@oddfoot}

\def\ceqno{\global\@fewtabfalse
    \ifcase\@eqcnt \def\@tempa{& & &}\or \def\@tempa{& &}
      \or \def\@tempa{&}
      \or\def\@tempa{}\fi\@tempa
{\rm(\theequation)}}

\def\aeqno#1{\global\@fewtabfalse
    \ifcase\@eqcnt \def\@tempa{& & &}\or \def\@tempa{& &}
      \or \def\@tempa{&}
      \or\def\@tempa{}\fi\@tempa
{\rm(\theequation,#1)}}

\def\label#1{\ifnum\draftcontrol=1
 \global\def\draftnote{$\scriptstyle #1$}\fi
 \@bsphack\if@filesw {\let\thepage\relax
   \def\protect{\noexpand\noexpand\noexpand}%
\xdef\@gtempa{\write\@auxout{\string
      \newlabel{#1}{{\@currentlabel}{\thepage}}}}}\@gtempa
   \if@nobreak \ifvmode\nobreak\fi\fi\fi
  \@esphack}

\def\alabel#1#2{\label{#1}\global\@fewtabfalse
    \ifcase\@eqcnt \def\@tempa{& & &}\or \def\@tempa{& &}
      \or \def\@tempa{&}
      \or\def\@tempa{}\fi\@tempa
{\hbox to 3cm{\phantom{\rm(\theequation,#2)}
\draftnote \hfil}\hskip -3cm {\rm(\theequation,#2)}}}

\def\clabel#1{\label{#1}\global\@fewtabfalse
    \ifcase\@eqcnt \def\@tempa{& & &}\or \def\@tempa{& &}
      \or \def\@tempa{&}
      \or\def\@tempa{}\fi\@tempa
{\hbox to 3cm{\phantom{\rm(\theequation)}
\draftnote \hfil}\hskip -3cm{\rm(\theequation)}}}

\def\eqnarray{\def\draftnote{{}}\global\@fewtabtrue
\stepcounter{equation}\let\@currentlabel=\theequation
\global\@eqnswtrue
\global\@eqcnt\z@\tabskip\@centering\let\\=\@eqncr
$$\halign to \displaywidth\bgroup\@eqnsel\hskip\@centering\@eqcnt\z@
  $\displaystyle\tabskip\z@{##}$&\global\@eqcnt\@ne
  \hskip 1\arraycolsep \hfil${##}$\hfil
  &\global\@eqcnt\tw@ \hskip 1\arraycolsep
$\displaystyle\tabskip\z@{##}$
\hfil  \tabskip\@centering&\global\@eqcnt\thr@@\llap{##}\tabskip\z@
\cr}

\def\endeqnarray{\@@eqncr\egroup
      \global\advance\c@equation\m@ne$$\global\@ignoretrue}

\def\@eqnnum{\hbox to 3cm{\phantom{\rm(\theequation)} \draftnote
                         \hfil}\hskip -3cm {\rm(\theequation)}}

\def\@@eqncr{\let\@tempa\relax
    \ifcase\@eqcnt \def\@tempa{& & &}\or \def\@tempa{& &}
      \or \def\@tempa{&}
      \or\def\@tempa{}
\fi\@tempa
\if@eqnsw
\if@fewtab\@eqnnum\fi
\stepcounter{equation}\fi\global
\@eqnswtrue\global\@eqcnt\z@\global\@fewtabtrue\cr}

\def\draftcite#1{\ifnum\draftcontrol=1#1\else{}\fi}

\def\@lbibitem[#1]#2{\item{}\hskip -3cm \hbox to 2cm
{\hfil$\scriptstyle\draftcite{#2}$}\hskip
1cm[\@biblabel{#1}]\if@filesw
     {\def\protect##1{\string ##1\space}\immediate
      \write\@auxout{\string\bibcite{#2}{#1}}}\fi\ignorespaces}

\def\@bibitem#1{\item\hskip -3cm \hbox to 2cm
{\hfil $\scriptstyle\draftcite{#1}$}\hskip 1cm
\if@filesw \immediate\write\@auxout
       {\string\bibcite{#1}{\the\value{\@listctr}}}\fi\ignorespaces}

\def\nsection#1{\section{#1}\setcounter{equation}{0}}
\font\tendl=msbm10  scaled \magstep1%double line
\font\sevendl=msbm7 scaled \magstep1
\font\fivedl=msbm5 scaled \magstep1
\font\tengl=eufm10  scaled \magstep1% gothic letters
\font\sevengl=eufm7 scaled \magstep1
\font\fivegl=eufm5 scaled \magstep1

\newfam\dlfam  % \dl is double line
\textfont\dlfam=\tendl \scriptfont\dlfam=\sevendl
\scriptscriptfont\dlfam=\fivedl
\newfam\glfam  % \gl is gothic letters
\textfont\glfam=\tengl \scriptfont\glfam=\sevengl
\scriptscriptfont\glfam=\fivegl

\def\draftdate{\number\month/\number\day/\number\year\ \ \ \hourmin }

\global\def\draftcontrol{0}
\catcode`\@=12
\def\tilde{\widetilde}

\documentclass[aps,amsfonts,amsmath,amssymb,floatfix,nofootinbib]{revtex4}
\usepackage{epsfig}
\usepackage{rotating}
\usepackage[centerlast]{subfigure}
\usepackage{bm}
\usepackage{amssymb}
\usepackage{comment}
\usepackage{xcolor}
\renewcommand{\thesection}{\arabic{section}}
\renewcommand{\theequation}{\thesection.\arabic{equation}}

\newcommand{\ii}{\mathrm{i}}
\newcommand{\be}{\begin{eqnarray}}
\newcommand{\en}{\end{eqnarray}\vs 0.5 cm}

\newcommand{\vs}{\vskip}

\newcommand{\Diff}{{Dif\hspace{-0.04cm}f_+S^1}}
\newcommand{\Difft}{{Dif\hspace{-0.04cm}f_+\hspace{-0.14cm}{}^{^\sim}\hspace{-0.05cm}S^1}}

\newcommand{\R}{{\mathbb R}}
\newcommand{\Z}{{\mathbb Z}}

\newcommand{\qq}{\begin{eqnarray}}

\newcommand{\ee}{{\rm e}}

\newcommand{\qqq}{\end{eqnarray}}

\newcommand{\Tr}{\hbox{Tr}}

\newcommand{\CA}{{\cal A}}
\newcommand{\CB}{{\cal B}}

\newcommand{\CD}{{\cal D}}
\newcommand{\CE}{{\cal E}}
\newcommand{\CF}{{\cal F}}
\newcommand{\CG}{{\cal G}}
\newcommand{\CH}{{\cal H}}
\newcommand{\CI}{{\cal I}}
\newcommand{\CJ}{{\cal J}}
\newcommand{\CK}{{\cal K}}
\newcommand{\CL}{{\cal L}}

\newcommand{\CP}{{\cal P}}
\newcommand{\CQ}{{\cal Q}}
\newcommand{\CR}{{\cal R}}
\newcommand{\CS}{{\cal S}}
\newcommand{\CT}{{\cal T}}
\newcommand{\CU}{{\cal U}}

\newcommand{\CW}{{\cal W}}
\newcommand{\CX}{{\cal X}}
\newcommand{\CY}{{\cal Y}}
\newcommand{\CZ}{{\cal Z}}

\newcommand{\m}{\hspace{0.025cm}}

\begin{document}
\title{\Large\bf{Full counting statistics of energy transfers
in inhomogeneous nonequilibrium states of
$(1\hspace{-0.1cm}+\hspace{-0.1cm}1)D$ CFT}}
\author{Krzysztof Gaw\c{e}dzki\footnote{chercheur \'em\'erite, email: kgawedzk@ens-lyon.fr} \,and
\,Karol K. Koz{\l}owski\footnote{email: karol.kozlowski@ens-lyon.fr}\vspace{0.2cm}}
\affiliation{Universit{\'e} de Lyon, ENS de Lyon,
Universit{\'e} Claude Bernard, CNRS,\\ Laboratoire de
Physique, F-69342 Lyon, France}

\begin{abstract}
\vskip 0.2truecm
\noindent Employing the conformal welding technique,
we find an exact expression for the Full Counting Statistics of energy
transfers in a class of inhomogeneous nonequilibrium states of a
(1+1)-dimensional unitary Conformal Field Theory.
The expression involves the Schwarzian action of a complex field
obtained by solving a Riemann-Hilbert type problem related
to conformal welding of infinite cylinders. On the way, we obtain
a formula for the extension of characters of unitary positive-energy
representations of the Virasoro algebra to 1-parameter groups
of circle diffeomorphisms and we develop techniques, based
on the analysis of certain classes of Fredholm operators, that allow
to control the leading asymptotics of such extensions for small real part
of the modular parameter $\,\tau$.
\end{abstract}

\maketitle

\nsection{Introduction}

\noindent $(1\hspace{-0.1cm}+\hspace{-0.1cm}1)D$ Conformal Field Theory
(CFT) provides an effective description of long-range physics in a
number of critical systems in one spatial dimension. The examples include  
electrons or cold atoms trapped in one-dimensional potential wells,
carbon nanotubes, quantum Hall edge currents or critical XXZ spin chains.
CFT permitted to explain the long-range equilibrium properties of such systems
that are driven by the low lying excitations but, more recently, it has
also been used to describe the nonequilibrium situations, like the evolution
after quantum quenches \cite{CaCa2} or in the partitioning protocol
after two haft-line systems prepared in different equilibrium states
are joined together \cite{BD2}.
In \cite{LLMM2} a smooth version of the partitioning
protocol was considered for nonlocal and local Luttinger model
\cite{Lutt,ML} of interacting fermions with the initial nonequilibrium
state possessing a built-in inverse-temperature profile
$\,\beta(x)\,$ interpolating smoothly between different constant values
on the left and on the right. More exactly, such a profile state
corresponds in finite box of length of order $\,L\,$ to the density matrix
proportional to $\,\exp[-G_L]\,$ where $\,G_L=\int\beta(x)\CE(x)\m dx\,$
with $\,\CE(x)\,$ standing for the energy density and the integral running
over the box. The evolution of the energy
density and current in such an initial state was computed in \cite{LLMM2}
in the $\,L\to\infty\,$ limit
in the perturbation theory in the difference of the asymptotic values of
the inverse-temperature profile, for the local version of the model to all
orders. The local
Luttinger model is a $(1\hspace{-0.1cm}+\hspace{-0.1cm}1)D$ CFT and
in \cite{GLM} the evolution of similar profile states was analyzed 
for a general unitary $(1\hspace{-0.1cm}+\hspace{-0.1cm}1)D$ CFT using 
global conformal symmetries of the theory\footnote{The restriction to unitary
CFTs is essentially of technical nature.}.  The latter  permitted
to reduce the arbitrary correlation functions of the energy-momentum
components or the primary fields in the nonequilibrium profile states
to equilibrium correlation functions, considerably generalizing the results
of \cite{LLMM2} on the local Luttinger model. 
\vskip 0.1cm

The present paper is, in a sense, a continuation of \cite{GLM}.
We show how global conformal symmetries may be used to provide
an exact expression for the generating function for Full Counting
Statistics (FCS) of energy transfers through a kink in the
profile of a profile state. The notion of FCS was introduced by
L. S. Levitov and G. B. Lesovik in \cite{LL}, where an exact formula
for the generating function for FCS of charge transport between
free-fermion channels was obtained. For finite volumes, the expression
involves Fredholm determinants containing scattering amplitudes \cite{ABGK}.
The definition of FCS requires a measurement protocol for the changes
of the conserved quantity which may be assimilated with its transfers.
The measurements may be indirect, performed on a device coupled to
the system \cite{LLL}, or direct, performed on the system in question,
\cite{MA}. Our approach to FCS of energy transfers in a profile state
will be based on the two-time measurement protocol that belongs to the
latter class. We shall extract the energy transfer through
the kink in $\,\beta(x)\,$ from two measurements, separated
by time $\,t$, \,of the finite box observable $\,G_L\,$ introduced
above. To associate the difference of the results of such
measurements with the energy transfer through the kink
encompassed by the spatial box, we shall have to impose boundary
conditions that guarantee that there is no energy transfer through
the edges of the box. Those are different from the periodic boundary
conditions used in \cite{GLM}. The finite-volume CFT with such boundary
conditions is chiral, i.e. its space of states carries a unitary,
positive-energy representation of a single Virasoro algebra with
generators $\,L_n\,$ and central charge $\,c$. \,The Virasoro
representation lifts to a projective unitary representation of the group
$\,\Diff\,$ of smooth, orientation-preserving
diffeomorphisms of the circle. We show that the generating function for
FCS of energy transfers extracted from the two-time measurements of
$\,G_L\,$ may be expressed in terms of the character
\qq
\Upsilon(f,\tau)=\Tr\Big(U(f)\,\ee^{2\pi\ii\tau(L_0-\frac{c}{24})}\Big)
\qqq
of the projective representation $\,f\mapsto U(f)\,$ of
$\,\Diff$, \,where $\,\tau\,$ is in the upper
half-plane.
In fact one only needs to know such a character on 
1-parameter subgroups $\,(f_s)\,$ of $\,\Diff$. \,Unlike
the Virasoro characters $\,\chi(\tau)\,$  defined by a similar
formula but with $\,U(f)\,$ absent, the characters $\,\Upsilon(f,\tau)\,$
have not been studied in detail. Inspired by the recent work
\cite{FH2}, which used conformal welding of boundaries of two discs
with a twist by the circle diffeomorphism $\,f_s\,$ to express the vacuum
matrix elements $\,\langle 0|U(f_s)|0\rangle$, \,see also \cite{Poly}, \,we 
obtain an exact formula for $\,\Upsilon(f_s,\tau)\,$ from conformal
welding with the twist by $\,f_s$ of the boundaries of complex annulus
$\,\{z\,|\,|\ee^{2\pi\ii\tau}|\leq|z|\leq 1\}$.
\,In fact, following the approach of \cite{Segal}, one of the present
authors (K.G.) has previously obtained a general formula for
$\,\Upsilon(f,\tau)\,$
involving Fredholm determinants of operators appearing in a Riemann-Hilbert
type problem underlying the same conformal welding construction \cite{KGr}.
The approach based on the idea of \cite{FH2}, that we describe here,
produces, however, a simpler integral expression for $\,\Upsilon(f_s,\tau)$.
This reduction resembles the use of a Riemann-Hilbert problem solution in
\cite{MA} in the context of the Levitov-Lesovik formula. In our case,
the generating function for finite-volume FCS is finally expressed by a
ratio of Virasoro characters at two different values of $\,\tau\,$ and
the exponential of an integral involving the Schwarzian derivative of a
solution of a Fredholm equation directly related to conformal welding of
the boundaries of annuli\footnote{The integral in question may be viewed
as a complexification of the Schwarzian action \cite{AlekShat} revived
recently in the context of the SYK model \cite{Kitaev}.}.
\vskip 0.1cm

In the infinite-volume limit, whose rigorous control is somewhat cumbersome,
the expression for the generating function for FCS simplifies to
the exponential of an integral involving the solution
of a Fredholm equation now related to conformal welding
of the boundaries of an infinite strip in the complex plane or,
equivalently, of the boundaries of two discs. The corresponding
Fredholm equation may
be studied numerically as discussed in \cite{ShM}. The infinite-volume
result exhibits a large degree of universality: the generating function
for FCS depends only on the profile $\,\beta(x)\,$ and on the central
charge $\,c\,$ that enters as an overall power, but not on the details
of the CFT. For large times $\,t$, \,the generating function for FCS should
take the large deviations form derived in \cite{BD0} for the partitioning
protocol. That form depends on $\,\beta(x)\,$ only through the asymptotic
values. It arises when conformal welding of the strip boundaries
involves the diffeomorphism that is a simple translation but we still
lack a sufficient rigorous control of the large $\,t\,$  asymptotics
of the finite-time Fredholm equation so that the large-deviations result
remains only heuristic at the moment. 
\vskip 0.1cm

The reference \cite{GLM} considered also nonequilibrium states 
with profiles for both the inverse temperature and the chemical
potential in CFTs with the $\,u(1)\,$ current algebra. The approach
to FCS discussed here may be extended to cover the $\,u(1)$-charge
transfers at least in states with chemical potential profile but
constant $\,\beta$. We postpone the study of such extensions
to a future publication.
\vskip 0.1cm

The present paper is organized as follows. In Sec.\,\ref{sec:box},
we describe the structure of $(1\hspace{-0.1cm}+\hspace{-0.1cm}1)D$ CFT
in a finite box with no energy-flux through the boundary.
Sec.\,\ref{sec:examples} gives three simple examples of such CFTs:
the free massless fermions (Sec.\,\ref{subsec:exA}), the free massless
compactified bosons (Sec.\,\ref{subsec:exB}), and the local Luttinger model
(Sec.\,\ref{subsec:exC}). In Sec.\,\ref{sec:Neqstates}, we construct
the finite-volume non-equilibrium profile states (Sec.\,\ref{subsec:neq}),
show how they may be related to equilibrium states (Sec.\,\ref{subsec:neqeq}),
and discuss their infinite-volume limit that coincides with the one
obtained in ref.\,\cite{GLM} which used periodic boundary conditions
(Sec.\,\ref{subsec:thlim}). Sec.\,\ref{sec:FCSprel} contains a
preliminary discussion of FCS of energy transfers in
the profiles states, describing the two-time measurement protocol
for FCS (Sec.\,\ref{subsec:twotime}) and defining the FCS generating
function (Sec.\,\ref{subsec:gfFCS}).
Sec.\,\ref{sec:FCSconfweld}, that constitutes the core of the paper,
relates FCS to conformal welding. First, we express the generating
function of finite-volume FCS via characters of
$\,\Diff\,$ (Sec.\,\ref{subsec:FCSchar}). Next, we explain a correspondence
between $\,\Diff\,$ and Virasoro characters that
originates from the isomorphism between tori which are conformally
welded from annuli with and without twist by a circle diffeomorphism
(Sec.\,\ref{subsec:reduct}). Then we discuss the connection between
conformal welding of tori and an inhomogeneous
Riemann-Hilbert problem (Sec.\,\ref{subsec:confweldRH}). Next, we
obtain a relation between the 1-point function of the Euclidian
energy-momentum tensor on the tori welded with and without twist
(Sec.\,\ref{subsec:1ptTEucl}). Such a relation gives rise to a formula
for $\,\Diff$-characters on 1-parameter subgroups
(Sec.\,\ref{subsec:on1parsub}) and for the correspondence between
modular parameters of tori welded with and without twist
(Sec.\,\ref{subsec:effmpar}). Finally, we apply the above formulae
to obtain an exact expression for the generating function
for finite-volume FCS (Sec.\,\ref{subsec:fvolFCS}). The infinite-volume
limit of that expression is discussed in Sec.\,\ref{sec:infvlim:heur}. First,
we study the infinite-volume behavior of the 1-parameter subgroups
of circle diffeomorphisms providing twists for conformal welding
of tori on which the finite-volume FCS formula was based
(Sec.\,\ref{subsec:largeLzeta}). Next, we discuss
conformal welding of cylinders to which conformal welding of tori
reduces in the infinite-volume limit (Sec.\,\ref{subsec:confweldcyl}).
Based on that, we extract, under some technical assumptions, our exact
infinite-volume formula for the generating function for FCS of energy
transfers (Sec.\,\ref{subsec:infvformula}) and check that it represents
correctly the first two moments of the energy transfer
(Sec.\,\ref{subsec:checks}). Sec.\,\ref{sec:longtime} examines
the long-time large-deviations asymptotics for FCS of energy transfers
and discusses its consequences that were first pointed out in \cite{BD0}.
Sec.\,\ref{sec:infvlim:proof} is devoted to a rigorous proof of the
technical assumptions used in Sec.\,\ref{sec:infvlim:heur} about the
the convergence of solutions of the Fredholm equations related to
conformal welding of tori and of cylinders. This is the most technical part
of the paper and it uses results developed in Appendix \ref{app:C} about
Fredholm operators of two classes relevant for our problem, their
determinants and their inverses. Finally, \,Sec.\,\ref{sec:concl}
\m lists our conclusions and other Appendices
\ref{app:A},\ref{app:B},\ref{app:D},\ref{app:E} and \ref{app:F}
establish few more technical results used on the way. 
\vskip 0.3cm

{\bf Acknowledgements.} \,K.G. thanks Chris Fewster and
Stefan Hollands for inspiring discussions and Jan Derezi\'{n}ski,
Stefan Hollands and Karl-Henning Rehren for an invitation to the BIRS 2018
workshop ``Physics and Mathematics of Quantum Field Theory'' that influenced
the work on the present paper.

\nsection{Minkowskian CFT in a finite box}
\label{sec:box}

\noindent Let us consider a Minkowskian CFT in the spatial 
box $\,[-{\frac{_1}{^4}L},{\frac{_1}{^4}L}]\,$ with 
a special type of boundary conditions that assure the following
gluing relations for the right- and left-moving components
$\,T_{+}(x^-)\,$ and $\,T_{-}(x^+)\,$ of the energy-momentum
tensor\footnote{The existence
of such symmetric boundary conditions constraints somewhat the class of
CFT models that we consider.}:
\qq
T_{+}(x^-)=T_{-}(x^+)\qquad{\rm for}\qquad x=\pm{\frac{_1}{^4}L}\,,
\label{gluer}
\qqq
where $\,x^\pm=x\pm vt\,$ are the light-cone
coordinates. The latter relations mean that there
is no energy flux through the boundary and they imply that $\,T_{+}(x)\,$
and $\,T_{-}(x)\,$ are $\,L$-periodic distributions with values in
self-adjoint operators in the Hilbert space $\,\CH_L\,$ of states of the
finite-box theory satisfying the relation
\qq
T_{-}(x)=T_{+}(-x\pm\frac{_1}{^2}L)\,.
\label{TpmL}
\qqq 
Such a theory is then chiral: there is just one independent $\,L$-periodic
component of the energy-momentum tensor for which we shall choose 
$\,T_{+}(x)\equiv T(x)$. \,The Fourier modes of $\,T(x)\,$
define the generators $\,L_n,\ n\in\mathbb{Z}$, \,of the Virasoro
algebra\footnote{The shift $\,x\,$ to $\,x+\frac{1}{4}L\,$
in the expansion, introduced for convenience, amounts to the
replacement of $\,L_n\,$ by $\,\ii^nL_n$.}
\qq
T(x)=\frac{_{2\pi}}{^{L^2}}\sum\limits_{n=-\infty}^\infty
\ee^{\frac{2\pi\ii n}{L}(x+\frac{1}{4}L)}\big(L_n-\frac{_c}{^{24}}\delta_{n,0}\big)\,,
\label{decomp}
\qqq
with $\,L_n=L_{-n}^\dagger\,$ satisfying (on a common dense domain)
the commutation relations
\qq
[L_n,L_m]=(n-m)L_{n+m}+\frac{_c}{^{12}}(n^3-n)\delta_{n,-m}\label{Vir}
\label{Virasoro}
\qqq
which are equivalent to 
\qq
[T(x),T(y)]=-2\ii\,\delta'_{L}(x-y)\,T(y)
+\ii\,\delta_{L}(x-y)\,T'(y)
+\frac{_{c\,\ii}}{^{24\pi}}\,\delta'''_{L}(x-y)\,,
\label{commper}
\qqq
where $\,c\,$ is the central charge of the theory and
$\,\delta_{L}\,$ is the $L$-periodized delta-function.
We assume that the Hilbert space
of states $\,\CH_L\,$ of the finite-box theory is a (possibly infinite)
direct sum of the unitary highest-weight representations of the Virasoro
algebra with fixed central charge $\,c\,$ containing the vacuum
representation exactly once.
\vskip 0.2cm

The energy density of the theory is defined by 
\qq
\CE(t,x)=v\big(T_{+}(x^-)+T_{-}(x^+)\big)=v\big(T(x^-)+T(-x^+-
\frac{_1}{^2}L)\big)\,,
\qquad
\label{SCE}
\qqq
and the self-adjoint Hamiltonian in the box is
\qq
H_L\,=\,\int_{-{\frac{_1}{^4}\hspace{-0.04cm}L}}^{
{\frac{_1}{^4}\hspace{-0.04cm}L}}\hspace{-0.15cm}\CE(t,x)\,dx
\,=\,v\int_{-{\frac{_3}{^4}\hspace{-0.04cm}L}}^{
{\frac{_1}{^4}\hspace{-0.04cm}L}}\hspace{-0.15cm}T(x^-)\,dx\,=\,
\frac{_{2\pi v}}{^L}\big(L_0-\frac{_c}{^{24}}\big).
\label{HL}
\qqq
It generates the right-moving dynamics of $\,T(x)$:
\qq
\ee^{\ii tH_L}T(x)\,\ee^{-\ii tH_L}\,=\,T(x^-)\,.
\label{TxTxm}
\qqq
We assume additionally that $\,\Tr\,\big(\ee^{-\beta_0H_L}\big)<\infty\,$ for
all $\,\beta_0>0$, \,a condition that is satisfied for a rich class of models
of CFT including the rational ones. The expectation values of observables
in the finite-box equilibrium state with inverse temperature $\,\beta_0\,$
are then defined by the formula
\qq
\big\langle A\big\rangle_{\beta_0;L}^{\rm eq}\,=\,\frac{\Tr\,\big(A\,\ee^{-\beta_0H_L}\big)}
{\Tr\,\big(\ee^{-\beta_0H_L}\big)}
\label{eqbetaL}
\qqq
(we set $\,\hbar=1=k_B$).
\vskip 0.2cm

Let $\,\Diff\,$ be the group of
smooth, orientation-preserving diffeomorphisms of the circle
$\,S^1_L=\mathbb R/L\mathbb Z\,$ and let
$\,\Difft\,$ be its universal cover. \,Elements of
$\,\Difft\,$ are smooth functions $\,f:\mathbb{R}\to\mathbb{R}\,$
such that $\,f'(x)>0\,$ and $\,f(x+L)=f(x)+L$. 
\,The group $\,\Difft\,$ is contractible and one has the central
extension of groups 
\qq
0\,\longrightarrow\,\mathbb Z\,\longrightarrow\,
\Difft\,\longrightarrow\,\Diff\,\longrightarrow\,1
\qqq
where $\,n\in\mathbb Z\,$ is represented
by the translation $\,x\mapsto x+nL$. \,(Direct sums of) unitary
highest-weight representations of the Virasoro algebra with central
charge $\,c\,$ lift to unitary projective representations
$\,f\mapsto U(f)\,$ of 
$\,\Difft\,$ [\onlinecite{GW2},\onlinecite{TL},\onlinecite{FH1}] \,such that,
\qq
U(f)\,T(x)\,U(f)^{-1}\,=\,f'(x)^2\,T(f(x))-\frac{c}{{24\pi}}(Sf)(x)\,,
\label{UT+U}
\qqq
where
\qq
(Sf)(x)=\frac{f'''(x)}{f'(x)}-\frac{3}{2}
\Big(\frac{f''(x)}{f'(x)}\Big)^2
\label{Schwarz}
\qqq
is the Schwarzian derivative of $\,f\,$ fulfilling the chain rule
\qq
S(f_1\circ f_2)=(f_2')^2(Sf_1)\circ f_2\,+\,Sf_2\,.
\label{Schwchain}
\qqq
If $\,\zeta:\mathbb R\rightarrow\mathbb R\,$ is a smooth function
satisfying $\,\zeta(x+L)=\zeta(x)\,$ defining a vector field
$\,-\zeta(x)\partial_x\,$ on $\,\mathbb R\,$ and if $\,f_s(x)\,$ is the
flow of the latter that forms a 1-parameter subgroup of
$\,\Difft$, \,i.e.
\qq
\partial_sf_s(x)=-\zeta(f_s(x))\,,\qquad f_0(x)=x\,,
\qqq
then
\qq
U(f_s)=c_{s,\zeta}\,\exp\Big[\ii s\int_{-{\frac{_3}{^4}\hspace{-0.04cm}L}}
^{{\frac{_1}{^4}\hspace{-0.04cm}L}}\hspace{-0.15cm}\zeta(x)T(x)dx\Big],
\label{exponen}
\qqq
where $\,|c_{s,\zeta}|=1$. \,In particular, if $\,\zeta(x)=L\,$ then
$\,f_s(x)=x-sL\,$ are translations that form the Cartan subgroup
of $\,\Difft\,$ and
\qq
U(f_s)=c_{s,L}\,\ee^{2\pi\ii s(L_0-\frac{c}{24})}\m.
\label{onCart}
\qqq
The projective factors in $\,U(f)\,$ may be fixed so that \cite{FH1}
\qq
U(f_1)\m U(f_2)=\exp\Big[\frac{\ii c}{48\pi}
\int_{-{\frac{_3}{^4}\hspace{-0.04cm}L}}^{{\frac{_1}{^4}\hspace{-0.04cm}L}}
\ln{(f_1\circ f_2)'(x)}\,d\ln{f_2'(x)}\Big]U(f_1\circ f_2)\,.
\label{multipl}
\qqq
The exponential term on the right-hand side of (\ref{multipl}) defines
the Bott 2-cocycle on $\,\Difft\,$ that corresponds
to the infinitesimal 2-cocycle given by the last terms on the
right-hand side of (\ref{Virasoro}) or (\ref{commper}) \cite{KhWe}.
\vskip 0.2cm

Eqs. (\ref{UT+U}) and (\ref{TpmL}) imply that in the special case
when
\qq
f(-x-\frac{_1}{^2}L)=-f(x)-\frac{_1}{^2}L
\label{fLsym}
\qqq
one also has the relation
\qq
U(f)\,T_{-}(x)\,U(f)^{-1}\,=\,f'(x)^2\,T_{-}(f(x))
-\frac{c}{{24\pi}}(Sf)(x)\,.
\label{UT-U}
\qqq
The projective representation $\,f\rightarrow U(f)\,$ in the Hilbert
space $\,\CH_L\,$ of the finite-box theory will be our main tool used
below.  
\vskip 0.2cm

The primary fields $\,\Phi(x,t)\,$ satisfy in the light-cone
re parameterization the commutation relations
\qq
[T(x),\Phi(y^-,y^+)]&=&-\ii\big(\Delta^+_\Phi\delta'_{L}(x-y^-)
+\Delta^-_\Phi\delta'_{L}(x+y^++\frac{_1}{^2}L)\big)\Phi(y^-,y^+)\cr\cr
&&+\ii\,\delta_{L}(x-y^-)\,\partial_-\Phi(y^-,y^+)
-\ii\,\delta_{L}(x+y^++\frac{_1}{^2}L)\,\partial_+\Phi(y^-,y^+)\,,
\qqq
where $\,(\Delta^+_\Phi,\Delta^-_\Phi)\,$ are the conformal weights of 
$\,\Phi$. \,Under the adjoint action of $\,U(f)\,$ for $\,f\in
\Difft\,$ they transform according to the rule
\qq
U(f)\m\Phi(x^-,x^+)\m U(f)^{-1}=f'(x^-)^{\Delta^+_\Phi}
f'(-x^+-\frac{_1}{^2}L)^{\Delta^-_\Phi}\,\Phi(f(x^-),-f(-x^+-\frac{_1}{^2}L)
-\frac{_1}{^2}L)
\qqq
that simplifies to
\qq
U(f)\m\Phi(x^-,x^+)\m U(f)^{-1}=f'(x^-)^{\Delta^+_\Phi}
f'(x^+)^{\Delta^-_\Phi}\,\Phi(f(x^-),f(x^+))
\label{UPhiU}
\qqq
if $\,f\,$ satisfies (\ref{fLsym}).

\nsection{Simple examples}
\label{sec:examples}

\noindent For illustration, we list three simple examples of models
of CFT with the structure as discussed above. Many other examples, e.g.,
the unitary minimal models or the WZW and coset theories, may be added
to that list.

\subsection{Free massless fermions}
\label{subsec:exA}

\noindent The classical action 
functional of anti-commuting Fermi fields has here the form
\qq
S[\psi_+,\psi_-]\,=\,\frac{_{2\ii v}}{^{\pi}}\int dt
\int_{-{\frac{_1}{^4}\hspace{-0.04cm}L}}^{{\frac{_1}{^4}\hspace{-0.04cm}L}}
\big[\psi_+^*
\partial_+\psi_+\,-\,
\psi_-^*\partial_-\psi_-\big]dx\,,
\label{Sf}
\qqq
where $\,\partial_{\pm}=\frac{1}{2}(\partial_x\pm v^{-1}\partial_t)$. \,We
impose on the fields the boundary conditions 
\qq
\psi_+(t,-\frac{_1}{^4}L)=\psi_-(t,-\frac{_1}{^4}L)\,,\qquad
\psi_+(t,\frac{_1}{^4}L)=-\psi_-(t,\frac{_1}{^4}L)\,.
\label{bdc}
\qqq
The quantized theory has the fermionic Fock space $\,\CF_{\hspace{-0.04cm}f}\,$ 
carrying the vacuum representation of CAR 
\qq
[c_p,c_{p'}]_{_{\hspace{-0.03cm}+}}=0=[c^\dagger_p, c^\dagger_{p'}]_{_{\hspace{-0.03cm}+}},
\qquad[c_p,c^\dagger_{p'}]_{_{\hspace{-0.03cm}+}}=\delta_{p,p'}
\qqq
with $\,p,p'\in\frac{1}{2}+\mathbb Z\,$ as the space of states $\,\CH_L$,
\,with the normalized Dirac vacuum $\,|0\rangle_{\hspace{-0.04cm}f}\,$
annihilated by $\,c_p\,$ and $\,c^\dagger_{-p}\,$ with $\,p>0$.
\,The quantum fermionic fields are
\qq
\psi_\pm(t,x)=\sqrt{\frac{_{\pi}}{^{L}}}\m
\sum\limits_{p}c_p\,\ee^{\pm\frac{2\pi\ii p}{L}(x^\mp+\frac{1}{4}L)}\,\equiv\,
\psi_\pm(x^\mp)
\label{ffields}
\qqq
and the energy momentum tensor components
\qq
T_{\pm}(x^\mp)=\pm\frac{_\ii}{^{2\pi}}:\big((\partial_\mp\psi^\dagger_\pm)\psi_\pm-
\psi_\pm^\dagger\partial_\mp\psi_\pm\big)(x^\mp):-\frac{_{\pi c}}{^{12L^2}}
\label{femtensor}
\qqq
satisfy (\ref{gluer}) and correspond
to $\,c=1\,$ Virasoro generators
\qq
L_n=\sum\limits_p(p-\frac{_n}{^2}):c_{p-n}^\dagger c_p:
\qqq
where the fermionic Wick ordering $\,:-:\,$ puts
the creators $\,c_p\,$ and $\,c^\dagger_{-p}\,$ with $\,p<0\,$
to the left of the annihilators, with a minus sign for each 
transposition. $\,\psi_\pm(x^\mp)\,$ are primary fields with
the conformal weights $\,(\frac{1}{2},0)\,$ and $\,(0,\frac{1}{2})$,
\,respectively, \,and so are their hermitian conjugates $\,\psi_\pm^\dagger
(x^\mp)$. \,Among the other primary fields of the theory
there are the chiral components of the $\,U(1)$ current
\qq
J_\pm(x^\mp)=\frac{_1}{^\pi}:(\psi^\dagger_\pm\psi_\pm)(x^\mp):
\label{fcurrent}
\qqq
with conformal weights $\,(1,0)\,$ and $\,(0,1)$, \,respectively.

\subsection{Compactified free massless bosons}
\label{subsec:exB}

\noindent The classical action functional of the massless
free field $\,\varphi(x,t)\,$ with values defined modulo
$\,2\pi\,$ is 
\qq \label{action_initial0}
S[\varphi]=\frac{_{r^2}}{^{4\pi}}\int
dt\int_{-{\frac{_1}{^4}\hspace{-0.04cm}L}}^{{\frac{_1}{^4}\hspace{-0.04cm}L}}
\big[(\partial_t
\varphi)^2-(\partial_x\varphi)^2\big]\,dx\,.
\qqq
The parameter $\,r\,$ has the interpretation of the radius
of compactification of the circle of values of the field.
We impose on the fields $\,\varphi(x,t)\,$ the Neumann
boundary conditions 
\qq
\partial_x\varphi(t,-\frac{_1}{^4}L)=0
=\partial_x\varphi(t,\frac{_1}{^4}L)\,.
\label{Nbc}
\qqq
The space of states $\,\CH_L\,$ of the quantized theory is then the tensor 
product $\,\CH_0\otimes\CF_b$. \,The second factor is the bosonic Fock
space $\,\CF_b\,$ carrying the vacuum representation of the CCR algebra
\qq
[\alpha_n,\alpha_m]=n\m\delta_{n,-m}
\label{alphaalpha}
\qqq
for $\,n\in\mathbb Z$, \,with the normalized vacuum state
$\,|0\rangle_b\,$ annihilated by $\,\alpha_n\,$ for $\,n\geq 0\,$
and with $\,\alpha_{-n}=\alpha_n^\dagger$. The mode $\,\alpha_0$, \,that commutes
with all $\,\alpha_n$, \,acts in the zero-mode space
$\,\CH_0\,$ spanned by the orthonormal vectors $\,|k\rangle, \,k\in\mathbb Z$,
\,by
\qq
\alpha_0|k\rangle=\frac{_{\sqrt{2}}}{^r}k\m|k\rangle\,.
\qqq
The quantum bosonic field is
\qq
\varphi(t,x)=\varphi_+(x^-)-\varphi_-(x^+)\,,
\label{bfields1}
\qqq
where
\qq
\varphi_\pm(x^\mp)=\pm\frac{_1}{^2}\varphi_0-\frac{_{\sqrt{2}\m\pi}}{^{rL}}\alpha_0
(x^\mp+\frac{_1}{^4}L)
\pm\ii\sum\limits_{n\not=0}\frac{_1}{^{\sqrt{2}rn}}\alpha_n\hspace{0.02cm}
\ee^{\pm\frac{2\pi\ii n}{L}(x^\mp+\frac{1}{4}L)}\,,
\label{bfields2}
\qqq
with $\,[\varphi_0,\alpha_n]=\ii\,\delta_{n0}\frac{{\sqrt{2}}}{r}$, 
so that the action of $\,\ee^{\pm\ii\varphi_0}\,$ on $\,\CH_0\,$ is well defined
with $\,\ee^{\pm\ii\varphi_0}|k\rangle=|k\pm1\rangle$.
\,The components of the energy-momentum tensor satisfying (\ref{gluer})
are
\qq
T_\pm(x^\mp)=\frac{_{r^2}}{^{2\pi}}:(\partial_\mp\varphi_\pm)^2(x^\mp):
-\frac{_{\pi}}{^{12L^2}}
\label{bemtensor}
\qqq
and they correspond to the $\,c=1\,$ Virasoro generators 
\qq
L_n=\frac{_1}{^2}\sum\limits_m:\alpha_{n-m}\alpha_m:\,,
\label{Lnbos}
\qqq
where the bosonic Wick ordering puts the creators $\,\alpha_n\,$ 
with $\,n<0\,$ to the left of annihilators $\,\alpha_n\,$ with $\,n>0$.
\,Among the primary fields, there are the chiral components
of \m a $\,U(1)\,$ current
\qq
J_\pm(x^\mp)=-\frac{_{r^2}}{^{2\pi}}(\partial_\mp\varphi_\pm)(x^\mp)=
\frac{_r}{^{\sqrt{2}L}}\sum\limits_n\alpha_n\hspace{0.02cm}
\ee^{\pm\frac{2\pi\ii n}{L}(x^\mp+\frac{1}{4}L)}
\label{bcurrent}
\qqq
with conformal weights $\,(1,0)\,$ and $\,(0,1)$, \,respectively.
\vskip 0.2cm

The bosonic theory with the compactification radius $\,r=\sqrt{2}\,$
is equivalent to the fermionic theory from the previous subsection.
The equivalence is established by the unitary isomorphism between
$\,\CF_{\hspace{-0.04cm}f}\,$ and $\,\CH_0\otimes\CF_b\,$ that maps
the fermionic vacuum
$\,|0\rangle_{\hspace{-0.04cm}f}\,$ to $\,|0\rangle\otimes|0\rangle_b\,$ \,and
intertwines the chiral components (\ref{fcurrent}) and 
(\ref{bcurrent}) of the $U(1)\,$ currents as well as those of the
energy-momentum tensor (\ref{femtensor}) and (\ref{bemtensor}).
The fermionic fields (\ref{ffields}) are, in turn, intertwined with
the bosonic vertex operators
\qq
\sqrt{\frac{_\pi}{^{L}}}:\ee^{\mp2\ii\varphi_\pm(x^\mp)}:\ \equiv\,
\sqrt{\frac{_\pi}{^{L}}}\,\ee^{-\ii\varphi_0\pm\frac{2\pi\ii}{L}(x^\mp+\frac{1}{4}L)
\alpha_0}\,\,\ee^{\,\sum\limits_{n<0}\frac{1}{n}\alpha_n\hspace{0.02cm}
\ee^{\pm\frac{2\pi\ii n}{L}(x^\mp+
\frac{1}{4}L)}}\ee^{\,\sum\limits_{n>0}\frac{1}{n}\alpha_n\hspace{0.02cm}
\ee^{\pm\frac{2\pi\ii n}{L}(x^\mp+\frac{1}{4}L)}}.
\qqq

\subsection{Local Luttinger model}
\label{subsec:exC}

\noindent The (local, spinless) Luttinger model
[\onlinecite{Tomonaga},\onlinecite{Lutt},\onlinecite{ML},\onlinecite{Voit},\onlinecite{Giam},\onlinecite{MM}] is obtained by perturbing the
Hamiltonian $\,H_L=\frac{2\pi v}{L}\big(L_0-\frac{1}{24}\big)\,$ of the
free massless fermions of Sec.\,\ref{subsec:exA} by the addition of
a singular local interaction term
\qq
H^{\rm int}_L=\int_{-{\frac{_1}{^4}\hspace{-0.04cm}L}}^{{\frac{_1}{^4}\hspace{-0.04cm}L}}\big[
g_2J_+J_-+\frac{_1}{^2}g_4(J_+^2+J_-^2)\big](x)\,dx\,+\,{\rm const.}
\qqq
with $\,J_\pm\,$ as in (\ref{fcurrent}) and $\,2\pi v+g_4>|g_2|$.
After replacing $\,H^{\rm int}_L\,$ by
its cutoff non-local version and conjugating with a cutoff-dependent
unitary \cite{ML}, one may remove the cutoff. What results is a CFT
that is equivalent to the bosonic massless free field with Neumann
boundary conditions considered in Sec.\,\ref{subsec:exB} with the
compactification radius given by the relation
\qq
\frac{_{r^2}}{^2}=\sqrt{\frac{2\pi v+g_4+g_2}{2\pi v+g_4-g_2}}\equiv K
\qqq
and the modified Fermi velocity
\qq
\tilde v=\frac{\sqrt{(2\pi v+g_4)^2-g_2^2}}{2\pi}\,.
\qqq
In particular, the chiral components of the energy momentum tensor
and of the $\,U(1)\,$ current of the Luttinger model are those of
the massless free field:
\qq
T_\pm(\tilde x^\mp)=\frac{_K}{^\pi}:(\tilde\partial_\mp\varphi_\pm)^2
(\tilde x^\mp):-\frac{_{\pi}}{^{12L^2}}\,,\qquad J_\pm(\tilde x^\mp)
=-\frac{_K}{^\pi}(\tilde\partial_\mp\varphi_\pm)(\tilde x^\mp)\,,
\qqq
for $\,\tilde x^\pm=x\pm\tilde v t\,$ and $\,\tilde\partial_\pm=
\frac{1}{2}\big(\partial_x\pm\frac{1}{\tilde v}\partial_t\big)$.
\,The fermionic fields of the Luttinger model are represented by the vertex
operators
\qq
\psi_+(x,t)=\psi_-(-x-\frac{_1}{^2}L,t)&=&A_{K,L}(x)
\,:\ee^{-\ii(K+1)\varphi_+(\tilde x^-)-\ii
(K-1)\varphi_-(\tilde x^+)}:\cr
&\equiv&A_{K,L}(x)\,\,\ee^{-\ii\varphi_0+\ii\frac{\pi\sqrt{K}}{L}\alpha_0
(\tilde x^-+\tilde x^++\frac{_1}{^2}L)
+\ii\frac{\pi}{\sqrt{K}L}\alpha_0(\tilde x^--\tilde x^+)}\cr
&&\times\ \ee^{\,\sum\limits_{n<0}\frac{1}{2\sqrt{K}n}\big((K+1)\,
\ee^{\frac{2\pi\ii n}
{L}(\tilde x^-+\frac{1}{4}L)}-(K-1)\,\ee^{-\frac{2\pi\ii n}{L}
(\tilde x^++\frac{1}{4}L)}\big)}\cr
&&\times\ 
\ee^{\,\sum\limits_{n>0}\frac{1}{2\sqrt{K}n}\big((K+1)\,\ee^{\frac{2\pi\ii n}
{L}(\tilde x^-+\frac{1}{4}L)}-(K-1)\,\ee^{-\frac{2\pi\ii n}{L}
(\tilde x^++\frac{1}{4}L)}\big)},
\qqq
where
\qq
A_{K,L}(x)=\frac{_1}{^{\sqrt{2}}}\Big(\frac{_{2\pi}}{^L}\Big)^{\frac{K^2+1}{4K}}
\big(2\cos(\frac{_{2\pi x}}{^L})\big)^{-\frac{K^2-1}{4K}}.
\qqq
The fields $\,\psi_+(x,t)\,$ and $\,\psi_+^\dagger(x,t)\,$ anti-commute
at different points and are primary fields with conformal weights
$\,\big(\frac{(K+1)^2}{8K},\frac{(K-1)^2}{8K})$.

\nsection{Nonequilibrium states with temperature profile}
\label{sec:Neqstates}

\begin{figure}[h]
\leavevmode
\begin{center}
\vskip -0.3cm
\includegraphics[width=6.5cm,height=2.5cm]{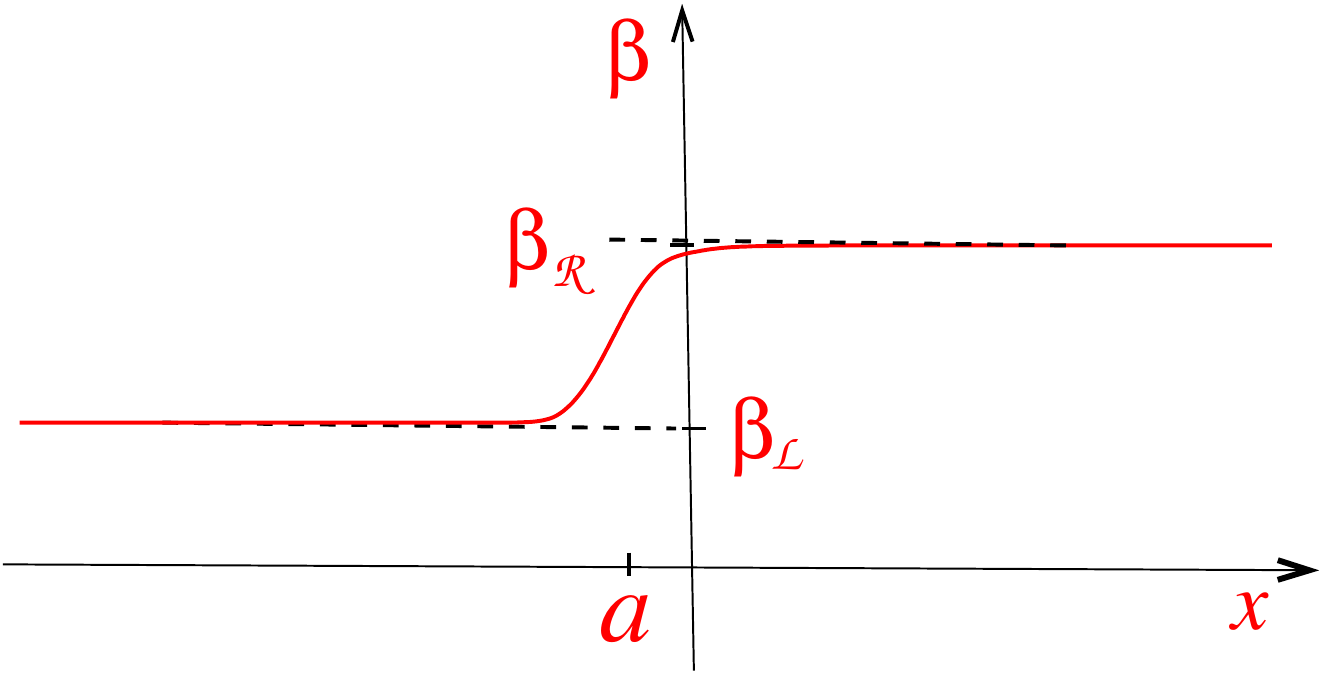}\\
\caption{An example of a profile $\,\beta(x)$}
\label{fig:profile}
\end{center}
\vskip -0.5cm
\end{figure}

\noindent The main aim of this paper, similarly to that of
[\onlinecite{LLMM2},\onlinecite{GLM}],
is to study certain aspects of the time evolution of nonequilibrium states
with preimposed smooth inverse-temperature kink-like profiles $\,\beta(x)>0\,$
such that
\qq
\beta(x)\,=\,\begin{cases}\,\beta_\CL\hspace{0.38cm}{\rm for}
\hspace{0.36cm}x\quad{\rm sufficiently\ negative}\m,\cr
\,\beta_\CR\quad{\rm for}\quad x\quad{\rm sufficiently\ positive}\m,
\end{cases}
\label{kink}
\qqq
for some positive constants $\,\beta_\CL,\beta_\CR$,
\,see Fig.\,\ref{fig:profile}.

\subsection{Finite-volume profile states} 
\label{subsec:neq}

\noindent We shall consider first such states in a finite box $\,[-\frac{1}{4}L,
\frac{1}{4}L]\,$ assuming the boundary conditions (\ref{gluer}) and
taking $\,L\,$ sufficiently
large so that $\,\beta(x)=\beta_\CL\,$ for $\,x\leq-{\frac{_1}{^4}
\hspace{-0.04cm}L}\,$ and
$\,\beta(x)=\beta_\CR\,$ for $\,x\geq {\frac{_1}{^4}\hspace{-0.04cm}L}$. 
\,Let $\,\beta_L(x)\,$ be the smooth $\,L$-periodic function  on
$\,\mathbb R\,$ satisfying
\qq
\beta_L(x)\,=\,\begin{cases}\,\beta(x)\hspace{1.87cm}{\rm for}\qquad
x\in[-{\frac{1}{4}\hspace{-0.04cm}L},{\frac{1}{4}\hspace{-0.04cm}L}]
\m,\cr\,\beta(-x-\frac{1}{2}L)\qquad{\rm for}\qquad
x\in[-{\frac{3}{4}\hspace{-0.04cm}L},-{\frac{1}{4}\hspace{-0.04cm}L}]
\m,\end{cases}
\label{betaL}
\qqq

\begin{figure}[h]
\leavevmode
\begin{center}
\vskip -0.3cm
\includegraphics[width=10cm,height=2.3cm]{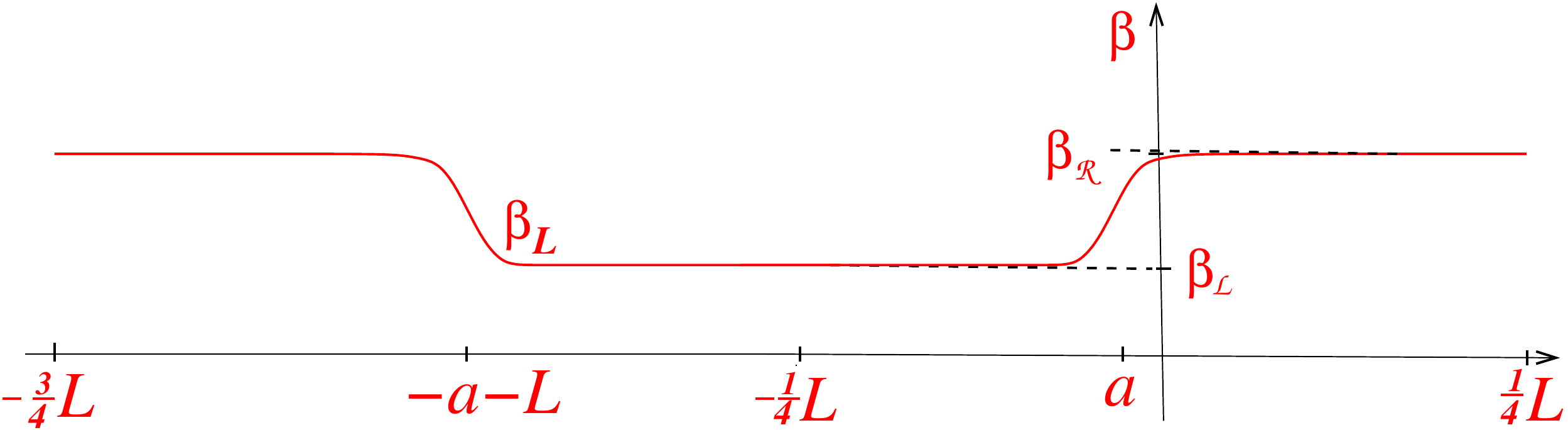}\\
\caption{The extended inverse-temperature profile}
\label{fig:reflprofile}
\end{center}
\vskip -0.3cm
\end{figure}

\noindent see Fig.\,\ref{fig:reflprofile}. The relations
\qq
\beta_L(x+L)=\beta_L(x)\,,\qquad\beta_L(-x-\frac{_1}{^2}L)=\beta_L(x)
\label{betabetaL}
\qqq
hold then for all real $\,x$. \,Let for the energy density $\,\CE(t,x)\,$
given by (\ref{SCE}),
\qq
G_L(t)\,=\,\int_{-{\frac{_1}{^4}\hspace{-0.04cm}L}}^{{\frac{_1}{^4}\hspace{-0.04cm}L}}
\hspace{-0.1cm}\beta(x)\,\CE(t,x)\,dx\,=\,v\int_{\CI_L}\hspace{-0.1cm}
\beta_L(x)\,T(x^-)\,dx\,=\,v\int_{\CI_L}\hspace{-0.1cm}\beta_L(x^+)\,T(x)\,dx\,,
\label{CG}
\qqq
where we used (\ref{TpmL}) and (\ref{betaL}) denoting by $\,\CI_L\,$
the extended interval $\,[-\frac{3}{4}L,\frac{1}{4}L]$. As will be shown
below, $\,G_L(t)\,$ is a bounded below self-adjoint operator
such that $\,\Tr\big(\ee^{-G_L(t)}\big)<\infty$, \,see also \cite{FH1}.
\,We shall consider the finite-box nonequilibrium state with
expectation values defined by 
\qq
\big\langle A\big\rangle^{\rm neq}_L\,=\,\frac{\Tr\,\big(A\,\ee^{-G_L(0)}\big)}
{\Tr\,\big(\ee^{-G_L(0)}\big)}\m.
\label{omegavarL}
\qqq

\subsection{Relation between nonequilibrium and equilibrium states}
\label{subsec:neqeq}

\noindent Let $\,h_L\,$ be a function on the real line defined by
\qq
h_L(x)\,=\,\beta_{0,L}\int_{-{\frac{_1}{^4}\hspace{-0.04cm}L}}^x\hspace{-0.15cm}
\beta_L(x')^{-1}dx'\,-\,
{\frac{_1}{^4}\hspace{-0.04cm}L}\qquad\text{for}\qquad
\beta_{0,L}^{-1}\,=\,2L^{-1}\int_{-{\frac{_1}{^4}\hspace{-0.04cm}L}}^{
{\frac{_1}{^4}\hspace{-0.04cm}L}}\hspace{-0.1cm}\beta(x)^{-1}dx\m.\ \quad
\label{someeq}
\qqq
Then
\qq
&&h_L'(x)\ =\,\frac{_{\beta_{0,L}}}{^{\beta_L(x)}},\label{fL'}\\
&&h_L(x+L)\ =\,\beta_{0,L}\int_{-{\frac{_1}{^4}\hspace{-0.04cm}L}}^x\hspace{-0.25cm}
\beta_L(x')^{-1}dx'
+\beta_{0,L}\int_{x}^{x+L}\hspace{-0.4cm}\beta_L(x')^{-1}dx'
-{\frac{_1}{^4}\hspace{-0.04cm}L}\cr
&&\hspace{2.1cm}=\beta_{0,L}\int_{-{\frac{_1}{^4}\hspace{-0.04cm}L}}^x\hspace{-0.25cm}
\beta_L(x')^{-1}dx'+L-\frac{_1}{^4}\hspace{-0.04cm}L
=h_L(x)+L\,,\qquad\label{fLper}\\
&&h_L(-x-\frac{_1}{^2}L)=\beta_{0,L}\int_{-{\frac{_1}{^4}\hspace{-0.04cm}L}}^{-x-\frac{_1}{^2}L}
\hspace{-0.4cm}\beta_L(x')^{-1}dx'
-{\frac{_1}{^4}\hspace{-0.04cm}L}=\beta_{0,L}
\int_{-{\frac{_1}{^4}\hspace{-0.04cm}L}}^{-x-\frac{_1}{^2}L}\hspace{-0.4cm}
\beta_L(-x'-\frac{_1}{^2}L)^{-1}dx'
-{\frac{_1}{^4}\hspace{-0.04cm}L}\cr
&&\hspace{2.1cm}=-\beta_{0,L}\int_{-{\frac{_1}{^4}\hspace{-0.04cm}L}}^{x}
\hspace{-0.25cm}\beta_L(x')^{-1}dx'-{\frac{_1}{^4}\hspace{-0.04cm}L}
=-h_L(x)-\frac{_1}{^2}L\,.
\label{fwif}
\qqq
In particular, $\,h_L\in\Difft$. \,It follows then from (\ref{CG}) and the
transformation rule (\ref{UT+U}) that
\qq
&&U(h_L)\,G_L(t)\,U(h_L)^{-1}\,=\,v
\int_{\CI_L}\hspace{-0.1cm}\beta_L(x^+)\,\Big(h'_L(x)^2\,T_+(h_L(x))
-\frac{_{c}}{^{24\pi}}(Sh_L)(x)\Big)dx\cr
&&=\,\gamma_L
\int_{\CI_L}\hspace{-0.1cm}\beta_L(x^+)\m\beta_L(x)^{-1}T_+(h_L(x))
\,dh_L(x)\,-\,C_{t,L}
\,=\,\int_{\CI_L}\hspace{-0.1cm}\zeta_{t,L}(y)\,T_+(y)\,dy
\,-\,C_{t,L}\,,
\label{UCGU}
\qqq
where we introduced the notation $\,\gamma_L=v\beta_{0,L}\,$ for the
combination of dimension of length that will frequently appear
below and where
\qq
C_{t,L}=\frac{_{cv}}{^{24\pi}}
\int_{\CI_L}\hspace{-0.1cm}\beta_L(x^+)\,(Sh_L)(x)\,dx
\label{CtL}
\qqq
is a number and to obtain the last equality in (\ref{UCGU}) we changed the
variable of integration to $\,y=h_L(x)\,$ setting
\qq
\zeta_{t,L}(y)=\gamma_L\frac{\beta_L(h_L^{-1}(y)+vt)}
{\beta_L(h_L^{-1}(y))}\m.
\label{zetaL}
\qqq
Note for the later use that
\qq
\zeta_{t,L}(y+L)=\zeta_{L,t}(y)\,,\qquad\zeta_{t,L}(-y-\frac{_1}{^2}L)
=\zeta_{-t,L}(y)\,.
\label{fluse}
\qqq
Since $\,\zeta_{0,L}(y)=\gamma_L$, \,it follows,
in particular, that
\qq
U(h_L)\,G_L(0)\,U(h_L)^{-1}&=&\beta_{0,L}H_L\,-\,C_{0,L}
\label{UUL}
\qqq
and that $\,G_L(0)\,$ is a bounded below self-adjoint
operator (and so is $\,G_L(t)\,$ because of (\ref{TxTxm})).
The identity (\ref{UUL}) allows to relate the non-equilibrium to equilibrium
expectation values:
\qq
\big\langle A\big\rangle^{\rm neq}_L\,=\,
\big\langle U(h_L)\,A\hspace{0.07cm}U(h_L)^{-1}
\big\rangle_{\hspace{-0.03cm}\beta_{0,L};L}^{\rm eq}\,.
\label{omegaLg}
\qqq
Although essentially tautological, this relation is the main result
of the present section. It means that the nonequilibrium state
is obtained from the equilibrium one by the composition with
the action of a $\,\Difft\,$ symmetry on observables.
in particular, using (\ref{UT+U}), (\ref{UT-U}) and (\ref{UPhiU}),
we infer from (\ref{omegaLg}) that
\qq
&&\Big\langle\prod\limits_iT_+(x_i)
\,\prod\limits_jT_-(x_j)\Big\rangle^{\rm neq}_L\cr
&&\,=\,
\Big\langle\prod\limits_i\Big(h'_L(x_i)^2\,T_+(h_L(x_i))
-\frac{_c}{^{24\pi}}
(Sh_L)(x_i)\Big)\,\prod\limits_j\Big(h'_L(x_j)^2\,T_-(h_L(x_j))-\frac{_c}{^{24\pi}}
(Sh_L)(x_j)\Big)\Big\rangle_{\hspace{-0.04cm}\beta_{0,L};L}^{\rm eq},
\qquad\label{omegaL1}\\ \cr
&&\Big\langle\prod\limits_i\Phi_i(x_i^-,x^+_i)\Big\rangle^{\rm neq}_L\,=\,
\prod\limits_i\Big(h'_L(x_i^-)^{\Delta^+_\Phi}\,h'_L(x_i^+)^{\Delta^-_\Phi}\Big)
\Big\langle\prod\limits_i\Phi_+\big(h_L(x_i^-),h_L(x_i^+)\big)\Big\rangle_{\hspace{-0.04cm}
\beta_{0,L};L}^{\rm eq}.
\label{omegaL2}
\qqq
Analogous relations were obtained in \cite{GLM} for CFT
in a periodic box.
\vskip 0.4cm

\noindent{\bf Remark.} \ Using more general $L$-periodic inverse-temperature
profiles $\beta_L(x)$ equal to $\beta_\CL$ around $x=-\frac{1}{4}L$ and 
$\beta_\CR$ around $\frac{1}{4}L$, one may similarly obtain expression
for the nonequilibrium expectations with $\,G_L(0)\,$ in (\ref{omegavarL})
replaced by
\qq
v\int_{-{\frac{_1}{^4}\hspace{-0.04cm}L}}^{
{\frac{_1}{^4}\hspace{-0.04cm}L}}\hspace{-0.15cm}
\big(\beta_+(x)\,T_+(x)+\beta_-(x)T_-(x)
\big)\,dx
\qqq
for $\,\beta_+(x)=\beta_L(x)\,$ and $\,\beta_-(x)=\beta_L(-x-\frac{_1}{^2}L)$,
\,i.e. for the right- and left-movers corresponding to different temperature
profiles with the same asymptotics. The boundary conditions (\ref{gluer})
do not admit, however, nonequilibrium states corresponding to different
profiles $\,\beta_\pm(x)\,$ possessing arbitrary asymptotic values that
were discussed in \cite{GLM} using periodic boundary conditions.

\subsection{Thermodynamic limit}
\label{subsec:thlim}

\noindent The thermodynamic limit $\,L\to\infty\,$ of the expectations
(\ref{omegaL1}) may be controlled similarly as in \cite{GLM}.
First, we observe that, for the fixed inverse-temperature profile
(\ref{kink}),
\qq
\beta_{0,L}^{-1}\,=\,2L^{-1}(\Delta\beta^{-1}_{\CL}+\Delta\beta_\CR^{-1})
\,+\,\frac{_1}{^2}(\beta_\CL^{-1}+\beta_\CR^{-1})\,,
\qqq
where
\qq
\Delta\beta^{-1}_\CL=\int_{-\infty}^0\hspace{-0.1cm}
\big(\beta(x)^{-1}-\beta_\CL^{-1}\big)\,dx\,,\qquad
\Delta\beta^{-1}_\CR=\int_0^\infty\hspace{-0.1cm}
\big(\beta(x)^{-1}-\beta_\CR^{-1}\big)\,dx\,.
\qqq
It follows that
\qq
\beta_{0,L}\,=\,\frac{_2}{^{\beta_\CL^{-1}+\beta_\CR^{-1}}}-
\frac{_8}{^{(\beta_\CL^{-1}+\beta_\CR^{-1})^2}}\,(\Delta\beta^{-1}_\CL
+\Delta\beta^{-1}_\CR)\,L^{-1}\ +\ O(L^{-2})\,,
\label{beta0Lbeta0}
\qqq
In particular,
\qq
\lim\limits_{L\to\infty}\,\beta_{0,L}
=\frac{_2}{^{\beta_\CL^{-1}+\beta_\CR^{-1}}}\equiv\beta_0\,.
\label{beta0}
\qqq
Similarly, for fixed $\,x$,
\qq
&&h_L(x)\,=\,\beta_{0,L}\Big(\Delta\beta^{-1}_\CL+\frac{_1}{^4}
\beta_\CL^{-1}{L}\,+\,
\int_0^x\hspace{-0.1cm}\beta(x')^{-1}dx'\Big)\,-\,{\frac{_1}{^4}
\hspace{-0.04cm}L}\cr
&&=\,\frac{_1}{^4}\frac{_{\beta_\CL^{-1}-\beta_\CR^{-1}}}{^{\beta_\CL^{-1}+\beta_\CR^{-1}}}L
+\frac{_{2(\Delta\beta^{-1}_\CL\beta_\CR^{-1}-\Delta\beta^{-1}_\CR\beta_{\CL}^{-1})}}
{^{(\beta^{-1}_\CL+\beta_\CR^{-1})^2}}
+\frac{_{2}}{^{\beta_\CL^{-1}+\beta_\CR^{-1}}}\int_0^x\hspace{-0.1cm}\beta(x')^{-1}dx'
\ +\ O(L^{-1})\,.
\label{hLlim}
\qqq
Since $\,\lim\limits_{L\to\infty}\beta_{0,L}=\beta_0\,$
and $\,\beta_L(x)=\beta(x)\,$ for $\,|x|\leq{\frac{_1}{^4}\hspace{-0.04cm}L}$,
\,it is enough to study the limit of the equilibrium expectations 
\qq
\Big\langle\prod\limits_{i}T_+(y_{i,L}))\,\prod\limits_j
T_-(y_{j,L})\Big\rangle_{\hspace{-0.04cm}\beta_{0,L};L}^{\rm eq}
\qqq
with $\,y_{k,L}=h_L(x_k)\,$ for $\,k=i\,$ or $\,k=j$, and with the products
running over subsets of the original indices. The latter equilibrium expectations are
the amplitudes of the Euclidean CFT on a cylinder of spatial length
$\,L\,$ and temporal
circumference $\,\gamma_L=v\beta_{0,L}$, \,see Fig.\,\ref{fig:zylinder}.
In the dual picture interchanging the roles
of the space and of the Euclidean time, they may be represented 
as matrix elements in the theory on the spatial circle of length
$\,\gamma_L\,$ with the variable $\,\frac{x}{v}\,$ playing the role
of the Euclidean time:

\begin{figure}[t]
\leavevmode
\begin{center}
\vskip -0.3cm
\includegraphics[width=7cm,height=2.3cm]{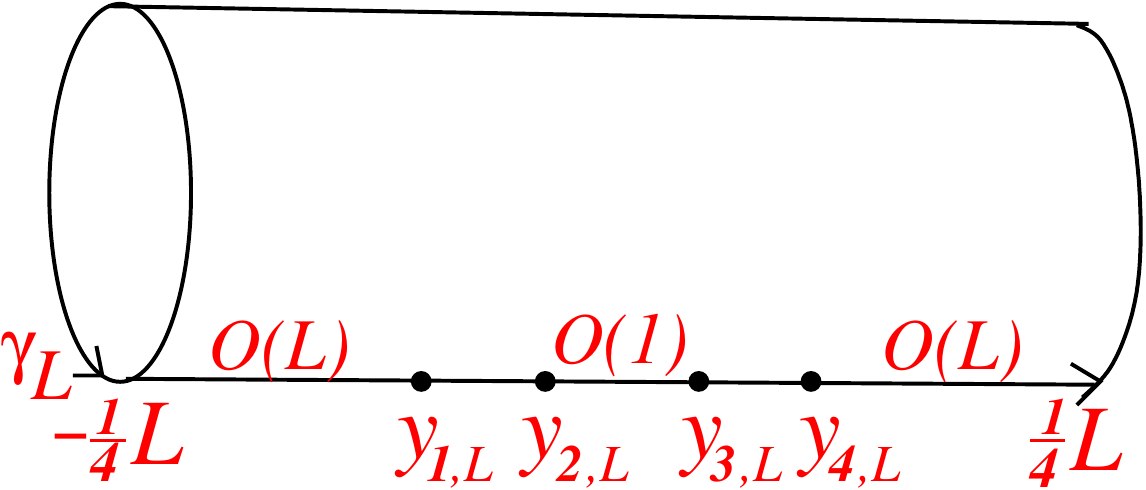}\\
\caption{Euclidian cylinder}
\label{fig:zylinder}
\end{center}
\vskip -0.3cm
\end{figure}

\qq
&&\Big\langle\prod\limits_{i}T_+(y_{i,L})\,\prod\limits_j
T_-(y_{j,L})\Big\rangle^{\rm eq}_{\beta_{0,L};L}\cr
&&=\,\frac{\big\langle\hspace{-0.1cm}
\big\langle B_\CL\big|\hspace{-0.06cm}\big|
\,\ee^{-{\frac{_1}{^{4v}}\hspace{-0.04cm}L}H_{\gamma_L}}\,
\CT\big(\prod\limits_i(-T_+(\ii y_{i,L}))\prod\limits_j(-T_-(-\ii y_{j,L}))
\big)\,\ee^{-{\frac{_1}{^{4v}}\hspace{-0.04cm}L}H_{\gamma_L}}\big|\hspace{-0.06cm}
\big|B_\CR\big\rangle\hspace{-0.1cm}\big\rangle}
{\big\langle\hspace{-0.1cm}\big\langle B_\CL
\big|\hspace{-0.06cm}\big|\,\ee^{-\frac{_1}{^v}LH_{\gamma_L}}\big|\hspace{-0.06cm}\big|
B_\CR\big\rangle\hspace{-0.1cm}\big\rangle},\quad
\label{omega4}
\qqq
where $\,\big|\hspace{-0.055cm}\big|B_\CL\big\rangle
\hspace{-0.1cm}\big\rangle\,$ and
$\,\big|\hspace{-0.055cm}\big|B_\CR\big\rangle\hspace{-0.1cm}\big\rangle\,$ are
the states in the (extended) space of states of the theory
on the circle of circumference $\,\gamma_L\,$
(an appropriate combinations of the so-called Ishibashi states 
\cite{Ishi}) that represent the conformal boundary conditions at the ends
of the interval $[-{\frac{_1}{^4}\hspace{-0.04cm}L},
{\frac{_1}{^4}\hspace{-0.04cm}L}]$, \,the operator $\,H_{\gamma_L}\,$ is
the Hamiltonian of that theory, 
\qq
T_\pm(x\pm\ii v\tau)\,=\,\ee^{\tau H_{\gamma_L}}T_\pm(x)\,\ee^{-\tau H_{\gamma_L}}\,,
\qqq
with $\,T_\pm(x)\,$ standing for the components of
the energy-momentum tensor of the theory on the circle, \,see
Appendix \ref{app:A}, \,and $\,\CT\,$ reorders the factors so that $\,y_{i,L}\,$
and $\,y_{j,L}\,$ increase from right to left.
\vskip 0.2cm

This type of finite-volume equilibrium
expectations is standard in CFT \cite{Cardy} and was discussed in detail
for the compactified free massless field in \cite{Gab} or \cite{GT}. 
In that case, the dual theory on the circle of length $\,\gamma_L\,$
used in the representation (\ref{omega4}) 
has the field with the periodic boundary condition 
$\,\varphi(t,x)=\varphi(t,x+\gamma_L)\,$
leading to independent left- and right-moving components and two commuting
sets of modes $\,\alpha_{n}\,$ and $\,\bar\alpha_{n}\,$ satisfying each
the relations (\ref{alphaalpha}). The zero-mode space
is generated by orthonormal vectors $\,|k,w\rangle\,$
with $\,k,w\in\mathbb Z$, $\,k\,$ corresponding to the momentum and $\,w\,$
the winding number of the fields. \,One has
\qq
&&\frac{_r}{^{\sqrt{2}}}(\alpha_0+\bar\alpha_0)|k,w\rangle=k|k,w\rangle\,,
\quad\frac{_1}{^{\sqrt{2}r}}(\alpha_0-\bar\alpha_0)|k,w\rangle=w
|k,w\rangle\,,\quad
\alpha_n|k,w\rangle=0=\bar\alpha_n|k,w\rangle\ \,{\rm for}\ \,n>0.\qquad\ \ 
\qqq
The dual-theory boundary states representing the Neumann boundary
conditions (\ref{Nbc}) of the original theory have the form \cite{Polch}
\qq
\big|\hspace{-0.055cm}\big|B_{\CL}\big\rangle\hspace{-0.1cm}\big\rangle
\,=\,\ee^{-\sum\limits_{n=1}^\infty\frac{1}{n}\hspace{0.02cm}
\alpha_{-n}\bar\alpha_{-n}}\sum\limits_{w\in\mathbb Z}
|0,w\rangle\,=\,\big|\hspace{-0.05cm}\big|B_{\CR}
\big\rangle\hspace{-0.1cm}\big\rangle.
\qqq
The action of $\,\alpha_{-n}\,$ or $\,\bar\alpha_{-n}\,$
for $\,n>0\,$ raises the eigenvalue of the Hamiltonian $\,H_{\gamma_L}\,$
of the theory by $\,\frac{2\pi n}{\beta_{0,L}}\,$ whereas
$\,H_{\gamma_L}|k,w\rangle=\frac{\pi}{\beta_{0,L}}
(r^{-2}k^2+r^2w^2-\frac{_1}{^6})$. \,The vacuum vector is $\,|0,0\rangle$.
\vskip 0.2cm

Coming back to the general case, let
\qq
\tilde y_{k,L}=h_L(x_k)-\frac{_1}{^4}\frac{_{\beta_\CL^{-1}-\beta_\CR^{-1}}}
{^{\beta_\CL^{-1}+\beta_\CR^{-1}}}L
\qqq
so that
\qq
\lim\limits_{L\to\infty}\,\tilde y_{k,L}=y_0+h(x_k)
\qqq
for $\,y_0\equiv\frac{{2(\Delta\beta^{-1}_\CL\beta_\CR^{-1}
-\Delta\beta^{-1}_\CR\beta_{\CL}^{-1})}}{{(\beta^{-1}_\CL+\beta_\CR^{-1})^2}}\,$
and
\qq
h(x)=\int_0^x\hspace{-0.1cm}\frac{_{\beta_0}}{^{\beta(x')}}\,dx'\m.
\label{h(x)}
\qqq
Note that 
\qq
\frac{_1}{^4}\hspace{-0.04cm}L-\frac{_1}{^4}\frac{_{\beta_\CL^{-1}-\beta_\CR^{-1}}}
{^{\beta_\CL^{-1}+\beta_\CR^{-1}}}L=\frac{_{\beta_\CR^{-1}}}{^{2(\beta_\CL^{-1}+\beta_\CR^{-1})}}L\,,
\qquad
\frac{_1}{^4}\hspace{-0.04cm}L+\frac{_1}{^4}\frac{_{\beta_\CL^{-1}-\beta_\CR^{-1}}}
{^{\beta_\CL^{-1}+\beta_\CR^{-1}}}L=\frac{_{\beta_\CL^{-1}}}{^{2(\beta_\CL^{-1}+\beta_\CR^{-1})}}L\m.
\qqq
If we replace $\,\beta_{0,L}\,$ by its limiting value $\,\beta_0\,$
on the right-hand side of (\ref{omega4}) and denote $\,v\beta_0=\gamma\,$
then
\qq
&&\frac{{\big\langle\hspace{-0.1cm}
\big\langle B_\CL\big|\hspace{-0.06cm}\big|\,\ee^{-{\frac{_1}{^{4v}}\hspace{-0.04cm}L}
H_{\gamma}}\,
\CT\Big(\prod\limits_i(-T_+(\ii y_{i,L}))\prod\limits_j(-T_-(-\ii y_{j,L}))
\Big)\,\ee^{-{\frac{_1}{^{4v}}\hspace{-0.04cm}L}H_{\gamma}}\big|\hspace{-0.06cm}\big|
B_\CR\big\rangle\hspace{-0.1cm}\big\rangle}}
{{\big\langle\hspace{-0.1cm}\big\langle B_\CL
\big|\hspace{-0.06cm}\big|\,\ee^{-\frac{_1}{^{2v}}LH_{\gamma}}
\big|\hspace{-0.06cm}\big|
B_\CR\big\rangle\hspace{-0.1cm}\big\rangle}}\cr
&&=\,\frac{{\big\langle\hspace{-0.1cm}
\big\langle B_\CL\big|\hspace{-0.06cm}\big|
\,\ee^{-\frac{_{\beta_\CR^{-1}}}{^{2(\beta_\CL^{-1}+\beta_\CR^{-1})v}}LH_{\gamma}}\,
\CT\Big(\prod\limits_i(-T_+(\ii\tilde y_{i,L}))
\prod\limits_j(-T_-(-\ii\tilde y_{j,L}))
\Big)\,\m\ee^{-\frac{_{\beta_\CL^{-1}}}{^{2(\beta_\CL^{-1}+\beta_\CR^{-1})v}}LH_{\gamma}}
\big|\hspace{-0.06cm}\big|B_\CR\big\rangle\hspace{-0.1cm}\big\rangle}}
{{\big\langle\hspace{-0.1cm}\big\langle B_\CL
\big|\ee^{-\frac{_1}{^{2v}}LH_{\gamma}}\big|\hspace{-0.06cm}\big|
B_\CR\big\rangle\hspace{-0.1cm}\big\rangle}}\cr\cr\cr
&&\mathop{\longrightarrow}\limits_{L\to\infty}\ 
\big\langle0\big|\,\CT\Big(\prod\limits_i(-T_+(\ii(y_0+y_{i})))
\prod\limits_j(-T_-(-\ii(y_0+y_{j})))
\Big)\big|0\big\rangle\cr\cr
&&\hspace{1cm}=\,\big\langle0\big|\,\CT\Big(\prod\limits_i(-T_+(\ii y_{i}))
\prod\limits_j(-T_-(-\ii y_{j}))
\Big)\big|0\big\rangle\ =\ 
\Big\langle\prod\limits_iT_+(y_i)\prod\limits_jT_-(y_j)
\Big\rangle^{\hspace{-0.05cm}\rm eq}_{\hspace{-0.05cm}\beta_0},
\qqq
where $\,y_k=h(x_k)\,$ for $\,k=i\,$ or $\,k=j\,$ and the expectation
on the right-hand
side is in the infinite-volume Gibbs state at inverse temperature
$\,\beta_0$. \,The limit was obtained using the fact that the leading
contribution to $\,\ee^{-CL\m H_{\gamma}}\big|\hspace{-0.06cm}
\big|B_{\CL,\CR}\big\rangle\hspace{-0.1cm}\big\rangle\,$ for $\,C>0\,$
and large $\,L\,$ comes from the vacuum vector
$\,\big|0\big\rangle\,$ of the theory on the circle of radius $\,\gamma$,
\,with the other contributions exponentially suppressed.
\,Finally, $\,\beta_{0,L}\,$ on the right hand side of (\ref{omega4})
may be turned to $\,\beta_0\,$ by rescaling
the Euclidian space coordinates by $\,C_L\equiv
\gamma/\gamma_L=1+O(L^{-1})$. \,Upon such a rescaling, the theory 
on the circle of length $\,\gamma_L\,$ is mapped to the one
on the circle of length $\,\gamma$, $\,T_\pm(\pm\ii y_{i,L})\,$ to
$\,C_L^2T_\pm(\pm\ii C_Ly_{i,L})$, and the boundary states
$\,\big|\hspace{-0.06cm}\big|B_{\CL,\CR}
\big\rangle\hspace{-0.1cm}\big\rangle\,$ to states with the
similar asymptotic behavior under the action of 
$\,\ee^{-CL\m H_{\gamma}}\,$ and the argument goes as before. 
Dropping the subscript $L$ in the infinite-volume limit of the equilibrium
and nonequilibrium expectations, we obtain from (\ref{omegaL1}) the
identity
\qq
&&\Big\langle\prod\limits_iT_+(x_i)\,
\prod\limits_jT_-(x_j)\Big\rangle^{\hspace{-0.04cm}\rm neq}\cr
&&\,=\,\Big\langle\prod\limits_i
\big(h'(x_i)^2\,T_+(h(x_i))-\frac{_c}{^{24\pi}}(Sh)(x_i)\big)\,
\prod\limits_j\big(h'(x_j)^2\,T_-(h(x_j))-\frac{_c}{^{24\pi}}(Sh)(x_j)\big)
\Big\rangle^{\hspace{-0.01cm}\rm eq}_{\hspace{-0.05cm}\beta_0}\m,\qquad
\label{omegainf1}
\qqq
where $\,\beta_0\,$ and $\,h\,$ are given by (\ref{beta0}) and (\ref{h(x)}).
This is the same infinite-volume relation that was obtained in \cite{GLM}
using in finite volume the periodic boundary conditions.
In particular, one gets
\qq
&&\Big\langle T_\pm(x)\Big\rangle^{\hspace{-0.04cm}\rm neq}=
\frac{{\pi c}}{{12(v\beta(x))^2}}-\frac{c}{{24\pi}}(Sh)(x)\,.
\label{Tinfty}\\
&&\Big\langle T_\pm(x_1)\,;\,T_\pm(x_2)\Big\rangle^{\hspace{-0.05cm}{\rm neq},\,c}\,=\,
\frac{{\pi^2 c}}{8(v\beta(x_1))^2(v\beta(x_2))^2
\sinh^4\big(\frac{\pi}{\gamma}(h(x_1)-h(x_2)\big)}\,,\label{TTinfty}\\
&&\Big\langle T_\pm(x_1)\,;\,T_\mp(x_2)\Big\rangle^{\hspace{-0.05cm}{\rm neq},\,c}\,=\,0\,,
\label{TTTinfty}\qqq
where $\,\big\langle\ \,;\ \big\rangle^{\hspace{-0.04cm}{\rm neq},\,c}$ denotes the
infinite-volume limit of the connected expectations.
\vskip 0.2cm

For (\ref{omegaL2}), the thermodynamic limit reduces again to that
for the equilibrium expectation values. For the equal-time correlations
or if the primary fields are chiral, the latter may be controlled the same
way as above. For non-chiral fields at non-equal times, one should first
analytically continue to imaginary times. For Euclidian points with purely
imaginary times the thermodynamic limit may again be controlled by passing
to the dual picture that leads to the vacuum expectation values. This
should yield the convergence for $\,L\to\infty\,$ of the analytic
continuations to imaginary times of the equilibrium correlators of the
primary fields and, in turn, of their boundary values with real times.
The thermodynamic limit of the latter may also be controlled directly for
the examples of CFTs and their primary fields discussed above. The end
result is the infinite-volume version 
\qq
\Big\langle\prod\limits_i\Phi_i(x_i^-,x^+_i)\Big\rangle^{\hspace{-0.04cm}\rm neq}\,
=\,\prod\limits_i\Big(h'(x_i^-)^{\Delta^+_\Phi}\,h'(x_i^+)^{\Delta^-_\Phi}\Big)
\Big\langle\prod\limits_i\Phi_+(h(x_i^-),h(x_i^+))
\Big\rangle^{\hspace{-0.05cm}\rm eq}_{\hspace{-0.05cm}\beta_0}
\label{omegainf2}
\qqq
of the identity (\ref{omegaL2}) \cite{GLM}.
\vskip 0.3cm

\noindent{\bf Remark.} \,The condition $\,\beta_0=\frac{2}{\beta_\CL^{-1}
+\beta_\CR^{-1}}\,$ in (\ref{h(x)}), (\ref{omegainf1}) and (\ref{omegainf2})
may be dropped using the scaling properties of the infinite-volume equilibrium
state. 
\vskip 0.4cm

\nsection{Full counting statistics of energy transfers: \,preliminaries}
\label{sec:FCSprel}
\subsection{Two-time measurement protocol}
\label{subsec:twotime}

\noindent The aim of the subsequent part of the paper is to describe
the statistics of energy transfers in the nonequilibrium states with
preimposed inverse-temperature profile with a kink as in Fig.\,1.
To access that statistics, we shall follow a two-time quantum
measurement protocol [\onlinecite{LL},\onlinecite{MA},\onlinecite{BD0}].
To this end, we consider in the setup of Sec.\,\ref{sec:Neqstates}
the observables 
\qq
G_L(t)\,=\,
\ee^{\ii t H_L}\,G_L(0)\m\,\ee^{-\ii t H_L}
\qqq
defined by (\ref{CG}) and possessing the spectral
decompositions\footnote{The operators
$\,G_L(t)$, \,that, by (\ref{UUL}), are unitarily equivalent to
$\,\beta_{0,L}H_L+{\rm const}.$, \,have discrete spectrum
with finite multiplicities.}
\qq
G_L(t)\,=\,\sum\limits_i g_i\m P_{i,L}(t)
\qqq
with $\,P_{i,L}(t)=\ee^{\ii tH_L}P_{i,L}(0)\,\ee^{-\ii tH_L}$.
\,If the inverse-temperature profile $\,\beta(x)\,$ is a narrow kink
with constant values
$\,\beta_\CL\,$ and $\,\beta_\CR\,$ to the left and to the right, respectively,
of a small interval $\,[a-\delta,a+\delta]\,$ then
\qq
G_L(t)=\beta_\CL E_\CL(t)+ \beta_\CR E_\CR(t)\,,
\label{approx}
\qqq
where the observable $\,E_\CL(t)\,$ measures the energy in the system
at time $\,t\,$ to the left of the kink and $\,E_\CR(t)\,$ the one
to the right of the kink (redistributing the small energy
contained within the kink appropriately).
\vskip 0.2cm

Suppose that we measure the observable $\,G_L(0)\,$
in the nonequilibrium state $\,\omega^{\rm neq}_{L}\,$ given by
(\ref{omegavarL}). The probability to obtain the result $\,g_i\,$ is
\qq
p_L(g_i)\,=\,\big\langle P_{i,L}(0)\big\rangle^{\rm neq}_L=\frac{\Tr\,\big(
P_{i,L}(0)\,\ee^{-G_L(0)}\big)}
{\Tr\,\big(\ee^{-G_L(0)}\big)}
\qqq
and, after the measurement, \,the nonequilibrium state is reduced to
the new state with expectations
\qq
\big\langle A\big\rangle_{i,L}\,=\,\frac{\Tr\,\big(P_{i,L}(0)\,A\,
P_{i,L}(0)\,\ee^{-G_L(0)}\big)}{\Tr\,\big(P_{i,L}(0)\,\ee^{-G_L(0)}\big)}\,
=\,\frac{\Tr\,\big(A\,P_{i,L}(0)\,\ee^{-G_L(0)}\big)}
{\Tr\,\big(P_{i,L}(0)\,\ee^{-G_L(0)}\big)}\,,
\label{omegai}
\qqq
where the second equality follows from the commutation of $\,P_{i,L}(0)\,$
with $\,G_L(0)$. \,We now let this state evolve for time $\,t\,$ and
then measure the observable $\,G_L(0)\,$ for the second
time. This is equivalent to the measurement of the observable
$\,G_L(t)\,$ in the state with expectations given by (\ref{omegai}).
\,The probability that the result of the second measurement is $\,g_j$,
under the condition that the first one was $\,g_i$, \,is then equal to
\qq
p_{t,L}(g_j|\,g_i)\,=\,\big\langle P_{j,L}(t)\big\rangle_{i,L}\,
=\,\frac{\Tr\,\big(P_{j,L}(t)\m P_{i,L}(0)\,\ee^{-G_L(0)}\big)}
{\Tr\,\big(P_{i,L}(0)\,\ee^{-G_L(0)}\big)}\,.
\qqq
Hence the probability to get the results $\,(g_i,g_j)\,$ in the two-time
measurement of $\,G_L(0)\,$ separated by time $\,t\,$ in the 
nonequilibrium state with expectation values (\ref{omegavarL}) is
\qq
p_{t,L}(g_i,g_j)\,=\,p_L(g_i)\,p_{t,L}(g_i|g_j)\,=\,
\frac{\Tr\,\big(P_{j,L}(t)\m P_{i,L}(0)\,\ee^{-G_L(0)}\big)}{\Tr\,
\big(\ee^{-G_L(0)}\big)}\,=\,\big\langle
P_{j,L}(t)\m P_{i,L}(0)\big\rangle^{\rm neq}_L\,.
\qqq

\subsection{The generating function for FCS of energy transfers}
\label{subsec:gfFCS}

\noindent We shall consider
\qq
\Delta E=\frac{g_j-g_i}{\Delta\beta}\,,
\label{Deltae}
\qqq
where $\,\Delta\beta\equiv \beta_\CR-\beta_\CL$,
\,as a measure of the net energy transfer through the kink during time
$\,t$. \,Indeed, the energy on the right of the kink should change
in time $\,t\,$ by $\,\Delta E\,$ and on the left of the
kink by $\,-\Delta E\,$ so that, by (\ref{approx}), $\,g_j-g_i\,$
should be equal to $\,\beta_\CR\Delta E-\beta_\CL\Delta E$.
Upon the identification (\ref{Deltae}), the PDF of $\,\Delta E$,
\,called Full Counting Statistics (FCS) \cite{QN} of energy transfers,
\,takes the form 
\qq
P_{t,L}(\Delta E)\,=\,\sum\limits_{i,j}\,\delta\Big(\Delta E-\frac{g_j-g_i}
{\Delta\beta}\Big)\,p_{t,L}(g_i,g_j)
\qqq
and its Fourier transform (called the FCS generating function)
becomes
\qq
\Psi_{t,L}(\lambda)&=&\int\ee^{\ii\lambda\Delta E}\,P_{t,L}(\Delta E)\,\,d(\Delta E)
\,=\,\sum\limits_{i,j}\ee^{\frac{\ii\lambda(g_j-g_i)}{\Delta\beta}}\,p_{t,L}(g_i,g_j)\cr
&=&\sum\limits_{i,j}\ee^{\frac{\ii\lambda(g_j-g_i)}{\Delta\beta}}\,\big\langle
P_{j,L}(t)\m P_{i,L}(0)\big\rangle^{\rm neq}_L
\,=\,\big\langle\ee^{\frac{\ii\lambda}{\Delta\beta}G_L(t)}\,
\ee^{-\frac{\ii\lambda}{\Delta\beta}G_L(0)}\big\rangle^{\rm neq}_L\,.
\label{F}
\qqq
In particular, the average of the energy transfer
\qq
\langle\Delta E\rangle_{t,L}\,=\,\frac{_1}{^{\ii}}\,
\partial_\lambda\ln{\Psi_{t,L}(0)}
\,=\,(\Delta\beta)^{-1}\,\big\langle G_L(t)-G_L(0)\big\rangle^{\rm neq}_{L}
\label{1stmom}
\qqq
and its variance
\qq
\langle\Delta E;\Delta E\rangle^c_{t,L}\,=\,-\partial_\lambda^2\m\ln{\Psi_{t,L}(0)}
\,=\,(\Delta\beta)^{-2}\,\big\langle G_L(t)-G_L(0);G_L(t)-G_L(0)
\big\rangle^{{\rm neq},\,c}_{L}\,.
\label{2ndmom}
\qqq
\vskip 0.2cm

$\,\Psi_{t,L}(\lambda)\,$ extends to an analytic function
in the interior of the strip $\,0\leq{\rm Im}(\lambda)\leq\Delta\beta\,$
if $\,\Delta\beta>0\,$ and $\,\Delta\beta\leq{\rm Im}(\lambda)\leq 0\,$
if $\,\Delta\beta<0$, \,with the boundary value
\qq
\Psi_{t,L}(-\lambda+\ii\Delta\beta)&=&
\frac{\Tr\,\big(\ee^{-\frac{\ii\lambda+\Delta\beta}{\Delta\beta}G_L(t)}
\,\ee^{\frac{\ii\lambda+\Delta\beta}
{\Delta\beta}G_L(0)}\,\ee^{-G_L(0)}\big)}
{\Tr\,\big(\ee^{-G_L(0)}\big)}\,=\,\frac{\Tr\,\big(\ee^{-G_L(t)}
\ee^{-\frac{\ii\lambda}{\Delta\beta}G_L(t)}\ee^{\frac{\ii\lambda}
{\Delta\beta}G_L(0)}\big)}{\Tr\,\big(\ee^{-G_L(0)}\big)}\cr
&&\hspace{-1.4cm}=\,
\frac{\Tr\,\big(\ee^{-G_L(0)}\ee^{-\frac{\ii\lambda}{\Delta\beta}G_L(0)}\ee^{\frac{\ii\lambda}
{\Delta\beta}G_L(-t)}\big)}{\Tr\,\big(\ee^{-G_L(0)}\big)}\,=\,
\frac{\Tr\,\big(\ee^{\frac{\ii\lambda}
{\Delta\beta}G_L(-t)}\,\ee^{-\frac{\ii\lambda}{\Delta\beta}G_L(0)}\,\ee^{-G_L(0)}\big)}
{\Tr\,\big(\ee^{-G_L(0)}\big)}\,=\,\Psi_{-t,L}(\lambda)\,.\qquad
\qqq
The above identity is a ``fluctuation relation'' between the FCS generating
functions for the direct and the time reversed dynamics.
If the CFT is time-reversal invariant so that 
there exists an anti-unitary involution or anti-involution $\,\Theta\,$
such that $\,\Theta\hspace{0.01cm}T_\pm(x)\Theta^{-1}=T_{\mp}(x)\,$
and, \,consequently, $\,\Theta\hspace{0.03cm}G_L(t)\Theta^{-1}=G_L(-t)\,$
then
\qq
\Psi_{-t,L}(\lambda)&=&
\frac{\Tr\,\big(\Theta\hspace{0.05cm}\ee^{-\frac{\ii\lambda}{\Delta\beta}G_L(t)}
\,\ee^{\frac{\ii\lambda}{\Delta\beta}G_L(0)}\,\ee^{-G_L(0)}\,
\Theta^{-1}\big)}
{\Tr\,\big(\ee^{-G_L(0)}\big)}\,=\,
\frac{\overline{\Tr\,\big(\ee^{-\frac{\ii\lambda}{\Delta\beta}G_L(t)}\,
\ee^{\frac{\ii\lambda}{\Delta\beta}G_L(0)}\,\ee^{-G_L(0)}\big)}}
{\Tr\,\big(\ee^{-G_L(0)}\big)}\cr
&=&\frac{\Tr\,\big(\ee^{-\frac{\ii\lambda}{\Delta\beta}G_L(t)}\,
\ee^{\frac{\ii\lambda}{\Delta\beta}G_L(0)}\,\ee^{-G_L(0)}
\big)^{\hspace{-0.05cm}\dagger}}{\Tr\,\big(\ee^{-G_L(0)}\big)}\,
=\,\frac{\Tr\,\big(\ee^{-G_L(0)}\,\ee^{-\frac{\ii\lambda}{\Delta\beta}G_L(0)}
\,\ee^{\frac{\ii\lambda}{\Delta\beta}
G_L(t)}\big)}
{\Tr\,\big(\ee^{-G_L(0)}\big)}\cr
&=&\frac{\Tr\,\big(\ee^{\frac{\ii\lambda}{\Delta\beta}
G_L(t)}\,\ee^{-\frac{\ii\lambda}{\Delta\beta}G_L(0)}\,\ee^{-G_L(0)}\big)}
{\Tr\,\big(\ee^{-G_L(0)}\big)}\,=\,\Psi_{t,L}(\lambda)
\qqq
and we infer the ``transient'' fluctuation relation for the
generating function $\,\Psi_{t,L}(\lambda)\,$
[\onlinecite{ES},\onlinecite{Kur},\onlinecite{BD1}]:
\qq
\Psi_{t,L}(\lambda)\,=\,\Psi_{t,L}(-\lambda+\ii\Delta\beta)
\label{CFR}
\qqq
In fact, in the Hilbert of the boundary theory such that
$\,T_{+}(x^-)=T_{-}(x^+)\,$ for $\,x=\pm\frac{1}{2}\hspace{-0.02cm}L$, \,there
always exists an anti-unitary involution $\,\Theta\,$
that does the job: \,it is sufficient to take $\,\Theta\,$ that preserves
the highest-weight vectors of the unitary irreducible representations
of the Virasoro algebra and that commutes with the Virasoro generators
$\,L_n$. \,Hence the transient fluctuation relation is always
valid in our case.

\nsection{FCS \,in finite volume and conformal welding}
\label{sec:FCSconfweld}
\subsection{FCS \,and $\,\Difft\,$ characters}
\label{subsec:FCSchar}

\noindent Using the relation (\ref{omegaLg}) between the nonequilibrium
and equilibrium expectations and the transformation
identity (\ref{UCGU}), we may rewrite the expression (\ref{F})
for the generating function for FCS of energy transfers in
the form
\qq
\Psi_{t,L}(\lambda)&=&\big\langle U(h_L)\,\m
\ee^{\frac{\ii\lambda}{\Delta\beta}G_L(t)}\,
\ee^{-\frac{\ii\lambda}{\Delta\beta}
G_L(0)}\hspace{0.03cm}U(h_L)^{-1}
\big\rangle^{\rm eq}_{\beta_{0,L};L}\cr\cr
&=&\big\langle\ee^{\frac{\ii\lambda}{\Delta\beta}
\int_{_{\CI_L}}\hspace{-0.1cm}\zeta_{t,L}(y)\,T(y)\,dy}\,
\ee^{-\frac{\ii\lambda}{\Delta\beta}\beta_{0,L}H_L}\,
\big\rangle^{\rm eq}_{\beta_{0,L};L}\,\ee^{-\frac{\ii\lambda}{\Delta\beta}(C_{t,L}-C_{0,L})}.
\label{FCSflat}
\qqq
In virtue of \m(\ref{HL}) and (\ref{exponen}),
\qq
&&\big\langle
\ee^{\frac{\ii\lambda}{\Delta\beta}
\int_{_{\CI_L}}\hspace{-0.1cm}\zeta_{t,L}(y)\,T(y)\,dy}\,
\ee^{-\frac{\ii\lambda}{\Delta\beta}\beta_{0,L}H_L}
\big\rangle^{\rm eq}_{\beta_{0,L};L}\cr
&&=\frac{\Tr\,\big(\ee^{\ii\frac{\lambda}{\Delta\beta}
\int_{_{\CI_L}}\hspace{-0.1cm}\zeta_{t,L}(y)
\,T(y)\,dy}\,\ee^{-\frac{2\pi\gamma_L}{L}
(1+\ii\frac{\lambda}{\Delta\beta})(L_0-\frac{c}{24})}\big)}
{\Tr\,\big(\ee^{-\frac{2\pi\gamma_L}{L}
(L_0-\frac{c}{24})}\big)}\,=\,
\frac{\Tr\,\big(U(f_{s,t,L})\,\m\ee^{2\pi\ii\tau_{s,L}(L_0-\frac{c}{24})}\big)}
{c_{s,\zeta_{t,L}}\,\Tr\,\big(\ee^{2\pi\ii\tau_{0,L}(L_0-\frac{c}{24})}\big)}
\label{charact0}
\qqq
where, as before, $\,\gamma_L=v\beta_{0,L}\,$ and
\qq
s=\frac{_\lambda}{^{\Delta\beta}}\,,\qquad\tau_{s,L}
=L^{-1}(\ii-s)\gamma_L
\label{stau}
\qqq
and $\,f_{s,t,L}\in\Difft\,$ describe the flow
of the vector field $\,-\zeta_{t,L}(y)\partial_y$. 
\,Note that the relation (\ref{fluse}) implies that
\qq
f_{s,t,L}(-y-\frac{_1}{^2}L)=-f_{-s,-t,L}(y)-\frac{_1}{^2}L\,.
\label{fluse2}
\qqq
The function
\qq
\chi(\tau)\,=\,\Tr\big(\ee^{2\pi\ii\tau(L_0-\frac{c}{24})}\big)
\label{Virchar}
\qqq
of $\,\tau\,$ from the complex upper half-plane $\,\mathbb H_+$,
\,where the trace is over the space of a Virasoro algebra
representation, \,defines the character of the latter. Such
characters are explicitly known for the (direct sums of) irreducible
highest-weight representations of the Virasoro algebra \cite{RC}.
By analogy, we shall call the function
\qq
\Upsilon(f,\tau)\,=\,\Tr\big(U(f)\,\ee^{2\pi\ii\tau(L_0-\frac{c}{24})}\big)
\label{Diffchar}
\qqq
of $\,f\in\Difft\,$ and $\,\tau\in\mathbb H_+\,$ the character
of the projective representation $\,U\,$ of $\,\Difft$.
\,We may then rewrite the relation (\ref{FCSflat}) as
\qq
\Psi_{t,L}(\lambda)\,=\,\frac{\Upsilon(f_{s,t,L}\m,\tau_{s,L})}
{c_{s,\zeta_{t,L}}\,\chi(\tau_{0,L})}\,\,\ee^{-\ii s\m(C_{t,L}-C_{0,L})}\,,
\label{PsichiUpsilon}
\qqq
where the characters pertain to the representation of the Virasoro
algebra in the space $\,\CH_L\,$ of states of the boundary CFT
and to its corresponding lift to $\,\Difft$.
\vskip 0.2cm

\,Unlike the Virasoro characters, the characters $\,\Upsilon(f,\tau)\,$
where not studied in detail and we shall try to fill this gap in
the next section for the special case that occurs on the
right hand side of (\ref{PsichiUpsilon}) for which $\,f=f_s\,$ belongs to
the flow of a vector field.

\subsection{Reduction of $\,\Difft\,$ characters to Virasoro ones}
\label{subsec:reduct}

\begin{figure}[b]
\begin{center}
\vskip -0.1cm
\leavevmode
\hspace*{-1.6cm}
{%
      \begin{minipage}{0.4\textwidth}
      \vspace{-0.05cm}          
        \includegraphics[width=3.3cm,height=3.1cm]
        {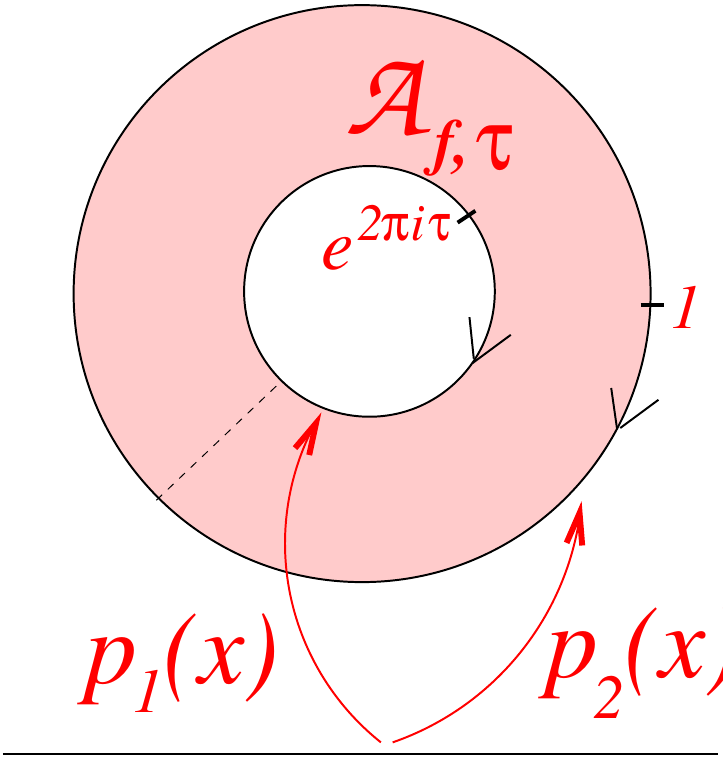}\\
      \vspace{-0.9cm} \strut
        \end{minipage}}\hspace*{-0.2cm}
{%
  \begin{minipage}{0.4\textwidth}
      \vspace{0.2cm}
        \includegraphics[width=3.2cm,height=3cm]
         {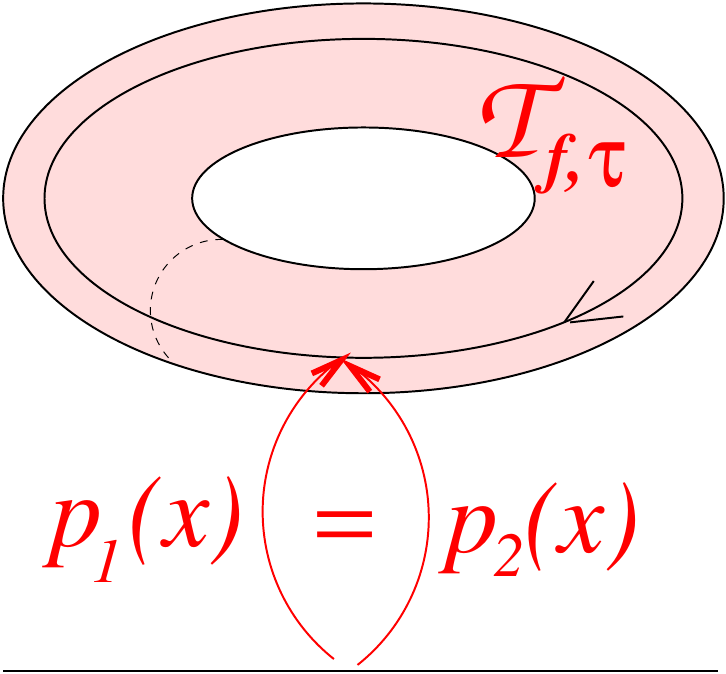}\\
         \vspace{-0.9cm} \strut
         \end{minipage}}\hspace*{-0.8cm}
         \vspace{0.55cm}
\caption{Annulus $\,\CA_{f,\tau}\,$ and torus $\,\CT_{f,\tau}$}
\label{fig:at}
\end{center}
\vskip -0.3cm
\end{figure}

\noindent The Virasoro character $\,\chi(\tau)\,$ of (\ref{Virchar}) is 
given by the trace of the operator $\,\ee^{2\pi\ii\tau(L_0-\frac{c}{24})}$.
\,If $\,\tau\,$ were real, such an operator would represent (up to a phase) a translation in the Cartan subgroup of $\,\Difft$, \,see
(\ref{onCart}). Instead, $\,\tau\,$ is taken with a positive imaginary
part in order to assure the convergence of the trace. In the simpler
context of finite-dimensional representations of compact groups, it is
enough to know the characters on the Cartan subgroup, \m as they
are class functions that are constant on conjugacy classes and each
conjugacy class contains an element in the Cartan subgroup.
When dealing with the characters $\,\Upsilon(f,\tau)\,$ of $\,\Difft$,
\,however, \,there
are complications. The first one is due to the insertion of the operator
$\,\ee^{2\pi\ii\tau(L_0-\frac{c}{24})}\,$ that represents an element in the
complexification of the Cartan subgroup of $\,\Difft\,$
and assures the convergence of the trace. The second complication
is due to the projectivity of the representation of $\,\Difft$.
Finally, the conjugacy classes in $\,\Difft\,$ rarely contain
elements in the Cartan subgroup. All these difficulties find an elegant
resolution in the setup advocated by G. Segal \cite{Segal} where
one replaces $\,\Diff\,$ by a semigroup of complex
annuli with parameterized boundaries and one studies its
projective representations. Within this approach, the operator
$\,U(f)\,\ee^{2\pi\ii\tau(L_0-\frac{c}{24})}\,$ is proportional to
the trace-class operator (the chiral amplitude) representing the
complex annulus
\qq
\CA_{f,\tau}=\big\{z\in\mathbb C\,\big|\,|q|\leq|z|\leq 1\big\}
\qqq
for $\,q=\ee^{2\pi\ii\tau}\,$ with the boundary components
parameterized by
\qq
p_1(x)=q\,\ee^{-\frac{2\pi\ii}{L}f(x)}=p_1(x+L)\,,\qquad p_2(x)
=\ee^{-\frac{2\pi\ii}{L}x}=p_2(x+L)\,,
\label{parametr}
\qqq
see Fig.\,\ref{fig:at}. The precise normalization of the chiral
amplitudes that allows to handle the projectivity of the representations
of the semigroup of annuli is fixed in \cite{Segal} using the theory
of determinant line bundles for Riemann surfaces with boundary and
we shall not dwell on it here. Within Segal's approach,
the counterpart of the ``class-function''property of the compact-group
characters is the fact that the trace of the chiral amplitude
associated to a complex annulus depends, up to a controllable
factor, only on the complex torus obtained from the annulus by 
``conformal welding'' that identifies the two boundary components
using their parameterizations. The complex structure of such a welded
torus is fixed by defining the local holomorphic functions on it as those
whose pullback to the annulus is smooth and holomorphic in the interior.

\vskip 0.2cm

In particular, identifying the boundaries of the annulus $\,\CA_{f,\tau}\,$
by setting $\,p_1(x)=p_2(x)$, \,one obtains a complex torus
$\,\CT_{f,\tau}$. \,The torus $\,\CT_{f,\tau}\,$ comes with a natural marking
$\,([a],[b])$, where
$\,[a]\,$ and $\,[b]\,$ are homology classes of 1-cycles $\,a\,$
and $\,b\,$ with intersection number 1 given by the curves
\qq
\CI_{_{\hspace{-0.01cm}L}}\ni x\,\mathop{\longmapsto}\limits^a\,\ p_1(x)=p_2(x)
\qquad\text{and}\qquad[0,1]\ni s\,\mathop{\longmapsto}\limits^b\,
\ee^{(1-s)(2\pi\ii\tau-\frac{2\pi\ii}{L}f(0))},
\label{cycles}
\qqq
\,respectively. The marked complex tori have the upper half-plane
$\,\mathbb H_+\,$ as the moduli space. This means that $\,\CT_{f,\tau}\,$ 
must be isomorphic to a more standard complex torus
$\,\CT_{f_0,\widehat\tau}\,$
with $\,f_0(x)\equiv x\,$ for certain $\,\widehat\tau\in\mathbb H_+\,$ that
is unique if we demand that the isomorphism $\,\CT_{f,\tau}
\cong\CT_{f_0,\widehat\tau}\,$ intertwines the markings. Note that the torus
$\,\CT_{f_0,\widehat\tau}\,$ may be also viewed as obtained by identifying
$\,0\not=z\in\mathbb C\,$ with $\,\widehat q\m z\,$ for
$\,\widehat q=\ee^{2\pi\ii\widehat\tau}$.
\vskip 0.2cm

Since $\,U(f)\,\ee^{2\pi\ii\tau(L_0-\frac{c}{24})}\,$ is proportional to the chiral
amplitude of $\,\CA_{f,\tau}\,$ and $\,\ee^{2\pi\ii\widehat\tau(L_0-\frac{c}{24})}\,$
to the chiral amplitude of $\,\CA_{f_0,\widehat\tau}$, \,with controllable
proportionality constants, \,and the traces of those amplitudes differ
by a controllable factor, \,it follows that we must have
the relation
\qq
\Upsilon(f,\tau)\,=\,C_{f,\tau}\,\chi(\widehat\tau)
\qqq
with a controllable coefficient $\,C_{f,\tau}$. \,Indeed, a closer
examination of Segal's theory shows \cite{KGr} that $\,C_{f,\tau}\,$
may be expressed by explicit Fredholm determinants and the vacuum
matrix element $\,\langle 0|U(f)|0\rangle$, \,where $\,|0\rangle\,$
is the highest-weight vector in the vacuum representation of
the Virasoro algebra.
\vskip 0.2cm

The matrix element $\,\langle 0|U(f_s)|0\rangle\,$
for the flow $\,f_s\,$ of a vector field, where $\,U(f_s)\,$ is given
by (\ref{exponen}), was studied extensively in the context
of ``energy inequalities'', see [\onlinecite{FH1},\onlinecite{FFR}].
For free fields, it may be expressed in terms of Fredholm determinants,
see e.g. \cite{BrDer}, and for other CFTs the free-field expression
should be raised to the power given by the central charge $\,c\m$.
Recently, C. J. Fewster and S. Hollands derived in \cite{FH2} an integral
expression for $\,\langle 0|U(f_s)|0\rangle\,$ using 
conformal welding of discs \cite{ShM}. Employing in a similar vein
conformal welding of tori
$\,\CT_{f,\tau}$, \,we shall obtain below an integral expression without
Fredholm determinants for the proportionality constant $\,C_{f_s,\tau}$.
Its derivation is the subject of the three subsequent subsections.

\subsection{Conformal welding of tori $\,\CT_{f,\tau}\,$ and an
inhomogeneous Riemann-Hilbert problem}
\label{subsec:confweldRH}

\noindent We shall need more information about the isomorphism
$\,\CT_{f,\tau}\cong\CT_{f_0,\widehat\tau}\,$ of the complex tori. In the
first step to construct such an isomorphism we shall look for a continuous
function
\qq
Y:\CA_{f,\tau}\longmapsto\mathbb C
\qqq
holomorphic in the interior of $\,\CA_{f,\tau}\,$ and possessing the boundary
values $\,Y_i=Y\circ p_i\,$ that satisfy the relation
\qq
Y_{12}\equiv Y_1-Y_2=f-f_0+L(\widehat\tau-\tau)\,.
\label{Y12}
\qqq

\begin{figure}[ht]
\leavevmode
\begin{center}
\vskip -0.3cm
\hspace*{-0.1cm}
\includegraphics[width=8cm,height=1.85cm]{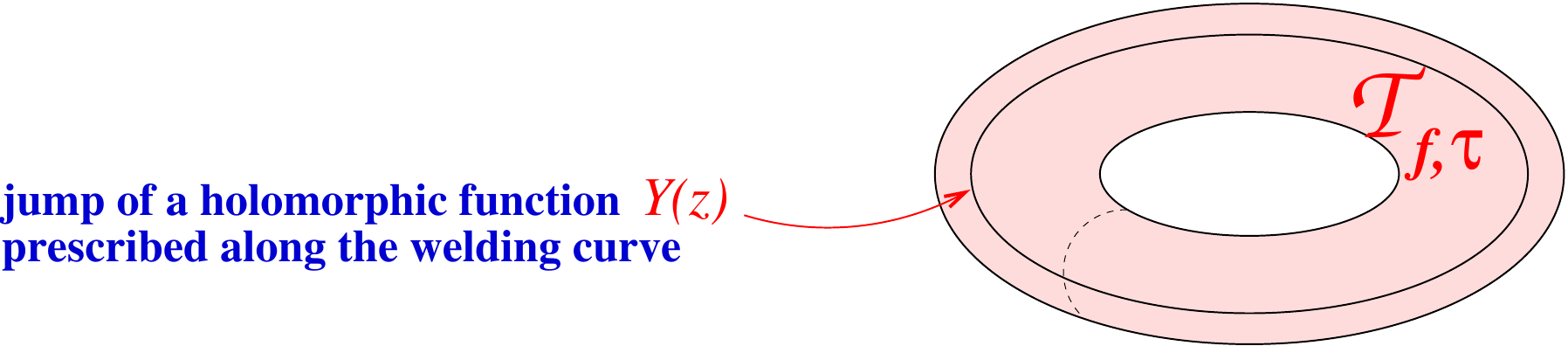}\\
\end{center}
\vskip -0.15cm
\caption{Inhomogeneous Riemann-Hilbert problem on $\,\CT_{f,\tau}$}
\label{fig:inRH}
\end{figure}

\noindent This may be viewed as an inhomogeneous Riemann-Hilbert problem
\cite{RH} on $\,\CT_{f,\tau}\,$ searching for a holomorphic function with
a prescribed jump along the $\,a$-curve of $\,\CT_{f,\tau}\,$ which was
obtained by welding the two edges of $\,\CA_{f,\tau}$, \,see
Fig.\,\ref{fig:inRH}. \,As we shall see below, this problem has a solution
$\,Y\,$ for a unique $\,\widehat\tau\in\mathbb H_+$,
\,assuming that $\,Y_i\in L^2(S^1_L)$. \,Besides
such a solution is unique up to an additive constant, and $\,Y_i\,$ are
automatically smooth. Given $\,Y$, \,the function
\qq
W(z)=z\,\ee^{\m\frac{2\pi\ii}{L}Y(z)}
\label{CW}
\qqq
on $\,\CA_{f,\tau}\,$ has the boundary values $\,W_i=W\circ p_i\,$
that satisfy
\qq
W_1=\widehat q\,W_2
\label{W1W2}
\qqq
so that it defines a holomorphic map from $\,\CT_{f,\tau}\,$ to
$\,\mathbb C^\times\hspace{-0.03cm}\big/(w\sim\widehat q^{\pm1} w)
\cong\CT_{f_0,\widehat\tau}$. That map provides the isomorphism $\,\CT_{f,\tau}
\cong\CT_{f_0,\widehat\tau}$.
\vskip 0.2cm

The solution of the inhomogeneous
Riemann-Hilbert problem searching for the function $\,Y\,$ follows
the standard strategy \cite{RH}. In the interior of $\,\CA_{f,\tau}$,
$\,Y(z)\,$ is expressed in terms of its boundary values $\,Y_i\,$ via
the Cauchy formula. Taking the limits $\,z\to p_i(x)\,$ in the latter
and using (\ref{Y12}), \,one obtains a Fredholm equation for, say, $\,Y_1$.
\,Its solution, together with $\,Y_2\,$ obtained from (\ref{Y12}),
\,and the Cauchy formula determine $\,Y$.
\vskip 0.2cm

Here are some more details. Let us first assume that $\,Y_i\,$ are
smooth. The Cauchy formula for $\,Y(z)\,$ with $\,z\,$ in the
interior of $\,\CA_{f,\tau}\,$ reads:
\qq
Y(z)\,=\,\frac{_1}{^{2\pi\ii}}\int_{\CI_L}\hspace{-0.1cm}
\Big(\frac{Y_1(y)\,dp_1(y)}{p_1(y)-z}-\frac{Y_2(y)\,dp_2(y)}
{p_2(y)-z}\Big)\,.
\label{CY}
\qqq
By sending $\,z\,$ to $\,p_i(x)$, \,one obtains
for the boundary values of $\,Y\,$ the relations
\qq
\frac{_1}{^2}Y_1(x)=\frac{_1}{^{2\pi\ii}}\,
PV\hspace{-0.1cm}\int_{\CI_L}\frac{Y_1(y)\,dp_1(y)}
{p_1(y)-p_1(x)}\,-\,\frac{_1}{^{2\pi\ii}}
\int_{\CI_L}\frac{Y_2(y)\,dp_2(y)}{p_2(y)-p_1(x)}
\,,\label{CY1=}\\
\frac{_1}{^2}Y_2(x)=\frac{_1}{^{2\pi\ii}}
\int_{\CI_L}\frac{Y_1(y)\,dp_1(y)}
{p_1(y)-p_2(x)}\,-\,\frac{_1}{^{2\pi\ii}}
\,PV\hspace{-0.1cm}\int_{\CI_L}\frac{Y_2(y)\,dp_2(y)}{p_2(y)-p_2(x)}
\,,\label{CY2=}
\qqq
where $\,PV\,$ stands for ``principal value''. We shall identify
$L$-periodic functions on $\,\mathbb R\,$
with functions on the circle $\,\R/L\Z=S^1_L\,$
and shall denote by $\,E_\pm\,$ the orthogonal projections
in $\,L^2(S^1_L)\,$ on the subspace
spanned by functions $\,e_n(x)=\frac{1}{\sqrt L}\ee^{-\ii p_nx}\,$
with $\,n>0\,$ and $\,n<0$, \,respectively,
for $\,p_n\equiv\frac{2\pi n}{L}$. \m Similarly, we shall denote by
$\,E_{0\pm}\,$ the orthogonal projections corresponding to $\,n\geq0\,$
and $\,n\leq0$. \,For a smooth $\,L$-periodic function $\,X$,
one has the relation
\qq
\frac{_1}{^2}\big((E_{0+}-E_-)X\big)(x)\,=\,-\frac{_1}{^{2\pi\ii}}\,
PV\hspace{-0.1cm}\int_{\CI_L}\frac{X(y)\,dp_2(y)}{p_2(y)-p_2(x)}\,.
\label{Pgeq<}
\qqq
Adding (\ref{Pgeq<}) for $\,X=Y_1\,$ to (\ref{CY1=}) and subtracting
it for $\,X=Y_2\,$ from (\ref{CY2=}), \,we obtain the identities:
\qq
E_{0+}Y_1=K_{11}Y_1+K_{12}Y_2\,,\qquad E_-Y_2=K_{21}Y_1\,,
\label{Pgeq<12}
\qqq
where 
\qq
&&(K_{11}X)(x)\,=\,\frac{_1}{^{2\pi\ii}}\int_{\CI_L}\hspace{-0.1cm}X(y)
\Big(\frac{dp_1(y)}{p_1(y)-p_1(x)}-\frac{dp_2(y)}{p_2(y)-p_2(x)}\Big),
\label{K11}\\
&&(K_{12}X)(x)\,=\,-\frac{_1}{^{2\pi\ii}}\int_{\CI_L}\hspace{-0.1cm}X(y)
\frac{dp_2(y)}{p_2(y)-p_1(x)}\,,\label{K12}\\
&&(K_{21}X)(x)\,=\,\frac{_1}{^{2\pi\ii}}\int_{\CI_L}\hspace{-0.1cm}X(y)
\frac{dp_1(y)}{p_1(y)-p_2(x)}\label{K21}
\qqq
are trace-class \cite{Simon} operators on $\,L^2(S^1_L)\,$ as they have smooth
kernels and $\,S^1_L\,$ is a compact manifold \cite{Sugi,Brisl}.
\m We also have the representations:
\qq
K_{11}=E_{0+}-F^{-1}E_{0+}F=-E_-+F^{-1}E_-F\,,\qquad K_{12}=F^{-1}QE_{0+}\,,
\qquad K_{21}=E_-Q^{-1}F\,,
\label{FQij}
\qqq
where $\,(FX)(x)=X(f^{-1}(x))\,$ and $\,Qe_n=q^ne_n$.
\,By a limiting argument,
the relations (\ref{Pgeq<12}) also hold if $\,Y_i\in L^2(S^1_L)$. 
\,Summing the equations (\ref{Pgeq<12}), substituting $\,Y_2=Y_1-Y_{12}\,$
and denoting
\qq
K=K_{11}+K_{12}+K_{21}\,,
\label{Kplus}
\qqq
we obtain the identity
\qq
(I-K)Y_1=(E_--K_{12})Y_{12}
\label{Freq}
\qqq
which is a Fredholm equation for $\,Y_1$, \,given $\,Y_{12}$.
\,It appears that, \,conversely, \,if
$\,Y_1\in L^2(S^1_L)\,$ is
a solution of (\ref{Freq}) for some $\,Y_{12}\in
L^2(S^1_L)\,$
then there exists a holomorphic function $\,Y\,$ on $\,\CA_{f,\tau}\,$
for which $\,Y_1\,$ and $\,Y_2=Y_1-Y_{12}\,$ are its boundary values.
Besides, if $\,Y_{12}\,$ is smooth then so is $\,(E_--K_{12})Y_{12}\,$
and $\,KY_1\,$ so that (\ref{Freq}) implies that $\,Y_1\,$ is also
smooth.
\vskip 0.2cm

The Fredholm operator $\,I-K\,$ has only constants in its
kernel. Indeed, if $\,(I-K)X_1=0\,$ for some $\,X_1\in L^2(S^1_L)\,$
then there exists a function $\,X\,$ holomorphic on
$\,\CA_{f,\tau}\,$ with smooth boundary values $\,X_1\,$ and $\,X_2=X_1$.
\,Such $\,X\,$ defines a holomorphic function on the torus
$\,\CT_{f,\tau}\,$ and must be constant. Since the index of the Fredholm
operator $\,I-K\,$ vanishes, \,it follows that the image of $\,I-K\,$
is of codimension one. The solubility of (\ref{Freq}) for $\,Y_1$,
\,given $\,Y_{12}$, \,requires a fine tuning of the constant contribution
to $\,Y_{12}$. \,Let $\,\omega\,$ be a holomorphic $\,1$-form on
$\,\CT_{f,\tau}\,$ fixed by the normalization condition
\qq
\int_a\omega=1\,.
\label{a}
\qqq
We shall identify $\,\omega\,$ with its pullback to $\,\CA_{f,\tau}\,$
that satisfies the relation $\,p_1^*\omega=p_2^*\omega$. 
\,If $\,Y_{12}\,$ is the difference of the boundary values of a
function $\,Y\,$ holomorphic in the interior of $\,\CA_{f,\tau}\,$ then,
by the Stokes theorem,
\qq
\int_{\CI_L}\hspace{-0.1cm}Y_{12}\m\,p_1^*\omega
=\int_{\CI_L}\hspace{-0.1cm}Y_1\,p_1^*\omega
-\int_{\CI_L}\hspace{-0.1cm}Y_2\,p_2^*\omega
=\int_{\CA_{f,\tau}}\hspace{-0.2cm}d(Y\omega)
\,=\,0\,.
\label{intcond}
\qqq
One may show that this is also a sufficient condition for the
solubility of (\ref{Freq}) for $\,Y_1$, \,given $\,Y_{12}\m$.
\,For $\,Y_{12}\,$ of \m(\ref{Y12}), \,it fixes
uniquely $\,\widehat\tau$. \,The form $\,\omega\,$ may be
constructed as the pullback by $\,W\,$ given by (\ref{CW})
of the holomorphic form
$\,-\frac{dw}{2\pi\ii w}\,$ on $\,\mathbb C^\times\,$ (that descends
to $\,\mathbb C^\times\hspace{-0.03cm}\big/
(w\sim\widehat q^{\pm1}w)$). \,It follows that it is given by the formula
\qq
\omega=-\frac{dz}{2\pi\ii z}-L^{-1}dY\,.
\label{omega}
\qqq
The integrability condition (\ref{intcond}) takes then for
$\,Y_{12}\,$ given by (\ref{Y12}) the form of an implicit equation
for $\,\widehat\tau\,$:
\qq
\widehat\tau=\tau-L^{-2}\int_{\CI_L}(f-f_0)(df-dY_1)\,.
\label{maybd2}
\qqq
One also obtains from (\ref{cycles}) and (\ref{omega}) the relation
\qq
\int_{b}\m\omega=\tau-L^{-1}f(0)
+L^{-1}Y_{12}(0)=\widehat\tau
\label{b}
\qqq
which, together with (\ref{a}), shows that
$\,\widehat\tau\in\mathbb H_+\,$ and that the isomorphism
$\,\CT_{f,\tau}\cong\CT_{f_0,\widehat\tau}\,$ defined by $\,W\,$ of
(\ref{CW}) intertwines the markings defined by (\ref{cycles}).
\vskip 0.2cm

It will be convenient to introduce the functions
\qq
&&X_1(x)=f(x)-Y_1(x)-L\tau\,,\cr\cr
&&X_2(x)=x-Y_2(x)=x-Y_1(x)+Y_{12}(x)=f(x)+L(\widehat\tau-\tau)
-Y_1(x)=X_1(x)+L\widehat\tau
\qqq
such that 
\qq
W_i(x)=\ee^{-\frac{2\pi\ii}{L}X_i(x)}\,.
\qqq
for the boundary values $\,W_i\,$ of $\,W$. 
The functions $\,X_i\,$ and their derivatives satisfy the relations
\qq
X_i(x+L)=X_i(x)+L\,,\qquad X_i'(x+L)=X_i'(x)\,,\qquad X'_1(x)=X'_2(x)\,.
\label{X'1=X'2}
\qqq
We shall need below the following identities.
\vskip 0.3cm

\noindent{\bf Lemma 1.} \ One has:
\qq
&&\int_{\CI_L}\hspace{-0.1cm}\big(X_i'(x)\big)^2dx
=\int_{\CI_L}\hspace{-0.1cm}\frac{_1}{^{f'(x)}}\,
\big(X_i'(x)\big)^2dx\,,\label{stform1}\\
&&\int_{\CI_L}\hspace{-0.1cm}(SX_i)(x)\,dx
=\int_{\CI_L}\hspace{-0.1cm}\frac{_1}{^{f'(x)}}\,
\big((SX_i)(x)-(Sf)(x)\big)\,dx\,,\label{stform2}
\qqq
where $\,S\,$ denotes the Schwarzian derivative,
\,see (\ref{Schwarz}).
\vskip 0.2cm

\noindent{\bf Proof of Lemma 1.} \ Consider first the 1-form 
$\,\eta_1(z)=\big(\frac{W'(z)}{W(z)}\big)^2z\m dz\,$ on $\,\CA_{f,\tau}$.
One has
\qq
\int_{\CI_L}\hspace{-0.1cm}p_2^*\eta_1
=\big(-\frac{_{2\pi\ii}}{^L}\big)
\int_{\CI_L}\hspace{-0.1cm}\big(X_{2}'(x)\big)^2dx\,.
\qqq
On the other hand,
\qq
\int_{\CI_L}\hspace{-0.1cm}p_1^*\eta_1
=\big(-\frac{_{2\pi\ii}}{^L}\big)
\int_{\CI_L}\hspace{-0.1cm}\frac{_1}{^{f'(x)}}\big(X_{1}'(x)\big)^2dx
\qqq
so that the identity (\ref{stform1}) follows from the holomorphicity
of the 1-form $\,\eta_1\,$ in the interior of $\,\CA_{f,\tau}\m$.
Similarly, \,consider the 1-form $\,\eta_2(z)=(SW)(z)\m z\m dz\,$
that is also holomorphic on the interior of $\,\CA_{f,\tau}$.
\,Now one has:
\qq
\int_{\CI_L}\hspace{-0.1cm}p_2^*\eta_2
=\big(-\frac{_{2\pi\ii}}{^L}\big)
\int_{\CI_L}\hspace{-0.1cm}\Big(-\frac{_{1}}{^{2}}\big(X'_2(x)\big)^2
-\frac{_{L^2}}{^{4\pi^2}}(SX_2)(x)+\frac{_1}{^2}\Big)dx
\qqq
and
\qq
&&\int_{\CI_L}\hspace{-0.1cm}p_1^*\eta_2
=\big(-\frac{_{2\pi\ii}}{^L}\big)
\int_{\CI_L}\hspace{-0.1cm}\frac{_1}{^{f'(x)}}
\Big(-\frac{_{1}}{^{2}}\big(X'_1(x)\big)^2-\frac{_{L^2}}{^{4\pi^2}}
(SX_1)(x)+\frac{_{L^2}}{^{4\pi^2}}(Sf)(x)+\frac{_1}{^2}f'(x)^2\Big)\,dx\cr
&&=\big(-\frac{_{2\pi\ii}}{^L}\big)\int_{\CI_L}\bigg(\frac{_1}{^{f'(x)}}
\Big(-\frac{_{1}}{^{2}}\big(X'_1(x)\big)^2
-\frac{_{L^2}}{^{4\pi^2}}(SX_1)(x)+\frac{_{L^2}}{^{4\pi^2}}
(Sf)(x)\Big)+\frac{_1}{^2}\bigg)\m dx\,.
\qqq
Comparing both integrals and using (\ref{stform1}) and the equality
$\,SX_1=SX_2$, \,we obtain (\ref{stform2}).

\hspace{12cm}$\square$

\subsection{1-point function of the Euclidian energy-momentum tensor}
\label{subsec:1ptTEucl}

\noindent The holomorphic component of the Euclidian energy-momentum
tensor is given by the formula\footnote{The replacement of $\,L_n\,$
by $\,i^nL_n\,$ in the usual formula for $\,T^E\,$ absorbs the
shift of $\,x\,$ introduced in (\ref{decomp}).}
\qq
T^E(z)=\sum\limits_{n=-\infty}^\infty i^nL_n\m z^{-n-2}\m.
\qqq
In view of (\ref{decomp}), the Minkowskian energy-momentum tensor
$\,T(x)\,$ is related to $\,T^E(z)\,$ for $\,|z|=1\,$ by the identity
\qq
T(x)=\frac{_{2\pi}}{^{L^2}}\Big(\ee^{-\frac{4\pi\ii}{L}x}\,
T^E(\ee^{-\frac{2\pi\ii}{L}x})-\frac{_c}{^{24}}\Big)
\label{T+T}
\qqq
that we shall use below several times.
We would like to find the 1-point function of $\,T^E(z)\,$
on the torus $\,\CT_{f,\tau}\,$ defined as
\qq
\big\langle\,T^E(z)\m\big\rangle_{_{\hspace{-0.04cm}\CT_{f,\tau}}}
\,=\,\frac{\Tr\Big(U(f)\,T^E(z)\,\m\ee^{2\pi\ii\tau(L_0-\frac{c}{24})}\Big)}
{\Tr\Big(U(f)\,\m\ee^{2\pi\ii\tau(L_0-\frac{c}{24})}\Big)}\m.
\qqq
$\langle\,T^E(z)\m\rangle_{_{\hspace{-0.04cm}\CT_{f,\tau}}}$ is holomorphic
in the interior of $\,\CA_{f,\tau}\,$ and has boundary values, \m at least
in the distributional sense, \m that we would like to relate. Since
the commutator with $\,L_0-\frac{2}{24}\,$ generates the dilations of
$\,T^E(z)\,$ so that.
\qq
q^2\m T^E(qz)\,=\,\ee^{2\pi\ii\tau(L_0-\frac{c}{24})}\,T^E(z)\,
\ee^{-2\pi\ii\tau(L_0-\frac{c}{24})}\m,
\qqq
we infer using (\ref{T+T}) and the transformation rule (\ref{UT+U}) that

\qq
&&f'(x)^2\,\Big(q^2\,\ee^{-\frac{4\pi\ii}{L}f(x)}\,\Big\langle\m T^E(q\,
\ee^{-\frac{2\pi\ii}{L}f(x)})\Big\rangle_{_{\hspace{-0.04cm}\CT_{f,\tau}}}
\hspace{-0.1cm}-\frac{_c}{^{24}}\Big)\cr\cr\cr
&&=\frac{f'(x)^2\,\Tr\bigg(U(f)\,
\Big(q^2\,\ee^{-\frac{4\pi\ii}{L}f(x)}\,T^E(q\,\ee^{-\frac{2\pi\ii}{L}f(x)})\,
-\,\frac{c}{24}\Big)\,\ee^{2\pi\ii\tau(L_0-\frac{c}{24})}\bigg)}
{\Tr\Big(U(f)\,\m\ee^{2\pi\ii\tau(L_0-\frac{c}{24})}\Big)}\cr
&&=\,\frac{f'(x)^2\,\Tr\bigg(U(f)\,\m\ee^{2\pi\ii\tau(L_0-\frac{c}{24})}\Big(
\ee^{-\frac{4\pi\ii}{L}f(x)}\,T^E(\ee^{-\frac{2\pi\ii}{L}f(x)})\,-\,\frac{c}{24}
\Big)\bigg)}
{\Tr\Big(U(f)\,\m\ee^{2\pi\ii\tau(L_0-\frac{c}{24})}\Big)}\cr
&&=\,\frac{f'(x)^2\,\Tr\Big(U(f)\,\m\ee^{2\pi\ii\tau(L_0-\frac{c}{24})}
\,\frac{L^2}{2\pi}T(f(x))\Big)}
{\Tr\Big(U(f)\,\m\ee^{2\pi\ii\tau(L_0-\frac{c}{24})}\Big)}\,
=\,\frac{\frac{L^2}{2\pi}\,\Tr\Big(f'(x)^2\,T(f(x))\,U(f)
\,\m\ee^{2\pi\ii\tau(L_0-\frac{c}{24})}\Big)}
{\Tr\Big(U(f)\,\m\ee^{2\pi\ii\tau(L_0-\frac{c}{24})}\Big)}\cr
&&=\,\frac{\frac{L^2}{2\pi}\,\Tr\bigg(\Big(f'(x)^2T(f(x))
-\frac{c}{24\pi}(Sf)(x)\Big)\,U(f)
\,\m\ee^{2\pi\ii\tau(L_0-\frac{c}{24})}
\bigg)}
{\Tr\Big(U(f)\,\m\ee^{2\pi\ii\tau(L_0-\frac{c}{24})}\Big)}
\,+\,\frac{_{cL^2}}{^{48\pi^2}}(Sf)(x)\cr
&&=\,\frac{\frac{L^2}{2\pi}\,\Tr\Big(U(f)\,T(x)\,\m\ee^{2\pi\ii\tau(L_0-\frac{c}{24})}
\Big)}
{\Tr\Big(U(f)\,\m\ee^{2\pi\ii\tau(L_0-\frac{c}{24})}\Big)}
\,+\,\frac{_{cL^2}}{^{48\pi^2}}(Sf)(x)\cr
&&=\,\frac{\Tr\bigg(
U(f)\,\Big(\ee^{-\frac{4\pi\ii}{L}x}T^E(\ee^{-\frac{2\pi\ii}{L}x})-\frac{c}{24}
\Big)\,\ee^{2\pi\ii\tau(L_0-\frac{c}{24})}\bigg)}
{\Tr\Big(U(f)\,\m\ee^{2\pi\ii\tau(L_0-\frac{c}{24})}\Big)}
\,+\,\frac{_{cL^2}}{^{48\pi^2}}(Sf)(x)\cr\cr\cr
&&=\,\ee^{-\frac{4\pi\ii}{L}x}\,\Big\langle\m T^E(\ee^{-\frac{2\pi\ii}{L}x})
\Big\rangle_{_{\hspace{-0.04cm}\CT_{f,\tau}}}
\hspace{-0.1cm}-\frac{_c}{^{24}}
\,+\,\frac{_{cL^2}}{^{48\pi^2}}\,(Sf)(x)\,.
\qqq
This is a linear inhomogeneous equation for the boundary values of the
1-point function $\,\langle\,T^E(z)\m\rangle_{_{\hspace{-0.04cm}\CT_{f,\tau}}}
\hspace{-0.1cm}$. \,Let us consider now the function
\qq
G(z)=\frac{_c}{^{12}}(SW)(z)\,,
\qqq
where $\,W(z)\,$ is given by (\ref{CW}).
By the chain rule for the Schwarzian derivative (\ref{Schwchain})
and (\ref{W1W2}),
\qq
&&(SW_{1})(x)=\big(q(-\frac{_{2\pi\ii}}{^L})f'(x)\,\ee^{-\frac{2\pi\ii}{L}f(x)}
\big)^2\,(SW)(q\,\ee^{-\frac{2\pi\ii}{L}f(x)})+\frac{_{2\pi^2}}{^{L^2}}\,f'(x)^2
+(Sf)(x)\cr
&&=\,(SW_{2})(x)=\big((-\frac{_{2\pi\ii}}{^L})\,\ee^{-\frac{2\pi\ii}{L}x}\big)^2\,
(SW)(\ee^{-\frac{2\pi\ii}{L}x})+\frac{_{2\pi^2}}{^{L^2}}
=-\frac{_{4\pi^2}}{^{L^2}}\big(
\ee^{-\frac{4\pi\ii}{L}x}(SW)(\ee^{-\frac{2\pi\ii}{L}x})-\frac{_1}{^2}\big)
\label{SWSW}
\qqq
from which we conclude that
\qq
f'(x)^2\Big(q^2\,\ee^{-\frac{4\pi\ii}{L}f(x)}\,G(
q\,\ee^{-\frac{2\pi\ii}{L}f(x)})-\frac{_c}{^{24}}\Big)
\,=\,\ee^{-\frac{4\pi\ii}{L}x}\,
G(\ee^{-\frac{2\pi\ii}{L}x})-\frac{_c}{^{24}}
+\frac{_{cL^2}}{^{48\pi^2}}\,(Sf)(x)\,,\qquad
\qqq
which is the same linear inhomogeneous equation that for the boundary
values of $\,\langle\,T^E(z)\m\rangle_{_{\hspace{-0.04cm}\CT_{f,\tau}}}$\hspace{-0.1cm}.
\,But the corresponding homogeneous equation for the
boundary values of a holomorphic function $\,F(z)\,$ on
the interior of $\,\CA_{f,\tau}\m$,
\qq
f'(x)^2\,q^2\,\ee^{-\frac{4\pi\ii}{L}f(x)}\,
F(q\,\ee^{-\frac{2\pi\ii}{L}f(x)})=\ee^{-\frac{4\pi\ii}{L}x}\,
F(\ee^{-\frac{2\pi\ii}{L}x})\,,
\qqq
has solutions that correspond to holomorphic quadratic differentials
$\,F(z)(dz)^2\,$ pulled back from $\,\CT_{f,\tau}\,$ that necessarily are
of the form $\,A(d\ln{W(z)})^2$, \,i.e.
\qq
F(z)=A\m\Big(\frac{W'(z)}{W(z)}\Big)^{\hspace{-0.05cm}2}
\label{As}
\qqq
for some constant $\,A$. \,This leads to the identity
\qq
\big\langle\,T^E(z)\m
\big\rangle_{_{\hspace{-0.04cm}\CT_{f,\tau}}}\hspace{-0.1cm}=\,G(z)+F(z)
\,=\,\frac{_c}{^{12}}\,(SW)(z)\,+\,F(z)\,.
\label{GtG}
\qqq
The parameter $\,A\,$ in (\ref{As}) may be fixed from the
transformation rule of the Euclidian energy-momentum expectations
between the isomorphic tori $\,\CT_{f,\tau}\,$ and
$\,\CT_{f_0,\widehat\tau}\,$ in Euclidian CFT which gives
\qq
F(z)&=&W'(z)^2\,\big\langle\,T^E(W(z))\m
\big\rangle_{_{\hspace{-0.04cm}\CT_{f_0,\widehat\tau}}}\,,
\qqq
see Eq.\,(5.124) in \cite{DFMS} or (2.15) in \cite{KGPr}. \,Since
\qq
&&\big\langle\,T^E(w)\m
\big\rangle_{_{\hspace{-0.04cm}\CT_{f_0,\widehat\tau}}}\hspace{-0.1cm}\,=\,
\frac{\Tr\Big(T^E(w)\,\m\ee^{2\pi\ii\widehat\tau(L_0-\frac{c}{24})}\Big)}
{\Tr\Big(\ee^{2\pi\ii\widehat\tau(L_0-\frac{c}{24})}\Big)}=
\frac{\Tr\Big(L_0\m w^{-2}\,\ee^{2\pi\ii\widehat\tau(L_0-\frac{c}{24})}\Big)}
{\Tr\Big(\ee^{2\pi\ii\widehat\tau(L_0-\frac{c}{24})}\Big)}\cr
&&=w^{-2}\bigg(\frac{\Tr\Big((L_0-\frac{c}{24})\,
\m\ee^{2\pi\ii\widehat\tau(L_0-\frac{c}{24})}
\Big)}
{\Tr\Big(\ee^{2\pi\ii\widehat\tau(L_0-\frac{c}{24})}\Big)}+\frac{_c}{^{24}}\bigg)=
w^{-2}\Big(\frac{_1}{^{2\pi\ii}}\m\partial_{\widehat\tau}\ln{\chi(\widehat\tau)}
+\frac{_c}{^{24}}\Big),\qquad
\qqq
we obtain then the identity
\qq
\big\langle\,T^E(z)\m
\big\rangle_{_{\hspace{-0.04cm}\CT_{f,\tau}}}\,=\,\frac{_c}{^{12}}\,(SW)(z)\,+\,
\Big(\frac{W'(z)}{W(z)}\Big)^{\hspace{-0.05cm}2}
\Big(\frac{_1}{^{2\pi\ii}}\,
\partial_{\widehat\tau}\ln{\chi(\widehat\tau)}+\frac{_c}{^{24}}\Big)
\label{fsz}
\qqq
that is the main result of the present subsection.

\subsection{$\Difft\,$ character on 1-parameter subgroups}
\label{subsec:on1parsub}

\noindent We shall use the identity (\ref{fsz}) to calculate
the logarithmic $\m s$-derivative of
\qq
c_{s,\zeta}^{-1}\,\Upsilon(f_s,\tau)=\Tr\Big(
\ee^{\ii s\int_{\CI_L}\hspace{-0.1cm}\zeta(x)T(x)dx}\,\m
\ee^{2\pi\ii\tau(L_0-\frac{c}{24})}\Big)\,\equiv\,
\widetilde\Upsilon(f_s,\tau)
\qqq
for the flow $\,f_s\,$ of the vector field $\,-\zeta(x)\partial_x$,
\,see (\ref{Diffchar}) and (\ref{exponen}). First note that 
\qq
&&\partial_s\ln{\widetilde\Upsilon(f_s,\tau)}\,=\,
\frac{\ii\int_{\CI_L}\zeta(x)\,\Tr\Big(U(f_s)\,T(x)\,\m
\ee^{2\pi\ii\tau(L_0-\frac{c}{24})}\Big)\m dx}
{\Tr\Big(U(f_s)\,\m\ee^{2\pi\ii\tau(L_0-\frac{c}{24})}\Big)}\,.
\label{pstUp}
\qqq
Denoting by $\,W_s\,$ the function defining the isomorphism
$\,\CT_{f_s,\tau}\cong\CT_{f_0,\widehat\tau_s}\,$ constructed as in
Sec.\,\ref{subsec:confweldRH} and by $\,W_{s;i}=\ee^{-\frac{2\pi\ii}{L}X_{s;i}}\,$
its boundary values such that $\,X_{s;1}=X_{s;2}-L\widehat\tau_s$, \,we infer
from (\ref{T+T}), (\ref{fsz}), (\ref{SWSW}) and (\ref{X'1=X'2}) that
\qq
&&\frac{\frac{L^2}{2\pi}\,
\Tr\Big(U(f_s)\,T(x)\,\m\ee^{2\pi\ii(L_0-\frac{c}{24})}\Big)}
{\Tr\Big(U(f_s)\,\m\ee^{2\pi\ii(L_0-\frac{c}{24})}\Big)}\,=\,
\ee^{-\frac{4\pi\ii}{L}x}\,\big\langle T^E(\ee^{-\frac{2\pi\ii}{L}x})\big
\rangle_{_{\hspace{-0.04cm}\CT_{f,\tau}}}\hspace{-0.1cm}-\,\frac{_c}{^{24}}\cr
&&=\,\frac{_c}{^{12}}\,\ee^{-\frac{4\pi\ii}{L}x}\,(SW_s)(\ee^{-\frac{2\pi\ii}{L}x})
\,+\,\ee^{-\frac{4\pi\ii}{L}x}\,\Big(\frac{W'_s(\ee^{-\frac{2\pi\ii}{L}x})}{
W_s(\ee^{-\frac{2\pi\ii}{L}x})}\Big)^{\hspace{-0.05cm}2}\Big(\frac{_1}{^{2\pi\ii}}\,
\partial_{\widehat\tau_s}\ln{\chi(\widehat\tau_s)}+\frac{_c}{^{24}}\Big)\,-\,
\frac{_c}{^{24}}\cr
&&=\,-\frac{_{cL^2}}{^{48\pi^2}}\,(SW_{s;2})(x)\,-\,\frac{_{L^2}}{^{4\pi^2}}
\,\big(\partial_x\ln{W_{s;2}(x)}\big)^2\,
\Big(\frac{_1}{^{2\pi\ii}}\,
\partial_{\widehat\tau_s}\ln{\chi(\widehat\tau_s)}+\frac{_c}{^{24}}\Big)\cr\cr
&&=\,-\frac{_{cL^2}}{^{48\pi^2}}\,(SX_{s;i})(x)+\frac{_1}{^{2\pi\ii}}\,
(X'_{s;i}(x))^2\,\partial_{\widehat\tau_s}\ln{\chi(\widehat
\tau_s)}\,.
\label{fluus}
\qqq
The substitution of (\ref{fluus}) to (\ref{pstUp}) gives then the relation
\qq
&&\partial_s\ln{\widetilde\Upsilon(f_s,\tau)}\,=\,-\ii\,
\frac{_c}{^{24\pi}}\int_{\CI_L}\hspace{-0.15cm}\zeta(x)\,(SX_{s;i})(x)\,dx\,+\,
L^{-2}\int_{\CI_L}\hspace{-0.15cm}\zeta(x)\,(X'_{s;i}(x))^2\m dx\ \,
\partial_{\widehat\tau_s}\ln{\chi(\widehat\tau_s)}\,.
\label{chardifff}
\qqq
In the next subsection, we shall show that
\qq
\partial_s\widehat\tau_s=\,L^{-2}\hspace{-0.05cm}\int_{\CI_L}\hspace{-0.15cm}
\zeta(x)\,(X'_{s;i}(x))^2\m dx\,.
\label{tbpins}
\qqq
Using this identity, we infer from (\ref{chardifff}) that
\qq
\Upsilon(f_s,\tau)\,=\,C_{f_s,\tau}\,\chi(\widehat\tau_s)
\qqq
with
\qq
C_{f_s,\tau}\,=\,c_{s,\zeta}\,\m
\exp\hspace{-0.04cm}\bigg[\hspace{-0.04cm}-\ii\frac{_c}{^{24\pi}}\int_0^s
ds'\int_{\CI_L}\hspace{-0.1cm}
\zeta(x)\m(SX_{s',i})(x)\,dx\m\bigg].
\qqq
This establishes the relation between the characters of
$\,\Difft\,$ on 1-parameter subgroups
and the Virasoro characters.
\vskip 0.3cm

\noindent{\bf Remark.} \ We could obtain formulae similar to
(\ref{chardifff}) and (\ref{tbpins}) if we replaced  $\,f_s\,$ by
$\,f\circ f_s\,$ for any $\,f\in\Difft$. \,This would lead to an expression
for $\,\Upsilon(f,\tau)\,$ involving the integration along any smooth curve
joining $\,f\,$ to $\,f_0$, \,but we shall not need
such a generalization here.

\subsection{The effective modular parameter}
\label{subsec:effmpar}

\noindent In this subsection, we shall obtain an integral expression
for the effective modular parameter $\,\widehat\tau_s\,$ such that
$\,\CT_{f_s,\tau}\cong\CT_{f_0,\widehat\tau_s}$. \,With the applications to
FCS in sight, \,we shall consider a slightly more general
situation with the initial modular parameter $\,\tau\,$ replaced
by $\,\tau_s=\tau-L^{-1}as\,$ for some real constant $\,a$. \,Let
$\,W_s:\CA_{f_s,\tau_s}\rightarrow\mathbb C^\times\,$ be the map constructed
as in Sec.\,\ref{subsec:confweldRH} defining the isomorphism
$\,\CT_{f_s,\tau_s}\cong\CT_{f_0,\widehat\tau_s}\,$ with the boundary values
$\,W_{s;i}=\ee^{-\frac{2\pi\ii}{L}X_{s;i}}\,$ such that $\,X_{s;1}(x)=X_{s;2}(x)
-L\widehat\tau_s$. Clearly, $\,X_{s;i}\,$ and $\,\widehat\tau_s\,$ depend
now also on $\,a$.
\vskip 0.4cm

\noindent{\bf Proposition.} \ The effective modular parameter
$\,\widehat\tau_s\,$ satisfies the relation
\qq
\partial_s\widehat\tau_s=\m L^{-2}\hspace{-0.05cm}\int_{\CI_L}\hspace{-0.12cm}
\big(\zeta(x)-a\big)\m(X'_{s;i}(x))^2\m dx
\label{conject}
\qqq
\vskip 0.3cm

\noindent{\bf Proof of Proposition.} \ Eq.\,(\ref{conject}) will be established by
calculating the infinitesimal change in $\,\widehat\tau_s$. \,Denote by
$\,\CA_s\,$ the annular region in $\,\mathbb C^\times\,$ that
is the image of $\,\CA_{f_s,\tau_s}\,$ by $\,W_s\,$ and has the boundary
components parameterized by
\qq
x\ \longmapsto\ W_s(q_s\hspace{0.03cm}
\ee^{-\frac{2\pi\ii}{L}f_s(x)})=\ee^{-\frac{2\pi\ii}{L}X_{s;1}(x)}\,,\qquad
x\ \longmapsto\ W_s(\ee^{-\frac{2\pi\ii}{L}x})=\ee^{-\frac{2\pi\ii}{L}X_{s;2}(x)}
\qqq
for $\,q_s=\ee^{2\pi\ii\tau_s}$.
\,Consider the map $\,U_{s,\sigma}=W_{s+\sigma}\circ
W_s^{-1}:\CA_s\rightarrow\mathbb C^\times\,$ that is holomorphic
on the interior of $\,\CA_s\,$ and that satisfies the relations
\qq
&&U_{s,\sigma}(\ee^{-\frac{2\pi\ii}{L}X_{s;1}(f_s^{-1}(f_{s+\sigma}(x)+a\sigma))})=
W_{s+\sigma}(q_{s+\sigma}\,\ee^{-\frac{2\pi\ii}{L}f_{s+\sigma}(x)})=\ee^{-\frac{2\pi\ii}{L}
X_{s+\sigma;1}(x)}\cr
&&=\widehat q_{s+\sigma}\,\ee^{-\frac{2\pi\ii}{L}X_{s+\sigma;2}(x)}
=\widehat q_{s+\sigma}\,W_{s+\sigma}(\ee^{-\frac{2\pi\ii}{L}x})
=\widehat q_{s+\sigma}\,U_{s,\sigma}(\ee^{-\frac{2\pi\ii}{L}X_{s;2}(x)})
\label{Xssigma12}
\qqq
for $\,\widehat q_s=\ee^{2\pi\ii\widehat\tau_s}$.
\,One may write
\qq
U_{s,\sigma}(w)=w\,\ee^{\frac{2\pi\ii}{L}V_{s,\sigma}(w)}\,,
\qqq
where $\,V\,$ is a function on $\,\CA_s\,$ with the boundary
values
\qq
&&V_{s,\sigma;1}(x)\equiv V_{s,\sigma}(\ee^{-\frac{2\pi\ii}{L}X_{s;1}(f_s^{-1}
(f_{s+\sigma}(x)+a\sigma))})=X_{s;1}(f_s^{-1}(f_{s+\sigma}(x)+a\sigma))
-X_{s+\sigma;1}(x)\,,\cr
&&V_{s,\sigma;2}(x)\equiv V_{s,\sigma}(\ee^{-\frac{2\pi\ii}{L}X_{s;2}(x)})
=X_{s;2}(x)-X_{s+\sigma;2}(x)
\qqq
so that
\qq
V_{s,\sigma;12}(x)\equiv V_{s,\sigma;1}(x)-V_{s,\sigma;2}(x)&=&
X_{s;1}(f_s^{-1}(f_{s+\sigma}(x)+a\sigma))-X_{s;2}(x)+L\widehat\tau_{s+\sigma}\cr
&=&X_{s;2}(f_s^{-1}(f_{s+\sigma}(x)+a\sigma))-X_{s;2}(x)
+L(\widehat\tau_{s+\sigma}-\widehat\tau_s)\,.
\qqq
Consider on $\,\CA_s\,$ the holomorphic form
\qq
\omega_{s,\sigma}=U_{s,\sigma}^{\m*}\big(-\frac{_{du}}{^{2\pi\ii u}}\big)=
-\frac{_{dw}}{^{2\pi\ii w}}-L^{-1}dV_{s,\sigma}\,.
\qqq
Then
\qq
&&\omega_{s,\sigma}\big(\ee^{-\frac{2\pi\ii}{L}X_{s;1}
(f_s^{-1}(f_{s+\sigma}(x)+a\sigma))}\big)\m=\,
\omega_{s,\sigma}\big(\ee^{-\frac{2\pi\ii}{L}X_{s;2}(x)}\big)=
L^{-1}\m\big(X'_{s;2}(x)\,dx\,-\,V'_{s,\sigma;2}(x)\big)\,dx\,,
\qqq
where the first equality follows from (\ref{Xssigma12}).
\,Besides, $\,\omega_{s,\sigma}\,$ satisfies the normalization condition
\qq
\int_{\CI_L}\hspace{-0.1cm}
\omega_{s,\sigma}\big(\ee^{-\frac{2\pi\ii}{L}X_{s;2}(x)}\big)\,=\,1\,.
\qqq
As $\,V_{s,\sigma}\,$ is a holomorphic on the interior of $\,\CA_s$,
\,we have the relation
\qq
0&=&\int_{\CA_s}\hspace{-0.1cm}d\m(V_{s,\sigma}\m\omega_{s,\sigma})\,=\,
\int_{\CI_L}\hspace{-0.1cm}V_{s,\sigma;1}(x)\,\omega_{s,\sigma}
\big(\ee^{-\frac{2\pi\ii}{L}X_{s;1}(f_s^{-1}(f_{s+\sigma}(x)+a\sigma))}\big)-
\int_{\CI_L}\hspace{-0.1cm}V_{s,\sigma;2}(x)\,\omega_{s,\sigma}
\big(\ee^{-\frac{2\pi\ii}{L}X_{s;2}(x)}\big)\cr
&=&\int_{\CI_L}\hspace{-0.1cm}V_{s,\sigma;12}(x)\,\omega_{s,\sigma}
\big(\ee^{-\frac{2\pi\ii}{L}X_{s;2}(x)}\big)\cr
&=&\int_{\CI_L}\hspace{-0.1cm}\Big(X_{s;2}(f_s^{-1}(f_{s+\sigma}(x)+a\sigma))
-X_{s;2}(x)\Big)\,\omega_{s,\sigma}
\big(\ee^{-\frac{2\pi\ii}{L}X_{s;2}(x)}\big)\,+\,
L(\widehat\tau_{s+\sigma}-\widehat\tau_s)\cr
&=&L^{-1}\int_{\CI_L}\hspace{-0.1cm}
\Big(X_{s;2}(f_s^{-1}(f_{s+\sigma}(x)+a\sigma))-X_{s;2}(x)
\Big)\hspace{-0.05cm}\Big(X'_{s;2}(x)-V_{s,\sigma;2}'(x)\Big)\m dx\,+\,
L(\widehat\tau_{s+\sigma}-\widehat\tau_s)\m.
\label{671}
\qqq
Differentiating the last line over $\,\sigma\,$ at $\,\sigma=0\,$
and using the fact that $\,V_{s,0}=0$, \,as well as the relation
\qq
(f_s^{-1})'(f_s(x))\,\zeta(f_s(x))=\zeta(x)
\qqq
that follows from the differentiation over $\,s\,$ of the identity
$\,f_s^{-1}(f_s(x))=x$, \,we infer from (\ref{671}) that
\qq
0&=&L^{-1}\hspace{-0.05cm}\int_{\CI_L}\hspace{-0.1cm}
X'_{s;2}(x)\,(f_s^{-1})'(f_s(x))\big(-\zeta(f_s(x))+a\big)\,X'_{s;2}(x)\,dx\,+\,
L\,\partial_s\widehat\tau_s\cr
&=&-\,L^{-1}\hspace{-0.05cm}\int_{\CI_L}\hspace{-0.1cm}(\zeta(x)
-\frac{_a}{f'_s(x)})\,(X'_{s;2}(x))^2\,+\,L\,\partial_s\widehat\tau_s\,.
\qqq
Together with (\ref{stform1}), \,this implies (\ref{conject}).

\hspace{12cm}$\square$
\vskip 0.3cm

\noindent{\bf Corollary.} \ Taking $\,a=0$, \,we obtain (\ref{tbpins}).

\subsection{Generating function of FCS in finite volume}
\label{subsec:fvolFCS}

\noindent We are now ready to eliminate the
$\,\Difft$-character $\,\Upsilon(f_{s,t,L},
\tau_{s,L})\,$ from the expression
(\ref{PsichiUpsilon}) for the finite-volume generating function $\,\Psi_{t,L}(\lambda)\,$
of \m FCS for energy transfers. \,To this end, we first
rewrite the formula (\ref{PsichiUpsilon}) in the form
\qq
\Psi_{t,L}(\lambda)\,=\,\frac{\widetilde\Upsilon(f_{s,t,L}\m,\tau_{s,L})}
{\chi(\tau_{0,L})}\,\,\ee^{-\ii s\m(C_{t,L}-C_{0,L})}
\label{PsichiUpsilon1}
\qqq
recalling that
\qq
&&s=\frac{_\lambda}{^{\Delta\beta}}\,,\qquad\tau_{s,L}=L^{-1}(\ii-s)\gamma_L\,,
\qquad\zeta_{t,L}(x)=\gamma_L\frac{\beta_L(h_{L}^{-1}(x)+vt)}
{\beta_L(h_{L}^{-1}(x))}
\qqq
and $\,f_{s,t,L}\,$ is the flow of the vector field
$\,-\zeta_{t,L}(x)\partial_x$. The only difference with respect to
the setting of Sec.\,\ref{subsec:on1parsub} is that
$\,\tau\,$ is now replaced by $\,\tau_{s,L}\,$ which corresponds
to the situation considered in Sec.\,\ref{subsec:effmpar} if we set
$\,a=\gamma_L\,$ there. \,Differentiating $\,\ln{\Psi_{t,L}}\,$ over
$\,\lambda$, \,we obtain the relation
\qq
&&\partial_\lambda\ln{\Psi_{t,L}(\lambda)}\cr
&&=\,\frac{_1}{^{\Delta\beta}}
\bigg(-\ii\,
\frac{_c}{^{24\pi}}\int_{\CI_L}\hspace{-0.1cm}\zeta_{t,L}(x)\,(SX_{s,t,L;i})(x)\,dx\,+\,
L^{-2}\hspace{-0.05cm}\int_{\CI_L}\hspace{-0.1cm}\zeta_{t,L}(x)\,(X'_{s,t,L;i}(x))^2\m dx\ \,
\partial_{\widehat\tau_{s,L,t}}\ln{\chi(\widehat\tau_{s,L,t})}\cr
&&\hspace{1.3cm}-\,2\pi\ii L^{-1}\gamma_L\frac{\Tr\Big(U(f_{s,L,t})
\,(L_0-\frac{c}{24})\,\m
\ee^{2\pi\ii\tau_{s,L}(L_0-\frac{c}{24})}\Big)}{\Tr\Big(U(f_{s,L,t})\,\m
\ee^{2\pi\ii\tau_{s,L}(L_0-\frac{c}{24})}\Big)}\,-\,\ii\,\big(C_{t,L}-C_{0,L}\big)\bigg),
\label{theform1}
\qqq
where the $\,1^{\rm st}$-line on the right-hand side came from
the $\,s$-dependence of $\,f_{s,t,L}\,$ in
$\,\ln{\widetilde\Upsilon(f_{s,t,L},\tau_{s,L})}\,$ given by (\ref{chardifff})
\m and the $\,1^{\rm st}\,$ term of the $\m2^{\rm nd}$-line from the
$\m s$-dependence of $\,\tau_{s,L}$. \,Functions $\,X_{s,t,L;i}\,$ pertain to
the boundary values
of the maps $\,W_{s,t,L}\,$ defining the isomorphisms
$\,\CT_{f_{s,t,L},\tau_{s,L}}\cong\CT_{f_0,\widehat\tau_{s,t,L}}$,
\,see Sec.\,\ref{subsec:on1parsub}.
\,Now, \m from \m (\ref{fluus}),
\qq
&&\frac{2\pi\ii L^{-1}\,\Tr\Big(U(f_{s,L,t})\,
(L_0-\frac{c}{24})\,\m
\ee^{2\pi\ii\tau_s(L_0-\frac{c}{24})}\Big)}{\Tr\Big(
\ee^{2\pi\ii\tau_s(L_0-\frac{c}{24})}\,U(f_{s,L,t})\Big)}
=\frac{\ii\,\Tr\Big(U(f_{s,L,t})\int_{\CI_L}\hspace{-0.1cm}T(x)\m dx\,\,
\ee^{2\pi\ii\tau_s(L_0-\frac{c}{24})}\Big)}{\Tr\Big(
\ee^{2\pi\ii\tau_s(L_0-\frac{c}{24})}\,U(f_{s,L,t})\Big)}\cr
&&=\,-\ii\,\frac{_{c}}{^{24\pi}}\int_{\CI_L}\hspace{-0.1cm}
(SX_{s,t,L;i})(x)dx\,+\,L^{-2}\hspace{-0.05cm}\int_{\CI_L}\hspace{-0.15cm}
(X'_{s,t,L;i}(x))^2\m dx\
\,\partial_{\widehat\tau_{s,L,t}}\ln{\chi(\widehat\tau_{s,L,t})}\,.
\qqq
Thus the net effect of the $1^{\rm st}$ term on the $2^{\rm nd}$ line on
the right-hand side of (\ref{theform1}) is to replace $\,\zeta_{t,L}\,$ in the
$1^{\rm nd}$ line by
\qq
\xi_{t,L}=\zeta_{t,L}-\gamma_L\,.
\label{xitL}
\qqq
Altogether, we obtain the identity
\qq
\partial_\lambda\ln{\Psi_{t,L}(\lambda)}&=
&\frac{_1}{^{\Delta\beta}}\bigg(-\ii\,\frac{_c}{^{24\pi}}\int_{\CI_L}\hspace{-0.1cm}\xi_{t,L}(x)\,
(SX_{s,t,L;i})(x)\,dx\cr
&&\hspace{0.9cm}+\,L^{-2}\hspace{-0.05cm}
\int_{\CI_{L}}\hspace{-0.1cm}\xi_{t,L}(x)\,(X'_{s,t,L;i}(x))^2\m dx
\ \,\partial_{\widehat\tau_{s,t,L}}
\ln{\chi(\widehat\tau_{s,t,L})}\,-\,\ii\m\big(C_{t,L}-C_{0,L}\big)\bigg).
\label{theformula}
\qqq
Moreover, \m taking $\,\zeta=\zeta_{t,L}\,$ and $\,a=\gamma_L\,$ in Proposition
of Sec.\,\ref{subsec:effmpar}, \m we infer from (\ref{conject}) that
\qq
\partial_s\widehat\tau_{s,t,L}=L^{-2}\hspace{-0.07cm}\int_{\CI_L}\hspace{-0.17cm}\xi_{t,L}(x)
\,(X_{s,t,L;i}'(x))^2\m dx 
\qqq
so that (\ref{theformula}) implies that
\qq
\Psi_{t,L}(\lambda)\,=\,\exp\hspace{-0.07cm}\bigg[\hspace{-0.04cm}
-\ii\,\frac{_c}{^{24\pi}}\int_0^{\frac{\lambda}{\Delta\beta}}
\hspace{-0.15cm}ds\int_{\CI_L}\hspace{-0.17cm}
\xi_{t,L}(x)(SX_{s,t,L;i})(x)\,dx\bigg]\,\,
\frac{\chi(\widehat\tau_{\frac{\lambda}{\Delta\beta},t,L})}
{\chi(\tau_{0,L})}\,\,\m\ee^{-\m\frac{\ii\lambda}{\Delta\beta}\big(C_{t,L}-C_{0,L}\big)}\,,\qquad
\label{PsitlambdaL}
\qqq
where
\qq
\widehat\tau_{\frac{\lambda}{\Delta\beta},t,L}\,=\,\tau_{0,L}\,+\,L^{-2}\hspace{-0.05cm}
\int_0^{\frac{\lambda}{\Delta\beta}}\hspace{-0.13cm}ds
\int_{\CI_L}\hspace{-0.17cm}\xi_{t,L}(x)\,(X_{s,t,L;i}'(x))^2\m dx
\label{hattaucw}
\qqq
and $\,C_{t,L}\,$ is given by (\ref{CtL}). \m This is our final formula
based on conformal welding for the finite-volume generating function
for FCS of energy transfers. Note that the right-hand side of
(\ref{PsitlambdaL}) depends on the spectrum of the CFT via the Virasoro
character $\,\chi$.

\nsection{Thermodynamic limit formula for FCS}
\label{sec:infvlim:heur}

\noindent We would like to compute the limit $\,L\to\infty\,$ of the
generating function $\,\Psi_{t,L}(\lambda)\,$ for FCS. 
Let us first consider the last factor
on the right hand side of (\ref{PsitlambdaL}). Using the symmetries
(\ref{betabetaL})  and (\ref{fwif}) of $\,\beta_L\,$ and $\,h_L\,$
and the behavior of the latter when $\,L\to\infty$, \,see (\ref{hLlim}),
\,we infer that
\qq
\lim\limits_{L\to\infty}\,\ee^{-\frac{\ii\lambda}{\Delta\beta}\big(C_{t,L}-C_{0,L}\big)}
&=&\lim\limits_{L\to\infty}\,\exp\Big[{-\frac{_{\ii\lambda}}{^{\Delta\beta}}
\frac{_{cv}}{^{24\pi}}
\int_{-{\frac{_3}{^4}\hspace{-0.04cm}L}}^{
{\frac{_1}{^4}\hspace{-0.04cm}L}}\hspace{-0.05cm}
(\beta_L(x^+)-\beta_L(x))\,(Sh_L)(x)\,dx}\Big]\cr\cr
&=&\exp\Big[{-\frac{_{\ii\lambda}}{^{\Delta\beta}}\m\frac{_{cv}}{^{24\pi}}
\int(\beta(x^+)+\beta(x^-)-2\beta(x))\,(Sh)(x)\,dx}\Big],
\label{ostat}
\qqq
where $\,h\,$ is given by (\ref{h(x)}) and (\ref{beta0}).
Note that the integral on the right hand side is concentrated
on the support of the kink in the profile $\,\beta(x)$.

\subsection{Large $\,L\,$ behavior of the vector field
$\,\zeta_{t,L}(y)\partial_y\,$ and of its flow}
\label{subsec:largeLzeta}

\noindent In order to study the limiting behavior of the other terms in
the formula (\ref{PsitlambdaL}) for $\,\Psi_{t,L}(\lambda)$, \,we shall need more
detailed information about the forms of the inverse function
$\,h_L^{-1}\,$ and of $\,\zeta_{t,L}\,$ given by (\ref{zetaL}). \,Let us assume
that $\,\beta(x)\,$ takes its asymptotic values outside the interval
$\,[a-\delta,a+\delta]\,$ containing the kink. Then on the interval
$\,[-\frac{1}{4}L,\frac{1}{4}L]$, \m the function $\,h_L^{-1}\,$
is linear to the left and to the right of the interval $\,[A_L,B_L]\,$
\qq
h_L^{-1}(y)\,=\,\begin{cases}\,\frac{\beta_\CL}{\beta_{0,L}}
(y+\frac{1}{4}L)-\frac{1}{4}L\qquad
\text{for}\quad-\frac{1}{4}L\leq y\leq A_L\m,\cr
\,\frac{\beta_\CR}{\beta_{0,L}}(y-\frac{1}{4}L)+\frac{1}{4}L\qquad
\text{for}\hspace{0.78cm}B_L\leq y\leq\frac{1}{4}L
\end{cases}
\qqq
for
\qq
A_L=h_L(a-\delta)=\frac{_{\beta_{0,L}}}{^{\beta_\CL}}(a-\delta)
-\big(1-\frac{_{\beta_{0.L}}}
{^{\beta_\CL}})\frac{_L}{^4}\m,\quad\ 
B_L=h_L(a+\delta)=\frac{_{\beta_{0,L}}}{^{\beta_\CR}}(a+\delta)
+\big(1-\frac{_{\beta_{0.L}}}
{^{\beta_\CR}})\frac{_L}{^4}\m,
\qqq

\begin{figure}[t]
\leavevmode
\begin{center}
\vskip -0.2cm
\includegraphics[width=7.4cm,height=5.6cm]{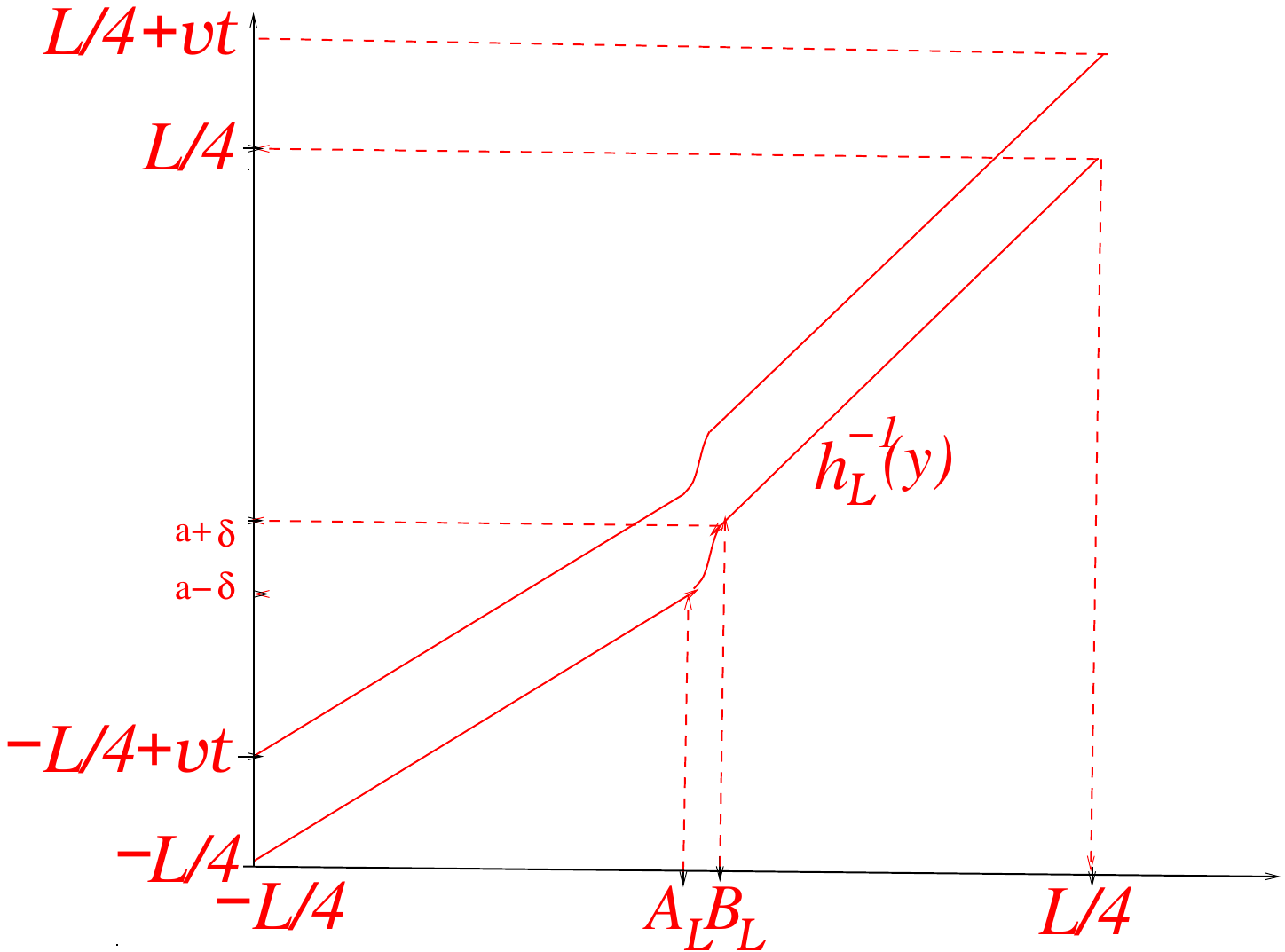}\\
\caption{$\,h_L^{-1}\,$ and $\,h_L^{-1}+vt$}
\label{fig:hL-3}
\end{center}
\vskip -0.3cm
\end{figure}

\noindent see Fig.\,\ref{fig:hL-3}, \,and its form stabilizes inside that
interval. More exactly, with the function $\,h(x)\,$ given by
(\ref{h(x)}) and (\ref{beta0}) and
\qq
A=h(a-\delta)\m,\qquad B=h(a+\delta)\m,
\label{A,B}
\qqq
and for
\qq
O_L^+=-A+A_L\m,
\label{OL+}
\qqq
we have the identity
\qq
h_L^{-1}\big(y+O_L^+\big)=h^{-1}\big(\frac{_{\beta_0}}{^{\beta_{0,L}}}y+
(1-\frac{_{\beta_0}}{^{\beta_{0,L}}})A\big)
\qqq
so that by (\ref{beta0Lbeta0})
\qq
\lim\limits_{L\to\infty}\,h_L^{-1}(y+O_L^+)=h^{-1}(y)\m,
\label{limhL-1}
\qqq
and the difference between $\,h_L^{-1}(y+O_L^+)\,$ and $\,h^{-1}(y)\,$
is $\,O(L^{-1})\,$ uniformly for $\,y\,$
in bounded sets and the same is true for all the $\,y$-derivatives
of that difference. The inverse function $\,h^{-1}\,$ is linear on the
left and on the right of
the interval $\,[A,B]\,$ and it maps this interval into
$\,[a-\delta,a+\delta]$. \,More precisely,
\qq
h^{-1}(y)\,=\,\begin{cases}\,\frac{\beta_\CL}{\beta_0}(y-A)+a-\delta\qquad
\,\text{for}\qquad y\leq A\,,\cr
\,\frac{\beta_\CR}{\beta_0}(y-B)+a+\delta\qquad
\text{for}\qquad y\geq B\,,
\end{cases}
\qqq
\vskip 0.2cm

An examination of the function $\,\zeta_{t,L}\,$ of (\ref{zetaL})
shows now that if $\,|x|\leq\frac{1}{4}L\,$ then
$\,\zeta_{t,L}(x)=v\beta_{0,L}=\gamma_L\,$
outside an interval $\,[A_L-O(|t|),B_L+O(|t|)]\,$
and that the form of $\,\zeta_{t,L}\,$ stabilizes inside this interval.
More precisely, 
\qq
\zeta_{t,L}(y+O_L^+)=\gamma_L\frac{\beta(h_L^{-1}(y+O_L^+)+vt)}
{\beta(h_L^{-1}(y+O_L^+))}
\qqq
for large $\,L\,$ so that
\qq
\lim\limits_{L\to\infty}\,\zeta_{t,L}(y+O_L^+)
=\gamma\m\frac{\beta\big(h^{-1}(y)+vt\big)}{\beta\big(h^{-1}(y)\big)}
\equiv\zeta^+_t(y)
\label{zetat}
\qqq
for $\,\gamma=v\beta_0\,$ with $\,\zeta^+_t(y)=\gamma\,$ outside an interval
$\,[A-O(|t|),B+O(|t|)]$. \,Similarly, for $\,\xi_{t,L}\,$
given by (\ref{xitL}),
\qq
\lim\limits_{L\to\infty}\,\xi_{t,L}(y+O_L^+)
=\zeta^+_t(y)-\gamma\equiv\xi^+_t(y)
\label{xit}
\qqq
with $\,\xi^+_t(y)\,$ vanishing outside $\,[A-O(|t|),B+O(|t|)]$.
\,In particular, if $\,vt\geq2\delta\,$ then
\qq
\hspace{-0.1cm}\xi^+_t(y)\,=\,\begin{cases}\,0\hspace{1.6cm}\text{for}
\hspace{3.28cm}y\leq A
-\frac{\beta_0}{\beta\CL}vt\m,\cr
\,\gamma\frac{\Delta\beta}{\beta_\CL}\hspace{1.05cm}\text{for}\quad A
-\frac{\beta_0}{\beta_\CL}(vt-2\delta)\leq y\leq A\m,\cr
\,0\hspace{1.58cm}\text{for}\hspace{2.57cm}B\leq y\m.
\end{cases}
\label{xiplus}
\qqq
\vskip 0.2cm

The functions $\,f_{s,t,L}\,$ forming the
flow of the vector field $\,-\zeta_{t,L}(y)\partial_y\,$ are equal
to $\,f_0-\gamma_Ls\,$ outside an interval
$\,[A_L-O(|t|+|s|),B_L+O(|t|+|s|)]\,$ when restricted to
$\,[-\frac{1}{4}L,\frac{1}{4}L]\,$ and their form stabilizes
inside that interval when $\,L\to\infty$. \,More exactly,
\qq
\lim\limits_{L\to\infty}\,\big(f_{s,t,L}\big(y+O_L^+)
-O_L^+\big)\,=\,f^+_{s,t}(y)\m,\label{conjug1}
\qqq
where the limiting functions $\,f^+_{s,t}\,$ form the flow of the vector field
$\,-\zeta^+_{t}(y)\partial_y$.
\,It will be convenient to introduce the shifted functions
\qq
g_{s,t,L}(y)=f_{s,t,L}(y)+\gamma_Ls
\label{gstl}
\qqq
equal to $\,y\,$ outside an interval
$\,[A_L-O(|t|+|s|),B_L+O(|t|+|s|)]\,$ when restricted to
$\,[-\frac{1}{4}L,\frac{1}{4}L]\,$ for which
\qq
\lim\limits_{L\to\infty}\,\big(g_{s,t,L}\big(y+O_L^+)
-O_L^+\big)\,=\,g^+_{s,t}(y)\,,
\label{conjug2}
\qqq
where the functions
$\,g^+_{s,t}(y)=f^+_{s,t}(y)+\gamma s\,$
are equal to $\,y\,$ outside an interval $\,[A-O(|t|+|s|),B+O(|t|+|s|)]$.
\,It is straightforward to see that the convergence in
(\ref{zetat}), (\ref{xit}), (\ref{conjug1}) and (\ref{conjug2})
is uniform on compacts with all derivatives and that it proceeds with speed 
$\,O(L^{-1})$.
\vskip 0.2cm

The symmetry relations (\ref{fluse}) and (\ref{fluse2}) imply then that
for
\qq
O_L^-=A-A_L-\frac{_1}{^2}L\,,
\label{OL-}
\qqq
\qq
&&\lim\limits_{L\to\infty}\,\zeta_{t,L}(y+O_L^-)\,=\,
\zeta^+_{-t}(-y)\equiv\zeta^-_{t}(y)\m,\label{zetat-}\\
&&\lim\limits_{L\to\infty}\,\xi_{t,L}(y+O_L^-)\,=\,
\xi^+_{-t}(-y)\equiv\xi^-_{t}(y)\m,\label{xit-}\\
&&\lim\limits_{L\to\infty}\,\big(f_{s,t,L}(y+O_L^-)
-O_L^-\big)\,=\,-f^+_{-s,-t}(-y)\equiv f^-_{s,t}(y)\m,\label{conjug3}\\
&&\lim\limits_{L\to\infty}\,\big(g_{s,t,L}(y+O_L^-)
-O_L^-\big)\,=\,f^-_{s,t}(y)+\gamma s\equiv g^-_{s,t}(y)\label{conjug4}
\qqq
with $\,f^-_{s,t}\,$ forming the flow of the vector field
$\,-\zeta^-_t(y)\partial_y$. \,Again the convergence 
is uniform on compacts with
all derivatives and speed $\,O(L^{-1})$. \,Similarly as for
$\,\xi^+_t\,$ and $\,g^+_{s,t}$, $\,\xi^-_t(y)=0\,$ and
$\,g^-_{s,t}(y)=y\,$ outside some bounded intervals. 
In particular, \,for $\,vt\geq 2\delta$,
\qq
\hspace{1cm}\xi^-_t(y)\,=\,\begin{cases}\,
0\hspace{1.6cm}\text{for}\hspace{3.93cm}y\leq-B-\frac{\beta_0}{\beta_R}vt\m,\cr
\,-\gamma\frac{\Delta\beta}{\beta_\CR}\hspace{0.78cm}\text{for}\hspace{0.5cm}
-B-\frac{\beta_0}{\beta_\CR}(vt-2\delta)
\leq y\leq -B\m,\cr
\,0\hspace{1.6cm}\text{for}\hspace{2.78cm}-A\leq y\m.
\end{cases}
\label{ximinus}
\qqq
\vskip 0.1cm

Note that the two different limits for $\,L\to\infty\,$
distinguished by the superscript $\,\pm\,$ were obtained in the
$\,L$-dependent frames with the centers at $\,O_L^\pm$, \,respectively,
\,where
\qq
O^+_L-O^-_L=-2h(a-\delta)+(2a-2\delta+\frac{_L}{^2})
\frac{_{\beta_{0,L}}}{^{\beta_\CL}}\,\equiv\,M_L
\label{Opm2}
\qqq
so that $\,O^\pm_L\,$ are separated by an $\,O(L)\,$ distance
from each other. For the later use, \,let us observe that
\qq
\lim\limits_{L\to\infty}\,L^{-1}M_L=\frac{_{\beta_\CR}}{^{\beta_\CL+\beta_\CR}}<1\,.
\label{ML/L}
\qqq
\vskip 0.2cm

\subsection{Conformal welding of cylinders}
\label{subsec:confweldcyl}

\noindent Below, for more clarity, we shall attempt to use capital Roman
letters for functions and operators in the finite-volume context
(whose dependence on $\,L\,$ will often be suppressed in the notation)
and capital script letters for functions and operators pertaining to
the infinite-volume context.
\vskip 0.2cm

A detailed analysis, \m that we shall perform in
Sec.\,\ref{sec:infvlim:proof}, \,will show that the derivatives
of the functions $\,X_{s,t,L;i}(x)\,$ appearing in Eqs.\,(\ref{PsitlambdaL})
and (\ref{hattaucw}) and obtained from conformal welding of the tori
$\,\CT_{f_{s,t,L},\tau_{s,L}}\,$ converge when considered in the 
frames centered in $\,O^\pm_L\,$ to the derivatives of functions
$\,\CX^\pm_{s,t;i}(x)\,$ that appear in the context of conformal welding
of cylinders that we shall describe now. 
\vskip 0.2cm

\begin{figure}[h]
\begin{center}
\vskip 0.1cm
\leavevmode
\hspace*{-0.8cm}
{%
      \begin{minipage}{0.4\textwidth}
      \vspace{0.0cm}          
        \includegraphics[width=3.9cm,height=2.8cm]
        {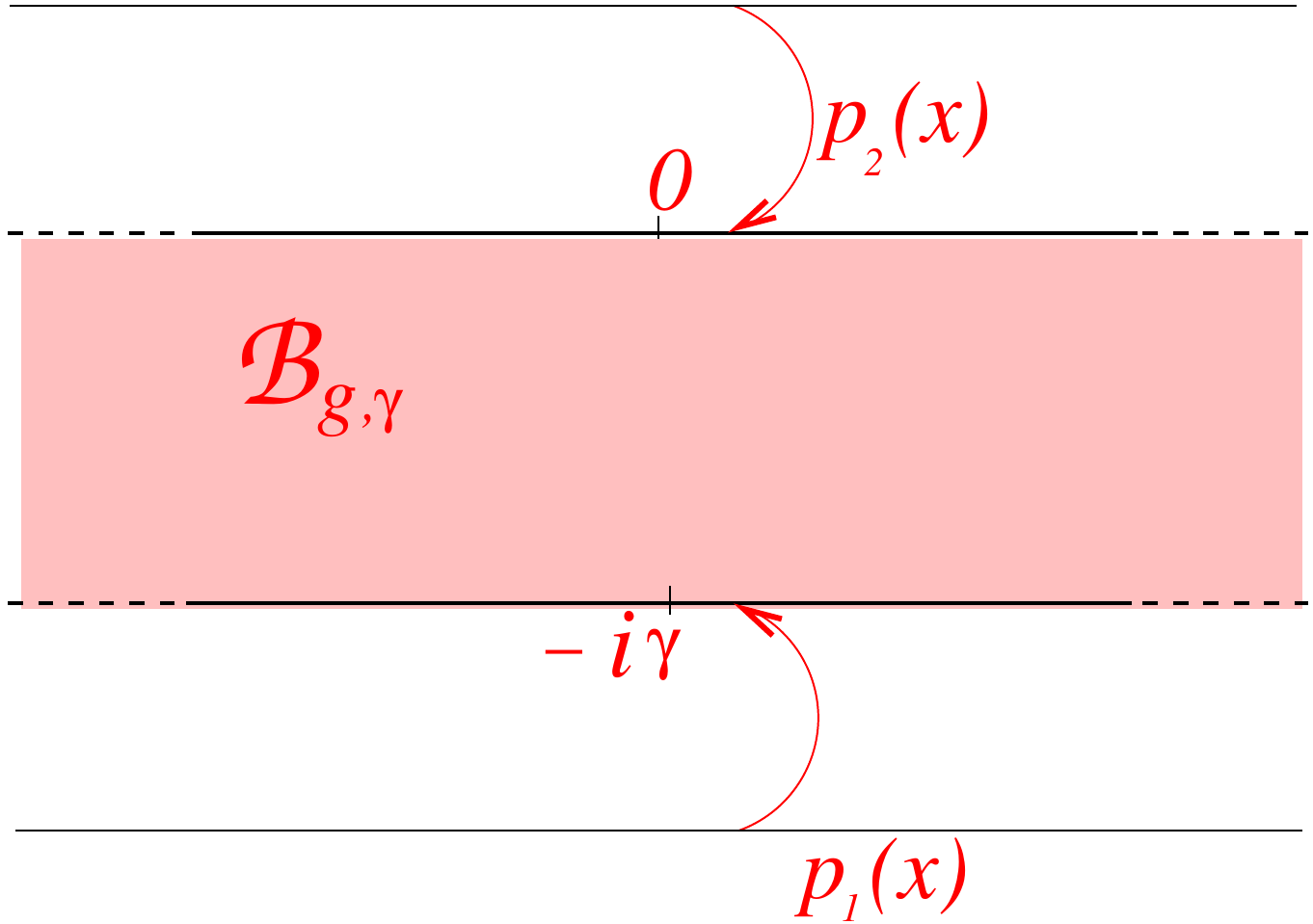}\\
      \vspace{-0.9cm} \strut
        \end{minipage}}\hspace*{-0.9cm}
{%
  \begin{minipage}{0.4\textwidth}
      \vspace{0.8cm}
        \includegraphics[width=4.2cm,height=1.5cm]
         {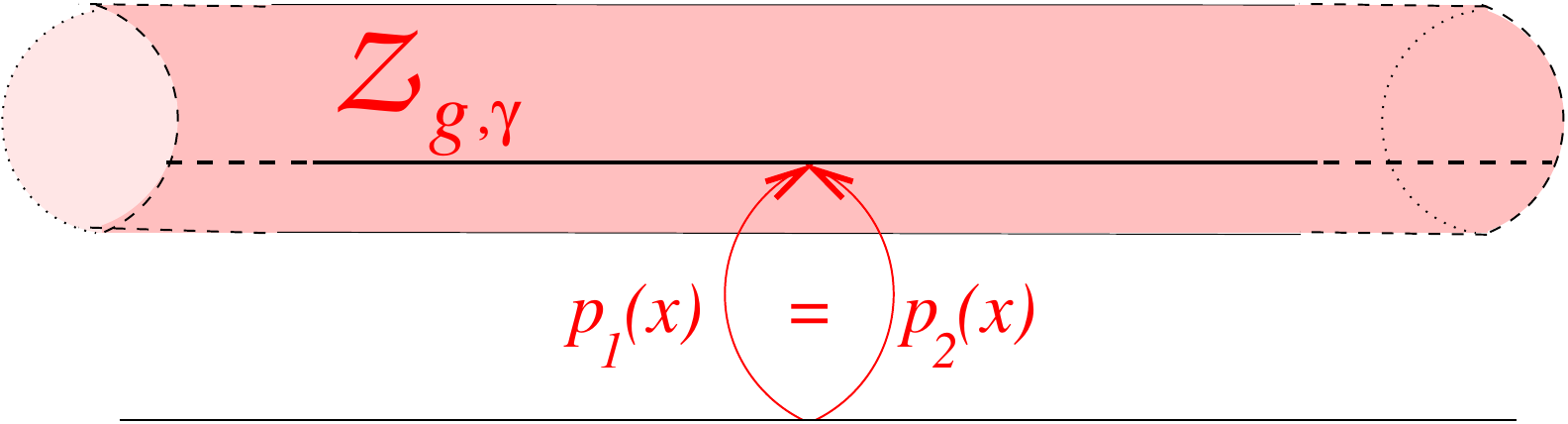}\\
         \vspace{-0.9cm} \strut
         \end{minipage}}\hspace*{-0.8cm}
         \vspace{0.55cm}
\end{center}
\vspace{-0.2cm}
\caption{Conformal welding of a cylinder}
\label{fig:strip_cylin}
\vspace{0.1cm}
\end{figure}

Consider for $\,\gamma=v\beta_0\,$ the infinite band 
\qq
\CB_{g,\gamma}\,=\,\big\{z\in\mathbb C\,\big|
\,-\ii \gamma\leq{\rm Im}(z)\leq0\big\}, 
\qqq
in the complex plane with the boundary components parameterized as
\qq
\mathbb R\ni x\,\longrightarrow\,p_1(x)=-\ii\gamma+g(x)\,,\qquad
\mathbb R\ni x\,\longrightarrow\,p_2(x)=x
\label{piCB}
\qqq
for a diffeomorphism $\,g:\mathbb R\rightarrow\mathbb R$, $\,g'(x)>0$,
equal to the identity outside a bounded interval.
Let $\,\CZ_{g,\gamma}\,$ be the complex cylinder obtained from
$\,\CB_{g,\gamma}\,$ by conformal welding that identifies
$\,p_1(x)\,$ with $\,p_2(x)$, \,see Fig.\,\ref{fig:strip_cylin}.

\noindent$\CZ_{g,\gamma}\m$ is isomorphic to
the standard cylinder $\,\CZ_{g_0,\gamma}\m$ for\footnote{$g_0=f_0\,$ are both
the identity diffeomorphism of $\,\R\,$ but $\,f_0\,$ was primarily viewed
as the unit of $\,\Difft$.} $\,g_0(x)\equiv x$, \,with
the isomorphism given by a function $\,\CX:\CB_{g,\gamma}\rightarrow\mathbb C\,$
holomorphic in the interior such that
its boundary values $\,\CX_i=\CX\circ p_i\,$ satisfy the relation
\qq
\CX_1(x)=\CX_2(x)-\ii\gamma\,.
\label{X1X2cyl}
\qqq
One may obtain such a function $\,\CX\,$ by conformal welding of discs
\cite{ShM}. To this end, let us first consider the map
\qq
\CB_{g,\gamma}\ni z\,\longmapsto\,u=\ee^{\frac{2\pi}{\gamma}z}
\qqq
that sends $\,\CB_{g,\gamma}\,$ onto $\,\mathbb C\,$ cut along
$\,\mathbb R_+\subset\mathbb C$, \,with the sides
of the cut parameterized by
\qq
\mathbb R\ni x\,\longmapsto\,\ee^{\frac{2\pi}{\gamma}g(x)}
+\ii0\equiv\tilde p_1(x)\qquad
\text{and}\qquad\mathbb R\ni x\,\longmapsto\,\ee^{\frac{2\pi}{\gamma}x}
-\ii0\equiv\tilde p_2(x)\,.
\qqq
Note that $\,\tilde p_1(x)\,$ and $\,\tilde p_2(x)\,$ agree for
$\,|x|\,$ large enough. Welding the sides of the cut by identifying
$\,\tilde p_1(x)=\tilde p_2(x)\,$ and adding the point $\,u=\infty$,
\,we obtain a closed Riemann surface $\,\CS_g\,$ of genus zero. 
Another way to describe the same surface may be obtained using the
map 
\qq
\,u\,\longmapsto\,\frac{{1-\ii u}}{{1+\ii u}}
\label{ufrakz}
\qqq
of $\,\mathbb C\cup\{\infty\}\cong\mathbb CP^1\,$ into itself that
sends the real line $\,\mathbb R\,$ to the unit circle
$\,S^1\,$ and $\,\mathbb R_+\subset\mathbb R\,$ into
the lower half of the circle. Then $\,S_g\,$ may be viewed as obtained from the two discs
into which the unit circle cuts $\,\mathbb C\cup\{\infty\}\,$
by welding them back together after twist by a circle diffeomorphism equal
to the identity on a neighborhood of the upper half of the circle. As discussed
\m e.g. in \cite{ShM}, one may construct an isomorphism of the surface
$\,S_g\,$ obtained this way with $\,\mathbb C\cup\{\infty\}\,$ by solving
an explicit Fredholm equation in $\,L^2(S^1)$. \,Composing that isomorphism
with the map (\ref{ufrakz}), \,one obtains a holomorphic map $\,\CW\,$ from
$\,\mathbb C\setminus\mathbb R_+\,$ to $\,\mathbb C\cup\{\infty\}\,$
whose boundary values satisfy the relation
\qq
\CW(\tilde p_1(x))=\CW(\tilde p_2(x))
\label{valsat}
\qqq
implying that $\,\CW\,$ is also holomorphic in neighborhoods of
$\,u=0\,$ and $\,u=\infty$. By composing $\,\CW\,$ with an appropriate
M\"obius transformation, we may also demand that $\,\CW(0)=0\,$ and
$\,\CW(\infty)=\infty\,$ so that
\qq
&&\CW(u)=\begin{cases}au+O(u^2)\qquad\text{\,for}
\qquad|u|\,
\ \text{small}\,,\cr
bu+O(1)\ \ \qquad\,\text{for}
\qquad|u|\,\ \text{large}\end{cases}\label{recentinf}
\qqq
for $\,a,b\not=0$. \,Then the map
\qq
\CB_{\gamma,g}\ni z\,\longmapsto\,\CX(z)\,
=\,\frac{_\gamma}{^{2\pi}}\ln{\CW(\ee^{\frac{2\pi}{\gamma} z})}\m,
\qqq
where $\,\ln(w)\,$ is chosen with a branch-cut along the image of
$\,\mathbb R_+\,$ by $\,\CW$, \,is holomorphic in the interior
of $\,\CB_{g,\gamma}\,$ and its boundary values satisfy the relation
(\ref{X1X2cyl}) so that $\,\CX\,$ defines the isomorphism $\,\CZ_{g,\gamma}
\cong\CZ_{g_0,\gamma}$. \,In particular, from (\ref{recentinf}) we infer
that
\qq
\CX_i'(x)=\frac{\gamma}{{2\pi}}\,\frac{\CW'(\tilde p_i(x))}{\CW(\tilde p_i(x))}
\m{\tilde p_i}^{\hspace{0.02cm}\prime}(x)\,
=\,\begin{cases}\,1+O(\ee^{\frac{2\pi}{\gamma}x})
\qquad\ \,\text{for}
\ \quad x\ll 0\m,\cr
\,1+O(\ee^{-\frac{2\pi}{\gamma}x})\hspace{-0.02cm}\qquad\text{for}
\ \quad x\gg 0\m.
\end{cases}
\label{expdec0}
\qqq
Besides, $\,\CX'_i(x)\not=0$. \,Writing
\qq
\CX_1(x)=g(x)-\CY_1(x)-\ii\gamma\m,
\qquad\CX_2(x)=x-\CY_2(x)=\CX_1(x)+\ii\gamma\m,
\label{X12Y12}
\qqq
so that
\qq
\CY_{12}(x)=\CY_1(x)-\CY_2(x)=g(x)-x\m,
\qqq
we infer from (\ref{expdec0}) that $\,\CY'_i(x)\,$ have an exponential decay
when $\,x\to\pm\infty$. \,Note that $\,\CY_i\,$ are the boundary
values for the function $\,\CY(z)=z-\CX(z)$.
\vskip 0.2cm

It is easy to obtain an integral equation for $\,\CY_1\,$ similar to
(\ref{Freq}) of Sec.\,\ref{subsec:confweldRH}. Indeed,
using the Cauchy formula,
\qq
\CY(z)=\frac{_1}{^{2\pi\ii}}\int\frac{g'(y)}{g(y)-\ii\gamma-z}\,\CY_1(y)\,dy
-\frac{_1}{^{2\pi\ii}}\int\frac{1}{y-z}\,\CY_2(y)\,dy
\qqq
for $\,z\,$ in the interior of $\,\CB_{g,\gamma}\,$ and sending $\,z\,$
to $\,p_i(x)$, \,one obtains the relations
\qq
&&\frac{_1}{^2}\CY_1(x)=
\frac{_1}{^{2\pi\ii}}\,PV\int\frac{g'(y)}{g(y)-g(x)}\,\CY_1(y)\,dy
-\frac{_1}{^{2\pi\ii}}\int\frac{1}{y-g(x)+\ii\gamma}\,\CY_2(y)\,dy\,,
\label{inft1}\\
&&\frac{_1}{^2}\CY_2(x)=
\frac{_1}{^{2\pi\ii}}\int\frac{g'(y)}{g(y)-\ii\gamma-x}\,\CY_1(y)\,dy
-\frac{_1}{^{2\pi\ii}}\,PV\int\frac{1}{y-x}\,\CY_2(y)\,dy
\label{inft2}
\qqq
that, with the help of the identity
\qq
\frac{_1}{^2}\big((\CE_+-\CE_-)\CX\big)(x)
=-\frac{_1}{^{2\pi\ii}}\,PV\int\frac{1}{y-x}\,\CX(y)\,dy\,,
\qqq
where $\,\CE_\pm\,$ are the orthogonal projections on functions
$\,\CX\in L^2(\mathbb R)\,$ with the Fourier transform
\qq
\widehat\CX(p)=\int\ee^{\ii px}\m\CX(x)\,dx 
\label{Fourtr}
\qqq
vanishing outside
$\,\mathbb R_{\pm}$, respectively,  may be rewritten in the form
\qq
\CE_+\CY_1=\CK_{11}\CY_1+\CK_{12}\CY_2\,,\qquad \CE_-\CY_2=\CK_{21}\CY_1\,,
\label{inft3}
\qqq
for
\qq
&&(\CK_{11}\CX)(x)=\frac{_1}{^{2\pi\ii}}\int\Big(\frac{g'(y)}{g(y)-g(x)}
-\frac{1}{y-x}\Big)\CX(y)\,dy\m,\label{CK11}\\
&&(\CK_{12}\CX)(x)=-\frac{_1}{^{2\pi\ii}}\int\frac{1}{y-g(x)+\ii\gamma}\,
\CX(y)\,dy\m,\label{CK12}\\
&&(\CK_{21}\CX)(x)=\frac{_1}{^{2\pi\ii}}\int\frac{g'(y)}{g(y)-x-\ii\gamma}\,
\CX(y)\,dy\m.\label{CK21}
\qqq
The summation of (\ref{inft3}) gives then rise to the integral equation
\qq
(\CI-\CK)\CY_1=(\CE_--\CK_{12})\CY_{12}\,,
\label{inteqinfty}
\qqq
where $\,\CI\,$ stands for the identity operator in $\,L^2(\R)\,$ and
\qq
\CK=\CK_{11}+\CK_{12}+\CK_{21}\,.
\label{opCK}
\qqq
\,We also have the representations:
\qq
\CK_{11}=-\CE_-+\CG^{-1}\CE_-\CG=\CE_+-\CG^{-1}\CE_+\CG\,,
\qquad \CK_{12}=\CG^{-1}\CQ\m\CE_+\,,\qquad\CK_{21}=\CE_-\CQ^{-1}\CG\,,
\label{CFCQij}
\qqq
where $\,(\CG\CX)(x)=\CX(g^{-1}(x))\,$ and
$\,\CQ\,\ee^{-\ii px}=\ee^{-\gamma p}\,\ee^{-\ii px}$.
Contrary to the operator $\,K\,$ of Sec.\,\ref{subsec:confweldRH},
the operator $\,\CK\,$ in $\,L^2(\R)\,$ is not trace-class
and the solution of Eq.\,(\ref{inteqinfty}) will require more care.
\vskip 0.2cm

Note that for $\,f=f_{s,t,L}\,$ and $\,\tau=\tau_{s,L}\,$ the kernels
of the operators $\,K_{ij}\,$ of \m(\ref{K11})-(\ref{K21}) 
\m become
\qq
&&K_{11}(x,y)=-\frac{1}{L}\Big(\frac{g'_{s,t,L}(y)}{1-\ee^{\frac{2\pi\ii}{L}
(g_{s,t,L}(y)-g_{s,t,L}(x))}}-\frac{1}{1-\ee^{-\frac{2\pi\ii}{L}(x-y)}}\Big)\,\\
&&K_{12}(x,y)=\frac{1}{L\Big(1-\ee^{\frac{2\pi\ii}{L}(y-g_{s,t,L}(x)
+\ii\gamma_L)}\Big)}\,,\\
&&K_{21}(x,y)=-\frac{g'_{s,t,L}(y)}{L\Big(1-\ee^{\frac{2\pi\ii}{L}(g_{s,t,L}(y)
-x-\ii\gamma_L)}\Big)}\,,
\qqq
where $\,g_{s,t,L}\,$ are given by (\ref{gstl}).
From the results of Sec.\,\ref{subsec:largeLzeta} it follows that
in the frames centered at $\,O^\pm_L\,$ of (\ref{OL+}) and (\ref{OL-})
they converge
pointwise to the kernels of operators $\,\CK_{ij}\,$ for $\,g=g^\pm_{s,t}\,$ when
$\,L\to\infty$. \,This renders plausible the convergence of the derivatives
of functions $\,X_{s,t,L;i}\,$ in the recentered frames to the derivatives
of the functions $\,\CX^\pm_{s,t;i}\,$ obtained from conformal
welding of the cylinders $\,\CZ_{g^\pm_{s,t},\gamma}$. \,Indeed, the corresponding
functions are determined in terms of the solution of, respectively,
Eq.\,(\ref{Freq}) and (\ref{inteqinfty}). A rigorous proof of such
a convergence, however, requires more subtle analysis that will be presented
in Sec.\,\ref{sec:infvlim:proof}.
\vskip 0.3cm

\subsection{Infinite-volume formula for the FCS generating function}
\label{subsec:infvformula}

\noindent We shall assume here the uniform convergence on
compacts of derivatives of the functions $\,X_{s,t,L;i}\,$ occurring
in (\ref{PsitlambdaL}) and (\ref{hattaucw}) in the frames with centers at
$\,O_L^\pm\,$ to derivatives of functions $\,\CX^\pm_{s,t;i}\,$
obtained from conformal welding of cylinders
$\,\CZ_{g^\pm_{s,t},\gamma}$, \,in agreement with the discussion of the last
subsection\footnote{We also assume that the above
convergence is uniform in $\,s\,$ for $\,s\,$ bounded.}.
For the first term on the right-hand side of (\ref{PsitlambdaL}),
\,we obtain then the limiting behavior
\qq
&&\lim\limits_{L\to\infty}\,\exp\hspace{-0.07cm}\bigg[\hspace{-0.04cm}
-\ii\,\frac{_c}{^{24\pi}}\int_0^{\frac{\lambda}{\Delta\beta}}
\hspace{-0.15cm}ds\int_{\CI_L}\hspace{-0.17cm}
\xi_{t,L}(x)(SX_{s,t,L;i})(x)\,dx\bigg]\cr
&&=\,\exp\hspace{-0.07cm}\bigg[\hspace{-0.04cm}
-\ii\,\frac{_c}{^{24\pi}}\sum\limits_\pm\int_0^{\frac{\lambda}{\Delta\beta}}
\hspace{-0.15cm}ds\int
\xi^\pm_{t}(x)(S\CX^\pm_{s,t;i})(x)\,dx\bigg].\quad
\label{pierw}
\qqq
\vskip 0.2cm

We are still left with the control of the $\,L\to\infty\,$ limit of
the ratio of Virasoro characters in (\ref{PsitlambdaL}). Note that
$\,\tau_{0,L}=\ii\m L^{-1}\gamma_L=O(L^{-1})\,$ and that
$\,\widehat\tau_{\frac{\lambda}{\Delta\beta},t,L}-\tau_{0,L}=O(L^{-2})\,$ as follows 
from (\ref{hattaucw}) and the relation
\qq
\lim\limits_{L\to\infty}\,\int_0^{\frac{\lambda}{\Delta\beta}}\hspace{-0.13cm}ds
\hspace{0.03cm}\int_{\CI_L}\hspace{-0.1cm}\xi_{t,L}(x)\,
(X'_{s,t,L;i}(x))^2\,dx
\,=\,\sum\limits_{\pm}\int_0^{\frac{\lambda}{\Delta\beta}}\hspace{-0.13cm}ds
\int\xi^\pm_{t}(x)\,({\CX^\pm_{s,t;i}}^{\hspace{-0.25cm}\prime}\hspace{0.14cm}(x))^2\,dx\,.
\qqq
The behavior when $\,\tau\to 0\,$ for the Virasoro character
$\,\chi(\tau)\,$ 
of the finite-volume theory may be obtained from the dual picture
or the modular properties of the characters of 
of the Virasoro algebra or its extensions: 
\qq
\chi(\tau)=
C\,\ee^{2\pi\ii\frac{c}{24\tau}}\big(1+o(e^{-\frac{\epsilon}{|\tau|}})\big),
\label{smalltau}
\qqq
where $\,C>0\,$ and $\,\epsilon>0\,$ are constants dependent on the theory
but independent of $\,\tau$. \,This holds e.g. for general rational unitary
CFTs where $\,\chi(\tau)\,$ is a finite sum of the characters of
a chiral algebra transforming linearly under $\,\tau\mapsto-\frac{1}{\tau}\,$
\cite{DFMS}, \,but also for toroidal compactifications of free fields, e.g.
for the massless bosonic field considered in Sec.\,\ref{subsec:exB} with any
radius of compactification. Assuming (\ref{smalltau}), we infer that
\qq
&&\lim\limits_{L\to\infty}\,\frac{\chi(\widehat\tau_{\hspace{-0.06cm}
\frac{\lambda}{\Delta\beta},t,L})}
{\chi(\tau_{0,L})}\,=\,\exp\Big[-\ii\frac{_{\pi c}}{^{12}}\,\lim\limits_{L\to\infty}
\frac{_{\hspace{-0.08cm}\widehat\tau_{\frac{\lambda}{\Delta\beta},t,L}-\,\m\tau_{0,L}}}{^{\tau_{0,L}^2}}\Big]\cr
&&=\,\exp\Big[\ii\m\frac{_{\pi c}}{^{12\gamma^{2}}}\,
\sum\limits_{\pm}\int_0^{\frac{\lambda}{\Delta\beta}}\hspace{-0.13cm}ds
\int\xi^\pm_{t}(x)\,({\CX^\pm_{s,t;i}}^{\hspace{-0.25cm}\prime}\hspace{0.12cm}
(x))^2\,dx\Big],
\label{postat}
\qqq
where, as before, $\,\gamma=v\beta_0$.
\,Collecting (\ref{pierw}), (\ref{postat}) and (\ref{ostat}), we obtain
for the infinite-volume generating function for FCS of energy transfers
the expression
\qq
\Psi_t(\lambda)\,\equiv\,\lim\limits_{L\to\infty}\,\Psi_{t,L}(\lambda)\,=\,
\prod\limits_{\pm}\Psi_t^\pm(\lambda)\,,
\label{Psitlambda0}
\qqq
where $\,\Psi^\pm_{t}(\lambda)\,$ are the contributions of the
right- and left-movers, \m respectively, \m given by the relations
\qq
&&\Psi^\pm_t(\lambda)\,=\,\exp\hspace{-0.06cm}\bigg[-\ii\,\frac{_c}{^{24\pi}}
\,\bigg(\int_0^{\frac{\lambda}{\Delta\beta}}\hspace{-0.15cm}ds\int
\xi^\pm_t(x)\m\Big((S\CX^\pm_{s,t;i})(x)-\frac{_{2\pi^2}}{^{\gamma^2}}\,
({\CX^\pm}'_{\hspace{-0.23cm}s,t;i}(x))^2\Big)\m dx\cr
&&\hspace{5.55cm}+\,\frac{_{\lambda\hspace{0.01cm}v}}{^{\Delta\beta}}\int
\big(\beta(x^\pm)-\beta(x)\big)\,(Sh)(x)\,dx\bigg)\bigg].
\label{Psitlambda}
\qqq
The last formulae show that in the infinite-volume, the generating function
$\,\Psi_t(\lambda)\,$ for the FCS of energy transfers in the non-equilibrium
profile states is universal depending only on the profile $\,\beta(x)\,$ and
the central charge $\,c\,$ of the CFT but not on the spectrum of the theory.
Besides, the central charge enters simply as an overall power.

\subsection{Simple checks: the first two moments of the energy transfer}
\label{subsec:checks}

\noindent In virtue of (\ref{1stmom}), the average energy transfer through
the kink is given in the infinite-volume limit by the expression
\qq
&&\langle \Delta E\rangle^{\ }_{\hspace{0.01cm}t}
=\lim\limits_{L\to\infty}\,\langle \Delta E\rangle^{\ }_{t,L}
=\lim\limits_{L\to\infty}\,
\frac{_1}{^{\Delta\beta}}\big\langle G_L(t)-G_L(0)
\big\rangle^{\rm neq}_{\,L}\cr
&&=\frac{_v}{^{\Delta\beta}}\int\Big((\beta(x^+)-\beta(x))
\big\langle T_+(x)\big\rangle^{\rm neq}+(\beta(x^-)-\beta(x))\big\langle
T_-(x)\big\rangle^{\rm neq}\Big)dx\cr
&&=\frac{_v}{^{\Delta\beta}}\int\Big((\beta(x^+)-\beta(x))\big(\frac{_{\pi c}}
{^{12(v\beta(x))^2}}-\frac{_c}{^{24\pi}}(Sh)(x)\big)
\,+\,(\beta(x^-)-\beta(x))\big(\frac{_{\pi c}}{^{12(v\beta(x))^2}}-\frac{_c}{^{24\pi}}(Sh)(x)\big)\Big)dx\cr
&&=\frac{_{\pi c}}{^{12\gamma^2\Delta\beta}}\sum\limits_{\pm}
\int\xi^\pm_t(y)\,dy
\,-\frac{_{cv}}{^{24\pi\Delta\beta}}\sum\limits_{\pm}\int\Big(\beta(x^\pm)-\beta(x)\Big)(Sh)(x)\,dx\,,\qquad
\qqq
where we have used (\ref{Tinfty}), (\ref{zetat}) and (\ref{zetat-}),
\m changing the integration variable to $\,y=h(x)\,$ in the part of
the integral. The result agrees with
$\,\frac{_1}{^\ii}\,\partial_\lambda\m\ln{\Psi_t(0)}\,$
for $\,\Psi_y(\lambda)\,$ given by (\ref{Psitlambda0}) and
(\ref{Psitlambda}) since $\,{\CX^\pm}'_{\hspace{-0.21cm}0,t;i}(x)=1$.
\vskip 0.2cm

For the variance of the energy transfer given by (\ref{2ndmom}),
we obtain in the infinite-volume limit the expression
\qq
&&\big\langle\Delta E;\Delta E\big\rangle^c_t=\lim\limits_{L\to\infty}\,
\big\langle\Delta E;\Delta E\big\rangle^c_{t,L}
=\lim\limits_{L\to\infty}\,\frac{_1}{{^{(\Delta\beta)^2}}}
\Big\langle G_L(t)-G_L(0)\,;\,G_L(t)-G_L(0)\Big\rangle^{\rm neq,\,c}_L\cr
&&=\frac{_{v^2}}{^{(\Delta\beta)^2}}\int\Big(\big(\beta(x_1^+)-\beta(x_1)\big)
\big(\beta(x_2^+)-\beta(x_2)\big)\Big\langle T_+(x_1)\,;\,T_+(x_2)
\Big\rangle^{{\rm neq},\,c}\cr
&&\hspace{1.7cm}+\,\big(\beta(x_1^-)-\beta(x_1)\big)
\big(\beta(x_2^-)-\beta(x_2)\big)\Big\langle T_-(x_1)\,;\,T_-(x_2)
\Big\rangle^{{\rm neq},\,c}\,\Big)\m dx_1dx_2\cr
&&=\frac{_{\pi^2 c}}{^{8(\Delta\beta)^2\gamma^4}}\sum\limits_{\pm}\int\bigg(
\frac{\xi^\pm_t(y_1)\,
\xi^\pm_t(y_2)}{\sinh^4\big(\frac{\pi}{\gamma}(y_1-y_2-\ii0)\big)}
\bigg)dy_1dy_2\,,
\label{2cum}
\qqq
where we used (\ref{TTinfty}), (\ref{TTTinfty}) and (\ref{zetat})
and the way the singularity $\,\propto\frac{1}{(y_1-y_2)^4}\,$
has been treated was obtained by writing
\qq
&&\Big\langle G_L(t)-G_L(0)\,;\,G_L(t)-G_L(0)\Big\rangle^{{\rm neq},\,c}_{\,L}\cr
&&=\,\lim\limits_{\epsilon\searrow0}
\,\Big\langle G_L(t)-G_L(0)\,;
\,\ee^{-\epsilon G_L(0)}\big(G_L(t)-G_L(0)\big)\,\ee^{\epsilon G_L(0)}
\Big\rangle^{{\rm neq},\,c}_L.\ \quad
\qqq
Expressing $\,\xi^\pm_t(y)\,$ by its Fourier transform,
\qq
\widehat{\xi^\pm_t}(p)=\int\ee^{\ii py}\,\xi^\pm_t(y)\,dy\,,
\qqq
and using the result (\ref{FTcalc}) from Appendix \ref{app:B}, we infer that
\qq
\big\langle\Delta E;\Delta E\big\rangle^c_t&=&
\frac{_{\pi^2 c}}{^{8(\Delta\beta)^2\gamma^4}}\frac{_1}{^{(2\pi)^2}}
\sum\limits_{\pm}
\int\Big(\frac{\ee^{-\ii(p_1y_1+p_2y_2)}}{\sinh^4\big(\frac{\pi}{\gamma}(y_1-y_2
-\ii0)\big)}\,\widehat{\xi^\pm_t}(p_1)\,\widehat{\xi_t^\pm}(p_2)\Big)\m
dp_1dp_2dy_1dy_2\cr
&=&\frac{_{c}}{^{48\pi^2(\Delta\beta)^2}}\sum\limits_{\pm}
\int
\frac{p\big(p^2+\frac{{4\pi^2}}{{\gamma^2}}\big)}{1-\ee^{-\gamma p}}
\,\widehat{\xi^\pm_t}(p)\,\widehat{\xi^\pm_t}(-p)\,dp\,.
\label{cummul21}
\qqq
In order to compare this to $\,-\partial_\lambda^2\m\ln{\Psi_t(0)}$,
\,we have to find the $\m1^{\rm st}\,$ order correction in $\,s\,$
given by $\,\partial_s{\CX^\pm}'_{\hspace{-0.2cm}0,t;i}(x)\,$ to
$\,{\CX^\pm}'_{\hspace{-0.2cm}0,t;i}=1$. \,Write
$\,\CX^\pm_{s,t;1}(x)=g^\pm_{s,t}(x)-\CY_{s;1}(x)-\ii\gamma$,
\,see (\ref{X12Y12}), recalling
that $\,\CY_{s;1}\,$ satisfies (\ref{inteqinfty}) for $\,g=g^\pm_{s,t}$.
\,Differentiating the latter equation with $\,\CY_{12}=g^\pm_{s,t}-g_0\,$
over $\,s\,$ at $\,s=0$, \,where, as before, $\,g_0(x)=x$, \,one gets:
\qq
(1-\CK^0)\partial_s\CY_{0;1}=(\CE_--\CK^0_{12})(-\xi^\pm_t)
\label{CK0}
\qqq
where $\,\CK^0_{ij}\,$ correspond to $\,g=g_0$. \,In terms of the Fourier
transforms, (\ref{CK0}) reads\footnote{$\theta(q)\,$ denotes
the Heaviside step function.}:
\qq
\big(1-\ee^{-\gamma|p|}\big)\partial_s\widehat{\CY_{0;1}}(p)=-
\big(\theta(-p)-\ee^{-\gamma p}\theta(p)\big)\widehat\xi^\pm_t(p)
\qqq
and we infer that
\qq
&&\partial_s{\CX^\pm}'_{\hspace{-0.23cm}0,t;1}(x)=-{\xi^\pm_t}^\prime(x)
-\frac{_\ii}{^{2\pi}}\int
\ee^{-\ii px}\,\frac{p(\theta(-p)-\ee^{-\gamma p}\theta(p))}{1-\ee^{-\gamma|p|}}
\,\widehat\xi^\pm_t(p)\,dp\,=\,\frac{_\ii}{^{2\pi}}\int
\frac{p\,\ee^{-\ii px}}{1-\ee^{-\gamma p}}\,\widehat\xi^\pm_t(p)\,dp\,,\cr
&&\partial_s(S\CX^\pm_{0,t;1})(x)=\partial_s{\CX^\pm}'''_{\hspace{-0.28cm}0,t;1}(x)=
-\frac{_\ii}{^{2\pi}}\int\frac{p^3\,\ee^{-\ii px}}{1-\ee^{-\gamma p}}\,
\widehat\xi^\pm_t(p)\,dp\,.
\qqq
Hence
\qq
&&-\partial_\lambda^2\m\ln{\Psi_t(0)}\,=\,\ii\,\frac{_c}{^{24\pi(\Delta\beta)^2}}
\sum\limits_{\pm}
\int\xi^\pm_t(x)\Big(\partial_s(S\CX^\pm_{0,t;1})(x)-\frac{_{4\pi^2}}{^{\gamma^2}}
{\CX^\pm}'_{\hspace{-0.1cm}0,t;1}(x)\m\partial_s{\CX^\pm}'_{\hspace{-0.2cm}0,t;1}(x)
\Big)\,dx\cr
&&=\,\frac{_c}{^{48\pi^2(\Delta\beta)^2}}\sum\limits_{\pm}
\int\xi^\pm_t(x)\Big(\int\frac{p\m(p^2+\frac{4\pi^2}{\gamma^2})
\,\ee^{-\ii px}}{1-\ee^{-\gamma p}}\,\widehat\xi^\pm_t(p)\,dp\Big)\,dx\cr
&&=\,\frac{_c}{^{48\pi^2(\Delta\beta)^2}}\sum\limits_{\pm}\int
\frac{p\big(p^2+\frac{{4\pi^2}}{{\gamma^2}}\big)}{1-\ee^{-\gamma p}}
\,\widehat\xi^\pm_t(p)\,\widehat\xi^\pm_t(-p)\,dp
\qqq
which agrees with (\ref{cummul21}).

\nsection{Long-time behavior of FCS}
\label{sec:longtime}

\noindent It is not difficult to understand heuristically
the long-time large-deviations type asymptotics of the right-hand side
of (\ref{Psitlambda}). Using the relations
(\ref{xiplus}) and (\ref{ximinus}), we observe that the functions
$\,\xi^\pm_t\m$ are, respectively, equal to constant values
$\,\gamma\frac{\Delta\beta}{\beta_\CL}\,$
and $\,-\gamma\frac{\Delta\beta}{\beta_\CR}\,$ for $\,\gamma=v\beta_0\,$
on intervals of length $\,\frac{\gamma}{\beta_\CL}t-O(1)\,$
and $\,\frac{\gamma}{\beta_\CR}t-O(1)$, \,and that they vanish
outside $\,O(1)\,$ extensions of those intervals. 
\,It follows that the shifted flows $\,g_{s,t}^\pm(x)=f_{s,t}^\pm(x)+\gamma s$,
which satisfy the equations
\qq
\partial_sg^\pm_{s,t}(x)=-\xi^\pm_t(g^\pm_{s,t}(x)-\gamma s)\,,\qquad g^\pm_{0,t}(x)=x\,,
\qqq
are equal, respectively, to $\,x-\gamma\frac{\Delta\beta}{\beta_\CL}s\equiv
\tilde g^+_s(x)\,$ and to
$\,x+\gamma\frac{\Delta\beta}{\beta_\CR}s\equiv\tilde g^-_s(x)\,$ on those
intervals shortened by $\,O(|s|)\,$ on either side. \,For estimating
the right-hand side of (\ref{Psitlambda}), we would like to
know the behavior of the functions $\,{\CX^\pm}'_{\hspace{-0.23cm}s,t;i}\,$
on the support of $\,\xi^\pm_t$. \,In the bulk of the support, we expect
$\,{\CX^\pm}'_{\hspace{-0.23cm}s,t;i}(x)\,$ to fast approach for large $\,t\,$
the functions obtained from conformal welding of the cylinders
$\,\CZ_{\tilde g^\pm_s\hspace{-0.05cm},\gamma}$. The holomorphic maps
$\,\widetilde\CX^\pm_s:\CB_{\tilde g^\pm_s\hspace{-0.05cm},\gamma}
\rightarrow\mathbb C\,$
generating the isomorphisms $\,\CZ_{\tilde g^\pm_s\hspace{-0.05cm},\gamma}
\cong\CZ_{g_0,\gamma}\,$
are in that case given simply by the multiplication by complex factors:
\qq
\widetilde\CX^+_s(z)
=\big(1-\ii\frac{_{\Delta\beta}}{^{\beta_\CL}}s\big)^{-1}z+C^+\,,\qquad
\widetilde\CX^-_s(z)
=\big(1+\ii\frac{_{\Delta\beta}}{^{\beta_\CR}}s\big)^{-1}z+C^-
\qqq
up to arbitrary additive constants $\,C^\pm$. \,Indeed,
\qq
\widetilde\CX^+_{s;1}(x)&=&\widetilde\CX^+_s(-\ii\gamma+\tilde g^+_s(x))=
\big(1-\ii\frac{_{\Delta\beta}}{^{\beta_\CL}}s\big)^{-1}
\big(-\ii\gamma+x-\gamma\frac{_{\Delta\beta}}{^{\beta_\CL}}s\big)+C^+\cr
&=&\big(1-\ii\frac{_{\Delta\beta}}{^{\beta_\CL}}s\big)^{-1}x+C^+-\ii \gamma=
\widetilde\CX^+_{s;2}(x)-\ii\gamma\,,\cr
\widetilde\CX^-_{s;1}(x)&=&\widetilde\CX^-_s(-\ii\gamma+\tilde g^-_s(x))=
\big(1+\ii\frac{_{\Delta\beta}}{^{\beta_\CR}}s\big)^{-1}
\big(-\ii\gamma+x+\gamma\frac{_{\Delta\beta}}{^{\beta_\CR}}s\big)+C^-\cr
&=&\big(1+\ii\frac{_{\Delta\beta}}{^{\beta_\CR}}s\big)^{-1}x+C^--\ii\gamma=
\widetilde\CX^-_{s;2}(x)-\ii\gamma
\qqq
so that
\qq
{\widetilde\CX^+_{s;i}}{}^{\hspace{-0.07cm}\prime}(x)
=\big(1-\ii\frac{_{\Delta\beta}}{^{\beta_\CL}}s\big)^{\hspace{-0.05cm}-1}\,,\qquad
{\widetilde\CX^-_{s;i}}{}^{\hspace{-0.07cm}\prime}(x)
=\big(1+\ii\frac{_{\Delta\beta}}{^{\beta_\CR}}s\big)^{\hspace{-0.05cm}-1}\,.
\qqq
A rigorous proof of the convergence of the derivatives
of functions $\,{\CX^\pm}_{\hspace{-0.22cm}s,t;i}\,$ to those of
$\,{\widetilde\CX^\pm}_{s;i}(x)\,$ in the bulk of the support
of $\,\xi^\pm_t\,$ requires a rather subtle control of the solutions
of the Fredholm equation that computes $\,{\CX^\pm}_{\hspace{-0.22cm}s,t;i}\,$
that we have not performed in detail. Assuming such a convergence, the
only contributions to the large-deviations rates
\qq
\Xi^\pm(\lambda)\,=\,\lim\limits_{t\to\infty}\,\frac{1}{t}\,\ln{\Psi^\pm_t(\lambda)}
\qqq
would come from the middle terms on the right-hand side of (\ref{Psitlambda})
leading to the formulae
\qq
&&\Xi^+(\lambda)\,=\,\ii\frac{_{\pi c}}{^{12\gamma^2}}\int_0^{\frac{\lambda}{\Delta\beta}}
\hspace{-0.13cm}\frac{_{\gamma}}{^{\beta_\CL}}\gamma\frac{_{\Delta\beta}}{^{\beta_\CL}}
\big(1-\ii\frac{_{\Delta\beta}}{^{\beta_\CL}}s\big)^{\hspace{-0.05cm}-2}\m ds\,=\,
\frac{_{\pi c}}{^{12}}\Big(\frac{_1}{^{\beta_\CL-\ii\lambda}}
-\frac{_1}{^{\beta_\CL}}\Big),\cr
&&\Xi^-(\lambda)\,=\,\ii\frac{_{\pi c}}{^{12\gamma^2}}
\int_0^{\frac{\lambda}{\Delta\beta}}
\hspace{-0.1cm}\frac{_\gamma}{^{\beta_\CR}}(-\gamma)\frac{_{\Delta\beta}}{^{\beta_\CR}}
\big(1+\ii\frac{_{\Delta\beta}}{^{\beta_\CR}}s\big)^{\hspace{-0.05cm}-2}\m ds\,=\,
\frac{_{\pi c}}{^{12}}\Big(\frac{_1}{^{\beta_\CR+\ii\lambda}}
-\frac{_1}{^{\beta_\CR}}\Big),
\qqq
in agreement with the Bernard-Doyon result \cite{BD0} obtained for
the partitioning protocol. For $\,c=1$, \,the latter expressions also
agree with the large-volume long-time limit of the Levitov-Lesovik formula
for two channels of free massless fermions with pure transmission that gives
\cite{LL2,LLL,MBHD}
\qq
\sum\limits_\pm\Xi^\pm(\lambda)=\frac{_1}{^{2\pi}}\int\ln\Big[1+
f_\CL(\omega)(1-f_\CR(\omega))(\ee^{\ii\lambda\omega}-1)+f_{\CR}(\omega)(1-f_{\CL}
(\omega))(\ee^{-\ii\lambda\omega}-1)\Big]\m d\omega\m,
\qqq
where $\,f_\CL(\omega)=\frac{1}{{\ee^{\beta_\CL\omega}+1}}\,$ and
$\,f_\CR(\omega)=\frac{1}{{\ee^{\beta_\CR\omega}+1}}\,$
are the Fermi functions for the right and left movers.
Observe that the functions $\,\Xi^\pm(\lambda)\,$
depend on the inverse-temperature profile only via its asymptotic values
$\,\beta_\CL,\beta_\CR\,$ exhibiting even more universality than $\,\Psi_t^\pm(\lambda)$. 
\,The Legendre transform
\qq
\CI(\sigma)\,=\mathop{{\rm sup}}\limits_{\nu\in ]-\beta_\CR,\beta_\CL[}
\hspace{-0.1cm}\Big(\nu\sigma\,-\,\frac{_{\pi c}}{^{12}}\Big(\frac{_\nu}
{^{\beta_\CL(\beta_\CL-\nu)}}-\frac{_\nu}{^{\beta_\CR(\beta_\CR+\nu)}}\Big)
\Big)
\qqq
with the asymptotic behavior
\qq
\CI(\sigma)=\begin{cases}\ \beta_\CL\sigma
-\sqrt{\frac{\pi c}{3}\sigma}\,+\,O(1)\quad
\hspace{0.46cm}\text{when}\quad \sigma\to\infty\cr
-\beta_\CR\sigma-\sqrt{-\frac{\pi c}{3}\sigma}\,+\,O(1)\quad
\text{when}\quad \sigma\to-\infty
\end{cases}
\qqq
determines the long-time large-deviations form of the of energy transfers
in the thermodynamic limit:
\qq
P_t(\Delta E)\,=\,\lim\limits_{L\to\infty}\,P_{t,L}(\Delta E)\ \
{_{\substack{_{\smile}\\ ^{\frown}}}}\ 
\,\exp\big[-t\hspace{0.03cm}\CI\big(\frac{_{\Delta E}}{^t}\big)\big].
\qqq
The rate function possesses the Gallavotti-Cohen symmetry \cite{GalCoh}:
$\,\CI(-\sigma)=\CI(\sigma)+\sigma\Delta\beta\,$ that follows here 
from the transient fluctuation relation (\ref{CFR}).
\vskip 0.1cm

Another way to characterize the behavior
of the FCS at large times is to note that
\qq
t\m\frac{_{\pi c}}{^{12}}\Big(\frac{_{\ii\lambda}}{^{\beta_\CL(\beta_\CL-\ii\lambda)}}
-\frac{_{\ii\lambda}}{^{\beta_\CR(\beta_\CR+\ii\lambda)}}\Big)=\,
t\frac{_{\pi c}}{^{12}}\int(\ee^{\ii\lambda q}-1)
\Big(\ee^{-\beta_\CL q}\theta(q)+\ee^{\beta_\CR q}\theta(-q)\Big)dq
\label{LK}
\qqq
is the logarithm of the Fourier transform
of the time-$t\,$ distribution of a L\'evy process \cite{Lawler} with
the jump rates
\qq
w(x,y)=\frac{_{\pi c}}{^{12}}\Big(\ee^{-\beta_\CL(y-x)}\theta(y-x)
+\ee^{-\beta_\CR(x-y)}\theta(x-y)\Big)
\qqq
that starts at $\,x=0$. \,The right hand side of (\ref{LK}) gives the
L\'evy-Khintchine representation for the infinitely divisible distribution
of such a process, see also \cite{BD0,BD2}.

\nsection{Thermodynamic limit of FCS: \m proof of convergence}
\label{sec:infvlim:proof}

\noindent This section fills the missing element in
the proof that the thermodynamic
limit of the FCS generating function $\,\Psi_{t,L}(\lambda)\,$
given by (\ref{PsitlambdaL}) and (\ref{hattaucw}) takes the
form (\ref{Psitlambda0})-(\ref{Psitlambda}). It is addressed
to readers not convinced by the heuristic arguments of
Sec.\,\ref{sec:infvlim:heur}.
\vskip 0.2cm

The functions $\,X_{s,t,L;i}\,$
that appear in (\ref{PsitlambdaL}) and (\ref{hattaucw})
satisfy the relations
\qq
X_{s,t,L;1}=g_{s,t,L}-Y_1-L\tau_{0,L}\,,\quad X_{s,t,L;2}=X_{s,t,L;1}
+L\widehat\tau_{s,t,L}\,,
\label{9.1}
\qqq
where $\,Y_1\,$ is the solution of the Fredholm equation (\ref{Freq})
related to conformal welding of tori described in
Sec.\,\ref{subsec:confweldRH}.
On the other hand, the functions $\,\CX^\pm_{s,t;i}\,$ that appear
in the formula (\ref{Psitlambda}) satisfy the relations 
\qq
\CX^\pm_{s,t;1}(x)=g^+_{s,t}(x)-\CY^\pm_1(x)-\ii\gamma\,,\qquad
\CX^\pm_{s,t;2}(x)=\CX^\pm_{s,t;1}+\ii\gamma\,,
\label{9.2}
\qqq
where $\,\CY^\pm_1\,$ are solutions of Eq.\,(\ref{inteqinfty})
for $\,g=g^\pm_{s,t}\,$ obtained in the context of conformal welding
of cylinders discussed in Sec.\,\ref{subsec:confweldcyl}.

We shall prove here the uniform convergence on compacts
with all derivatives of the functions $\,X'_{s,t,L;1}\,$ viewed in
the frames centered at $\,O^\pm_L\,$ of (\ref{OL+}) and (\ref{OL-})
to the functions $\,X^\pm_{s,t;i}\hspace{-0.25cm}{}'\ \,$.
We often suppress the dependence on $\,(s,t,L)\,$ in
the notation for the finite-volume quantities (like for $\,Y_1\,$ in
(\ref{9.1})) and on $\,(s,t)\,$ in the infinite volume (like for
$\,\CY^\pm_1\,$ in (\ref{9.2})). All the estimates established below
are uniform in $\,s=\frac{\lambda}{\Delta\beta}\,$ belonging to bounded
sets.

\subsection{Recasting equation for $\,Y_1$}
\label{subsec:eqY1}

\noindent The main point is to control the behavior for large $\,L\,$
of the derivatives of functions $\,Y_1\,$
solving Eq.\,(\ref{Freq}) in the frames centered at $\,O^\pm_L$.
To this end, it will be convenient to rewrite the Fredholm
equation (\ref{Freq}) in which $\,K=K_{11}+K_{12}+K_{21}\,$ with
\qq
K_{11}=E_{0+}-G^{-1}E_{0+}G=-E_-+G^{-1}E_-G\m,\qquad K_{12}=G^{-1}QE_{0+}\m,
\qquad K_{21}=E_-Q^{-1}G
\label{forKij}
\qqq
for $\,(G^{-1}X)(x)=X(g_{s,t,L}(x))\,$ and
$\,Q\m e_n=\ee^{-\frac{2\pi n}{L}\gamma_L}e_n$, \,see (\ref{FQij}), \,and where
\qq
Y_{12}(x)=g_{s,t,L}(x)-x+L(\widehat\tau_{s,t,L}-\tau_{0,L})\,.
\label{Y122}
\qqq
\vskip 0.1cm

First, we shall eliminate from (\ref{Freq}) the constant mode contributions
involving the toroidal modular parameters. Applying the orthogonal
projector $\,E^\perp_0=E_++E_-\,$ to the both sides of (\ref{Freq}),
we obtain the relation
\qq
E_0^\perp(I-K)Y_1=E_0^\perp(I-K)E_0^\perp Y_1=E_0^\perp(E_--K_{12})(E_0^\perp
Y_{12}+E_0Y_{12})=
E_0^\perp(E_--K_{12})E_0^\perp Y_{12}\label{11100}
\qqq
since $\,E_0^\perp K_{ij}E_0=0$. Recall that the kernel of $\,I-K\,$ is composed
of constants, the range of $\,I-K\,$ has codimension 1, and the solvability of
(\ref{Freq}) fixes uniquely the constant contribution to $\,Y_{12}$,
\,see (\ref{intcond}), with $\,E_--K_{12}\,$ mapping constants into
constants. Thus the range of $\,I-K\,$ cannot contain nonzero constants.
It follows that the operator $\,E_0^\perp(I-K)E_0^\perp\,$ is invertible
on the range of $\,E_0^\perp\,$ and that the function $\,E_0^\perp Y_1\,$
is uniquely determined
by (\ref{11100}) from $\,E_0^\perp Y_{12}$. \,Note from (\ref{forKij})
that $\,K^0=QE_{0+}+E_-Q^{-1}\,$ is the operator $\,K\,$ for $\,g_{s,t,L}\,$
replaced by the identity diffeomorphism. Explicitly,
$\,K^0e_n=\ee^{-\frac{2\pi|n|}{L}\gamma_L}e_n$. \,Factoring out the part related
to $\,K^0$, \,we shall rewrite
\qq
E_0^\perp(I-K)E_0^\perp
=E_0^\perp\Big(I-(K-K^0)E_0^\perp\big(E_0^\perp(I-K^0)E_0^\perp\big)^{-1}\Big)
E_0^\perp(I-K^0)E_0^\perp\m,
\qqq
where
\qq
K-K^0&=&-E_-+G^{-1}E_-G+(G^{-1}-I)QE_{0+}+E_-Q^{-1}(G-I)\cr\cr
&=&E_{0+}-G^{-1}E_{0+}G+(G^{-1}-I)QE_{0+}+E_-Q^{-1}(G-I)\m.
\qqq
In the matrix notation corresponding to the decomposition
$\,L^2_0(S^1_L)\equiv E_0^\perp L^2(S^1_L)=E_+L^2(S^1_L)\oplus E_-L^2(S^1_L)$,
\qq
E_0^\perp(I-K)E_0^\perp\,=\,
\left(\begin{matrix}E_++\Sigma_{++}&\ \Sigma_{+-}\cr \Sigma_{-+}&\ E_-
+\Sigma_{--}\end{matrix}\right)\left(\begin{matrix}(I-Q)_{++}&\ 0\cr
0&\ (I-Q^{-1})_{--}\end{matrix}\right),
\label{97}
\qqq
\vskip -0.3cm
\noindent where
\vskip -0.6cm
\qq
&&\Sigma_{++}=-\Big((G^{-1})_{+-}G_{-+}+\big((G^{-1}-I)Q\big)_{++}
\Big)(I-Q)_{++}^{\,-1}\,,\label{Sigma++}\\ \cr
&&\Sigma_{+-}=(G^{-1})_{++}G_{+-}(I-Q^{-1})^{\,-1}_{--}\,,\label{Sigma+-}\\ \cr
&&\Sigma_{-+}=-\Big((G^{-1})_{--}G_{-+}+(G^{-1}Q)_{-+}
+(Q^{-1}G)_{-+}\Big)(I-Q)_{++}^{\,-1}\,,\label{Sigma-+}\\ \cr
&&\Sigma_{--}=\Big((G^{-1})_{-+}G_{+-}-\big(Q^{-1}(G-I)\big)_{--}\Big)
(I-Q^{-1})^{\,-1}_{--}
\label{Sigma--}
\qqq
in the notation $\,D_{++}\equiv E_+DE_+$, $\,D_{+-}\equiv E_+DE_-$, \,etc.
\,Similarly,
\qq
E_0^\perp(E_--K_{12})E_0^\perp=\left(\begin{matrix}
-(G^{-1}Q)_{++}&\ 0\cr-(G^{-1}Q)_{-+}&\ E_-\end{matrix}\right).
\qqq
Hence (\ref{11100}) becomes
\qq
\left(\begin{matrix}E_++\Sigma_{++}&\ \Sigma_{+-}\cr \Sigma_{-+}&\ E_-
+\Sigma_{--}\end{matrix}\right)\left(\begin{matrix}Z_{+}\cr
Z_-\end{matrix}\right)=
\left(\begin{matrix}
-(G^{-1}Q)_{++}Y_{12}\cr-(G^{-1}Q)_{-+}Y_{12}+E_-Y_{12}
\end{matrix}\right).
\label{Frdmeq}
\qqq
for
\qq
Z=Z_++Z_-=(I-K^0)Y_1\qquad\text{or}\qquad Z_{+}=(I-Q)_{++}Y_1\,,
\quad\ Z_-=(I-Q^{-1})_{--}Y_1\,.
\label{Y1Z+-}
\qqq
Recall that $\,K\,$ is trace-class in $\,L^2(S^1_L)\,$ so that
from (\ref{97}) it follows that the operator $\,\Sigma
=\big(\substack{\Sigma_{++}\,\Sigma_{+-}
\\ \Sigma_{-+}\,\Sigma_{--}}\big)\,$ on $\,L^2_0(S^1_L)\,$ is also trace-class
and that $\,I+\Sigma\,$ is invertible on that space.
\vskip 0.2cm

To control the $\,L\to\infty\,$ limit, we shall need a better
control of those operators. Note that, for $\,L\,$ sufficiently large,
the diffeomorphism $\,g_{s,t,L}\,$ is equal to the identity except
on two disjoint intervals, one inside $]-\frac{1}{4}L,\frac{1}{4}L[\,$ and
the other inside $]-\frac{3}{4}L,-\frac{1}{4}L[$. \,Let us set
\qq
&&g^+_{s,t,L}(x)=\begin{cases}g_{s,t,L}(x+O^+_L)-O^+_L\quad\text{for}\quad
x+O^+_L\in]-\frac{1}{4}L,\frac{1}{4}L[\,,\cr
\,x\qquad\text{otherwise}\,,\end{cases}\label{gstL+}\\
&&g^-_{s,t,L}(x)=\begin{cases}g_{s,t,L}(x+O^-_L)-O^-_L\quad\text{for}\quad
x+O^-_L\in]-\frac{3}{4}L,-\frac{1}{4}L[\,,\cr
\,x\qquad\text{otherwise}\,,\end{cases}\label{gstL-}
\qqq
so that
\qq
g_{s,t,L}(x+O^+_L)-O^+_L-x=g^+_{s,t,L}(x)-x+g^-_{s,t,L}(x+M_L)-M_L-x\,,
\label{separat}
\qqq
where $\,M_L\,$ is given by (\ref{Opm2}). When
restricted to $\,\CI_L$, $\,g^\pm_{s,t,L}\,$ may be
viewed as diffeomorphisms of $\,S^1_L=\R/L\Z\,$ and,
\,when considered on the whole line, \,as diffeomorphisms of $\,\R$.
In the latter case, it follows
from the analysis of Sec.\,\ref{subsec:largeLzeta} that
$\,g^\pm_{s,t,L}(x)=g^\pm_{s,t}(x)=x\,$ outside an $\m L$-independent
bounded set and that for $\,\ell=0,1,\dots$,
\qq
|\partial_x^\ell g^\pm_{s,t,L}(x)-\partial_x^\ell g^\pm_{s,t}(x)|\leq L^{-1} C_\ell
\label{unifconv}
\qqq
uniformly in $\,x$.
\vskip 0.2cm

Let $\,T_b\,$ and $\,(G^\pm)^{\mp1}\,$ be the translation and substitution
operators acting on $\,L^2(S^1_L)\,$ by
\qq
(T_bX)(x)=X(x-b)\,,\qquad((G^\pm)^{-1} X)(x)=X(g^\pm_{s,t,L}(x))\,.
\label{TbG}
\qqq
Given an operator $\,D\,$ acting on $\,L^2(S^1_L)$, \,we shall
denote by $\,\widetilde{D}\,$ the operator $\,T_{O^+_L}^{-1}D\,T_{O^+_L}$,
\,i.e. the operator $\,D\,$ viewed in the
frame centered at $\,O^+_L$. \,The relation (\ref{separat}) implies that
\qq
{\widetilde G}^{-1}-I=(G^+)^{-1}-I+T_{M_L}^{-1}((G^-)^{-1}-I)\m T_{M_L}
\qqq
and, \,similarly,
\qq
\widetilde G-I=G^+-I+T_{M_L}^{-1}(G^--I)T_{M_L}\,.
\qqq
Since $\,T_b\,$ commutes with $\,E_\pm$, \,we infer that
\qq
&&({\widetilde G}^{\mp1}-I)_{++}=((G^+)^{\mp1}-I)_{++}+T_{M_L}^{-1}
((G^-)^{\mp1}-I)_{++}\m
T_{M_L}\,,\label{tG++}\\
&&({\widetilde G}^{\mp1}-I)_{--}=((G^+)^{\mp1}-I)_{--}+T_{M_L}^{-1}
((G^-)^{\mp1}-I)_{--}\m
T_{M_L}\,,\label{tG--}\\
&&({\widetilde G}^{\mp1})_{+-}=((G^+)^{\mp1})_{+-}+T_{M_L}^{-1}((G^-)^{\mp1})_{+-}\m
T_{M_L}\,,\label{tG+-}\\
&&({\widetilde G}^{\mp1})_{-+}=((G^+)^{\mp1})_{-+}+T_{M_L}^{-1}((G^-)^{\mp1})_{-+}\m
T_{M_L}\,.\label{tG-+}
\qqq
In the frame centered at $\,O^+_L$, the relation (\ref{Frdmeq}) becomes
\qq
(I+\widetilde\Sigma)\widetilde Z
=-\Big(({\widetilde G}^{-1}Q)_{++}
+({\widetilde G}^{-1}Q)_{-+}-E_-\Big)\widetilde Y_{12}\,,
\label{become}
\qqq
where
\qq
\widetilde Z=T_{O^+_L}^{-1}Z\,,\qquad \widetilde Y_{12}=T_{O^+_L}^{-1}Y_{12}
\label{ZTZ}
\qqq
and, \,by (\ref{Y122}) and (\ref{separat}), \,we may take
\qq
\widetilde Y_{12}=Y^+_{12}+T_{M_L}^{-1}Y^-_{12}
\qqq
for
\qq
Y^+_{12}=g^+_{s,t,L}-g_0\,,
\qquad Y^-_{12}=g^-_{s,t,L}-g_0
\label{Y12+-}
\qqq
dropping constant terms from $\,Y_{12}\,$ that do not change 
the right hand side of (\ref{become}).
The operator $\,\widetilde\Sigma=T_{O^+_L}^{-1}\Sigma\m T_{O^+_L}\,$ on
$\,L^2_0(S^1_L)\,$ is obtained from the relations
(\ref{Sigma++})-(\ref{Sigma--})
by replacing $\,G^{\pm1}\,$ by $\,\widetilde G^{\pm1}$.
\vskip 0.2cm

We have to deal with the fact that the contribution
of the right- and left-movers corresponding to diffeomorphisms
$\,g^\pm_{s,t,L}\,$ are mixed together in finite volume. Let $\,\Sigma^+\,$
be the operator obtained from $\,\widetilde\Sigma\,$ by replacing all
inputs with $\,\widetilde G^{\mp1}\,$ by the first terms on the right-hand
side of (\ref{tG++})-(\ref{tG-+}) and let $\,T_{M_L}^{-1}\Sigma^-T_{M_L}\,$
be obtained similarly but using the second terms on the right-hand side of
(\ref{tG++})-(\ref{tG-+}). Consider first the decoupled
equations\footnote{The superscripts $\,\pm\,$ pertain to the right-
and left-movers whereas the subscripts $\,\pm\,$ correspond to
components in the range of projectors $\,E_\pm$.}
\qq
(I+{\Sigma}^\pm)Z^\pm=
-\Big(((G^\pm)^{-1}Q)_{++}
+((G^\pm)^{-1}Q)_{-+}-E_-\Big)Y^\pm_{12}\m.\label{ZY12+-}
\qqq
They may be solved for $\,Z^\pm\,$ because $\,I+\Sigma^\pm$, \,similarly
as $\,I+\Sigma$, \,are invertible on $\,L^2_0(S^1_L)$.
\,Exhibiting the corrections due to the coupling between the right- and
left-movers, we shall write
\qq
\widetilde\Sigma\,=\,\Sigma^++T_{M_L}^{-1}\Sigma^-T_{M_L}
+\delta\widetilde\Sigma\,,\label{tilS}
\qqq
where
\qq
&&\delta\widetilde\Sigma_{++}=-\Big(((G^+)^{-1})_{+-}T_{M_L}^{-1}(G^-)_{-+}T_{M_L}
+T_{M_L}^{-1}((G^-)^{-1})_{+-}T_{M_L}(G^+)_{-+}\Big)(I-Q)_{++}^{\,-1}\,,
\label{deltaSigma++}\\ \cr
&&\delta\widetilde\Sigma_{+-}=\Big(((G^+)^{-1}-I)_{++}T_{M_L}^{-1}(G^-)_{+-}T_{M_L}+
T_{M_L}^{-1}((G^-)^{-1}-I)_{++}T_{M_L}(G^+)_{+-}\Big)(I-Q^{-1})^{\,-1}_{--}\,,
\label{deltaSigma+-}\quad\\ \cr
&&\delta\widetilde\Sigma_{-+}=-\Big(((G^+)^{-1}-I)_{--}T_{M_L}^{-1}(G^-)_{-+}T_{M_L}+
T_{M_L}^{-1}((G^-)^{-1}-I)_{--}T_{M_L}(G^+)_{-+}\Big)(I-Q)_{++}^{\,-1}\,,
\label{deltaSigma-+}\\ \cr
&&\delta\widetilde\Sigma_{--}=\Big(((G^+)^{-1})_{-+}T_{M_L}^{-1}(G^-)_{+-}T_{M_L}+
T_{M_L}^{-1}((G^-)^{-1})_{-+}T_{M_L}(G^+)_{+-}\Big)(I-Q^{-1})_{--}^{\,-1}\,.
\label{deltaSigma--}
\qqq
and
\qq
\widetilde Z\,=\,Z^++T_{M_L}^{-1}Z^-+\delta\widetilde Z\,.
\label{tilZ}
\qqq
The decoupled contributions (\ref{ZY12+-})
to (\ref{become}) will now cancel out resulting in the equation 
\qq
\big(I+\widetilde\Sigma\big)
\delta\widetilde Z
&=&-\Big(T_{M_L}^{-1}\Sigma^-T_{M_L}
+\delta\widetilde\Sigma\Big)Z^+-\Big(\Sigma^+
+\delta\widetilde\Sigma\Big)T_{M_L}^{-1}Z^-\cr
&&-\Big(T_{M_L}^{-1}(((G^-)^{-1}-I)Q)_{++}T_{M_L}
+T_{M_L}^{-1}((G^-)^{-1}Q)_{-+}T_{M_L}\Big)Y^+_{12}\cr
&&-\Big((((G^+)^{-1}-I)Q)_{++}
+((G^+)^{-1}Q)_{-+}\Big)T_{M_L}^{-1}Y^-_{12}
\label{coupleq}
\qqq
for $\,\delta\widetilde Z$. \,Below, we shall show that
this equation implies that $\,\delta\widetilde Z\,$ tends
to zero in an appropriate sense when $\,L\to\infty$.
This will establish the factorization of the right- and left-movers
contributions in the thermodynamic limit.  
\vskip 0.2cm

Following similar steps as in finite volume, \,the infinite-volume equation
(\ref{inteqinfty}) with $\,g=g^\pm_{s,t}\,$ may be recast upon writing
\qq
\CI-\CK=\big(\CI-(\CK-\CK^0)(\CI-\CK^0)^{-1}\big)(\CI-\CK^0)
\qqq
into the form
\qq
(\CI+\varSigma^\pm)\CZ^\pm=-\Big(((\CG^\pm)^{-1}\CQ)_{++}\m
+((\CG^\pm)^{-1}\CQ)_{-+}\m-\CE_-\Big)\CY^\pm_{12}
\label{limvvarS}
\qqq
where $\m\,\varSigma^\pm=\big(\begin{matrix}_{\varSigma^\pm_{++}}
&_{\varSigma^\pm_{+-}}\cr^{\varSigma^\pm_{-+}}&^{\varSigma^\pm_{--}}\end{matrix}\big)\,\m$
act in $\,L^2(\R)=\CE_+L^2(\R)\oplus\CE_-L^2(\R)\,$ and have
the components $\,\varSigma_{++}=\CE_+\varSigma\m\CE_+$, \,etc., \,given by 
Eqs.\,(\ref{Sigma++})-(\ref{Sigma--}) in which $\,G^{\pm1}\,$ is replaced by
the operators $\,(\CG^\pm)^{\pm1}\,$ such that
$\,((\CG^\pm)^{-1}\CX)(x)=\CX(g^\pm_{s,t}(x))$, \,and where
\qq
\CZ^{\pm}=(\CI-\CK^0)\CY^\pm_1\,,\qquad \CY^\pm_{12}=g^\pm_{s,t}-g_0\,.
\label{1amd2}
\qqq
Above, $\,\CK^0=\CQ\m\CE_++\CE_-\CQ^{-1}\,$ denotes the operator
$\,\CK\,$ of (\ref{opCK}) for $\,g\,$ equal to the identity diffeomorphism
$\,g_0$.

\subsection{Fast-decay and Schwartz type operators}
\label{subsec:fdSchw}

\noindent We shall consider operators on $\,L^2(\R)\,$ acting in the momentum
space representation by
\qq
(\widehat{\CD\CX})(p)\,=\,\frac{_1}{^{2\pi}}\int\widehat\CD(p,q)
\,\widehat\CX(q)\,dq
\,dx\m.
\qqq
for $\,\widehat X(p)\,$ given by (\ref{Fourtr}).
\vskip 0.2cm

\noindent{\bf Definition 1.} \,We shall call $\,\CD\,$ of fast-decay
type if for any $\,k=0,1,\dots\,$
there exists a constant $\,C_k\,$ such that
\qq
|\widehat\CD(p,q)|\leq\frac{C_k}{(1+p^2)^k(1+q^2)^k}\,.
\label{fast-dec}
\qqq
Let $\,\widehat\CJ,\,\widehat\CJ'\,$ be open subsets of $\,\mathbb R$.
We shall call $\,\CD\,$ of Schwartz $\widehat\CJ\times\widehat\CJ'$ type
if $\,(i)$ $\,\widehat\CD(p,q)=0\,$ for $\,(p,q)\notin\widehat\CJ\times
\widehat\CJ'$, $\,(ii)$ $\,\widehat\CD(p,q)\,$ is smooth on
$\widehat\CJ\times\widehat\CJ'$, and \,$(iii)$
for any $\,\ell_1,\ell_2,k=0,1,\dots\,$ there exists a constant
$\,C_{\ell_1,\ell_2,k}\,$ such that 
\qq
|\partial_{p}^{\ell_1}\partial_{q}^{\ell_2}\widehat\CD(p,q)|\leq
\frac{C_{\ell_1,\ell_2,k}}{(1+p^2)^k(1+q^2)^k}
\qqq
on $\,\widehat\CJ\times\widehat\CJ'$,
\vskip 0.3cm

\noindent{\bf Definition 2.} \,Let $\,\CD_L,\,\CD\,$ be
operators of fast-decay type. We shall
say that $\,\CD_L\,$ converge to $\,\CD\,$ with speed $\,L^{-1}\,$ if
for any $\,k=0,1,\dots\,$ there exists a constant $\,C_k\,$ such that
\qq
|\widehat\CD_L(p,q)-\widehat\CD(p,q)|\leq\frac{L^{-1}C_k}
{(1+p^2)^k(1+q^2)^k}
\qqq
for all $\,L\geq L_0\,$ where $\,L_0\,$ is $\,k$-independent.
Similarly, the convergence with speed $\,L^{-1}\,$ of
Schwartz $\widehat\CJ\times\widehat\CJ'$ type operators is defined
by demanding that for any $\,\ell_1,\ell_2,k=0,1,\dots\,$ there exists
a constant $\,C_{\ell_1,\ell_2,k}\,$ such that
\qq
\big|\partial_{p}^{\ell_1}\partial_{q}^{\ell_2}\big(\widehat\CD_L(p,q)
-\widehat\CD(p,q)\big)\big|\leq
\frac{L^{-1}C_{\ell_1,\ell_2,k}}{(1+p^2)^k(1+q^2)^k}
\qqq
on $\,\widehat\CJ\times\widehat\CJ'\,$ for all $\,L\geq L_0$.
\vskip 0.3cm

\noindent Note that the Schwartz-type operators are of fast decay type,
\,that the latter are Hilbert-Schmidt \cite{Simon},
\m and that the Schwartz-type convergence
implies the fast-decay one.
\vskip 0.3cm

\noindent{\bf Definition 3.} We shall call a function
$\,X\in  L^2_0(S^1_L)\,$ the $L$-periodization of a function
$\,\CX\,$ on $\,\R\,$ if $\,\sqrt{L}\,\big\langle e_n\big|X\big\rangle=
\widehat\CX(p_n)\,$ for $\,0\not=n\in\Z$, \,where $\,e_n(x)=\frac{1}{\sqrt{L}}
\ee^{-\ii p_nx}\,$ and $\,p_n\equiv\frac{2\pi n}{L}$. \,Similarly, we shall
call an operator $\,D\,$
on $\,L^2_0(S^1_L)\,$ with the matrix elements $\,D_{mn}=
\langle e_m|D|e_n\rangle\,$ the $L$-periodization of an
operator $\,\CD\,$ on $\,L^2(\R)\,$ with momentum-space kernel
$\,\widehat{\CD}(p,q)\,$ if $\,L\m D_{mn}=\widehat{\CD}(p_m,p_n)\,$
for $\,0\not=m,n\in\Z$. 
\vskip 0.3cm

\noindent{\bf Definition 4.} We shall call an operator $\,D\,$ on
$\,L^2_0(S^1_L)\,$ of fast-decay or
Schwartz $\,\widehat\CJ\times\widehat\CJ'\,$ type if $\,D\,$ is the
$L$-periodization of an operator $\,\CD\,$ on $\,L^2(\R)\,$ of
the corresponding type. If $\,D_L\,$
are the $L$-periodization of fast-decay or Schwartz
$\widehat\CJ\times\widehat\CJ'$ type operators
$\,\CD_L\,$ on $\,L^2(\R)\,$ converging as such to operator $\,\CD\,$
with speed $\,L^{-1}$, \,we shall say that $\,D_L\,$ converge to
$\,\CD\,$ with speed $\,L^{-1}\,$ as fast-decay or Schwartz
$\widehat\CJ\times\widehat\CJ'$ type operators.
\vskip 0.3cm

In Appendix \ref{app:C} we present basic general results about the
fast-decay and Schwartz type operators and the related Fredholm
operators, their determinants and their inverses that we shall frequently
evoke in the sequel.

\subsection{Schwartz-type convergence results}
\label{subsec:momest}

\noindent Let $\,g:\mathbb R\rightarrow\mathbb R\,$ be a diffeomorphism
that is equal to the identity outside a bounded subset of $\,\mathbb R\,$
and let $\,\CG\,$ be the substitution operator on $\,L^2(\mathbb R)$,
\qq
(\CG^{-1}\CX)(x)=\CX(g(x))\m.
\qqq
We shall denote by $\,\CP\,$ the operator $\,\ii\partial_x\,$ and by
$\,\CQ\,$ the operator $\,\ee^{-\gamma_\#\hspace{-0.03cm}\CP}\,$
for $\,\gamma_\#=\gamma_L\,$ or $\,\gamma_\#=\gamma$, \,as
specified below, \,with $\,\gamma_L=v\beta_{0,L}\,$ and $\,\gamma=v\beta_0$.
\,As before, $\,\CE_\pm\,$ will denote the orthogonal projections
in $\,L^2(\mathbb R)\,$ on functions with Fourier transform vanishing
outside $\,\mathbb R_\pm\,$ and we shall use the shorthand notation
$\,\CD_{++}=\CE_+\CD\m\CE_+$, \,etc.
\vskip 0.3cm

\noindent{\bf Lemma 2.} \,The following operators on $\,L^2(\mathbb R)\,$
are of Schwartz $\R_\sigma\times\R_{\sigma'}$ type:
\begin{itemize}
\item $\,\CD_1=(\CG^{-1}\CP^{-1})_{+-}=((\CG^{-1}-\CI)\CP^{-1})_{+-}\,$ \ for
$\,(\sigma,\sigma')=(+,-)$\m,
\item $\,\CD_2=(\CG^{-1}\CP^{-1})_{-+}=((\CG^{-1}-\CI)\CP^{-1})_{-+}\,$ \ for
$\,(\sigma,\sigma')=(-,+)$\m,
\item $\,\CD_3=((\CG^{-1}-\CI)\CQ\CP^{-1})_{++}\,$\hspace{2.3cm} for
$\,(\sigma,\sigma')=(+,+)$\m,
\item $\,\CD_4=(\CQ^{-1}(\CG^{-1}-\CI)\CP^{-1})_{--}\,$\hspace{1.92cm} for
$\,(\sigma,\sigma')=(-,-)$\m,
\item $\,\CD_5=(\CG^{-1}-\CI)_{++}(\CG\CP^{-1})_{+-}\,$\hspace{1.92cm} for
$\,(\sigma,\sigma')=(+,-)$\m,
\item $\,\CD_6=(\CG^{-1}-\CI)_{--}(\CG\CP^{-1})_{-+}\,$\hspace{1.92cm} for
$\,(\sigma,\sigma')=(-,+)$\m,
\end{itemize}
The same claims hold for the above operators with $\,\CG\,$ 
and $\,\CG^{-1}\,$ interchanged.
\vskip 0.3cm

The proof of Lemma 2 based on straightforward estimates is given
in Appendix \ref{app:D}. Applying Lemma 2 to the case when
$\,g=g^\pm_{s,t,L}\,$ of (\ref{gstL+})
and (\ref{gstL-}) with $\,\CQ=\ee^{-\gamma_L\hspace{-0.03cm}\CP}\,$
or \m to $\,g=g_{s,t}^\pm\,$
with $\,\CQ=\ee^{-\gamma\CP}$, \,we infer that the corresponding operators
$\,\CD^\pm_{i,L}\,$ and $\,\CD^\pm_i\,$ for $\,i=1,\dots,6\,$ are of Schwartz
type. A straightforward modification of the proof of Lemma 2, see
Appendix \ref{app:D}, together with the estimates (\ref{unifconv}),
show that $\,\CD^\pm_{i,L}\,$ converge to $\,\CD^\pm_{i}\,$ with
speed $\,L^{-1}\,$ as operators of Schwartz-type. \,For $\,i=5,6$,
\,we introduce an additional modification of $\,\CD^\pm_{i,L}$, \,described
at the end of the proof of Lemma 2 in Appendix \ref{app:D}, which does
not change the above properties. Now, let $\,D^\pm_{i,L}\,$ be the
$L$-periodization of the operators $\,\CD^\pm_{i,L}$. \,In view of Definition 4,
the operators $\,D^\pm_{i,L}\,$ converge with speed $\,L^{-1}\m$ to
$\,\CD^\pm_i\,$ as operators of Schwartz type. Explicitly, $\,D^\pm_{i,L}\,$
are the operators\footnote{The modification
of operators $\,\CD^\pm_{i,L}\,$ for $\,i=5,6\,$ just mentioned was done
to assure the stated form of their $L$-periodization.}
\qq
&&D^\pm_{1,L}=((G^\pm)^{-1}P^{-1})_{+-}\,,\hspace{2.3cm}
D_{2,L}^\pm=((G^\pm)^{-1}P^{-1})_{-+}\,,\\
&&D^\pm_{3,L}=((G^\pm)^{-1}-I)_{++}(QP^{-1})_{++}\,,
\qquad D^\pm_{4,L}=(Q^{-1})_{--}((G^\pm)^{-1}-I)_{--}(P^{-1})_{--}\,,\\
&&D^\pm_{5,L}=((G^\pm)^{-1}-I)_{++}(G^\pm P^{-1})_{+-}\,,\hspace{0,48cm}
D^\pm_{6,L}=((G^\pm)^{-1}-I)_{--}(G^\pm P^{-1})_{-+}\,,
\qqq
where $\,P\ee_n=p_n\ee_n\,$ and 
$\,(G^\pm)^{-1}\,$ are the operators of substitution of $\,g^\pm_{s,t,L}\,$
acting on $\,L^2_0(S^1_L)$. \,On the other hand,
\qq
&&\CD^\pm_1=((\CG^\pm)^{-1}\CP^{-1})_{+-}\,,
\hspace{2.34cm}\quad\CD^\pm_2=((\CG^\pm)^{-1}\CP^{-1})_{-+}\,,\\
&&\CD^\pm_3=((\CG^\pm)^{-1}-\CI)_{++}(\CQ\CP^{-1})_{++}\,,\qquad
\quad\CD^\pm_4=(\CQ^{-1})_{--}((\CG^\pm)^{-1}-\CI)_{--}(\CP^{-1})_{--}\,,\\
&&\CD^\pm_{5}=((\CG^\pm)^{-1}-\CI)_{++}(\CG^\pm\CP^{-1})_{+-}\,,\hspace{0,9cm}
\CD^\pm_{6}=((\CG^\pm)^{-1}-\CI)_{--}(\CG^\pm\CP^{-1})_{-+}\,,
\qqq
where $\,(\CG^\pm)^{-1}\,$ are the operator of substitution of
$\,g^\pm_{s,t}\,$ acting on $\,L^2(\R)$. 
\vskip 0.2cm

Recall from Sec.\,\ref{subsec:eqY1} that the operators $\m\,\Sigma^\pm=
\big(\begin{matrix}_{\Sigma^\pm_{++}}&_{\Sigma^\pm_{+-}}\cr^{\Sigma^\pm_{-+}}&^{\Sigma^\pm_{--}}
\end{matrix}\big)\,\m$ on $\,L^2_0(S^1_L)\,$ have the components
that are given by Eqs.\,(\ref{Sigma++})-(\ref{Sigma--}) with $\,G^{\pm1}\,$
replaced by $\,(G^\pm)^{\pm1}\,$ and that their $\,L=\infty\,$ version acting
in $\,L^2(\R)\,$ are the operators 
$\m\,\varSigma^\pm=\big(\begin{matrix}_{\varSigma^\pm_{++}}
&_{\varSigma^\pm_{+-}}\cr^{\varSigma^\pm_{-+}}&^{\varSigma^\pm_{--}}\end{matrix}\big)\,\m$
with the components given by Eqs.\,(\ref{Sigma++})-(\ref{Sigma--}) in which
$\,G^{\pm1}\,$ is replaced by
$\,(\CG^\pm)^{\pm1}\,$ and $\,Q\,$ by $\,\CQ=\ee^{-\gamma\CP}$. Let
$\,\R_{\not=0}=\R\setminus\{0\}$.
\vskip 0.4cm

\noindent{\bf Proposition 1.} \,The operators $\,\Sigma^\pm\,$ and
$\,\varSigma^\pm\,$ are of Schwartz $\R_{\not=0}\times\R_{\not=0}$ type and
$\,\Sigma^\pm\,$ converge with speed $\,L^{-1}\m$ to
$\,\varSigma^\pm\,$ as such.
\vskip 0.3cm

\noindent{\bf Remark.} \,Schwartz $\R_{\not=0}\times\R_{\not=0}$ type
operators have momentum-space kernels that are smooth away from
$\,\{0\}\times\R\cup\R\times\{0\}\,$ but may be discontinuous across
that set.
\vskip 0.3cm

\noindent{\bf Proof of Proposition 1.} \,The claim follows from
the above results and their version with diffeomorphisms
$\,g^\pm_{s,t,L}\,$ and $\,g^{\pm}_{s,t}\,$ replaced by
their inverses, \,together with Propositions C2 and C3 of Appendix \ref{app:C}.
For example, in
\qq
\Sigma^\pm_{++}=-\Big(\big(G^\pm)^{-1}\big)_{+-}G^\pm_{-+}+\big(((G^\pm)^{-1}-I)Q\big)_{++}\Big)(I-Q)^{-1}_{++}\,,
\qqq
the operators $\,\big((G^\pm)^{-1}\big)_{+-}G^\pm_{-+}(I-Q)^{-1}_{++}\,$
are the $L$-periodization of $\,\CD^\pm_{1,L}\CP\m\CD^\pm_{2,L}\big(\frac{\CP}
{\CI-\ee^{-\gamma_L\hspace{-0.03cm}\CP}}\big)_{\hspace{-0.03cm}{++}}$, \,where in $\,\CD^\pm_{2,L}\,$ one should
invert $\,g^\pm_{s,t,L}$. \,Those operators converge with speed $\,L^{-1}\,$ as
operators of Schwartz $\R_+\times\R_+$ type to
$\,\CD^\pm_{1}\CP\m\CD^\pm_{2}\big(\frac{\CP}{\CI-\ee^{-\gamma\CP}}
\big)_{\hspace{-0.03cm}++}$,
\,where in $\,\CD^\pm_2\,$ one should invert
$\,g^\pm_{s,t,}$. \,Similarly, $\,\big(((G^\pm)^{-1}-I)Q\big)_{++}(I-Q)^{-1}_{++}\,$
is the $L$-periodization of $\,\CD^\pm_{3,L}\big(\frac{\CP}{\CI
-\ee^{-\gamma_L\hspace{-0.03cm}\CP}}
\big)_{\hspace{-0.03cm}++}\,$ that converges to
$\,\CD^\pm_{3}\big(\frac{\CP}{\CI-\ee^{-\gamma\CP}}
\big)_{\hspace{-0.03cm}++}$.
\,In
\qq
\Sigma^\pm_{-+}=-\Big(\big((G^\pm)^{-1}-I\big)_{--}G^\pm_{-+}+G^\pm_{-+}
+\big((G^\pm)^{-1}Q\big)_{-+}+\big(Q^{-1}G^\pm\big)_{-+}\Big)(I-Q)^{-1}_{++}\,,
\qqq
the operators $\,\big((G^\pm)^{-1}-I\big)_{--}G^\pm_{-+}(I-Q)^{-1}_{++}\,$
are the $L$-periodization of $\,\CD^\pm_{6,L}\big(\frac{\CP}
{I-\ee^{-\gamma_L\hspace{-0.03cm}\CP}}\big)_{++}$. They converge with speed $\,L^{-1}\,$
to $\,\CD^\pm_6\big(\frac{\CP}{I-\ee^{-\gamma\CP}}\big)_{++}\,$ as operators
of Schwartz $\R_-\times\R_+$ type. Similarly, $\,G^\pm_{-+}(I-Q)^{-1}_{++}\,$
are the $L$-periodization of $\,\CD^\pm_{2,L}\big(\frac{\CP}{\CI
-\ee^{-\gamma_L\hspace{-0.03cm}\CP}}
\big)_{++}$, \,where in $\,\CD_{2,L}^\pm\,$ one should invert $\,g^\pm_{s,t,L}$.
They converge to $\,\CD^\pm_2\big(\frac{\CP}{\CI-\ee^{-\gamma\CP}}\big)$,
where in $\,\CD^\pm_2\,$ one should invert $\,g^\pm_{s,t}$. 
$\,\big((G^\pm)^{-1}Q\big)_{-+}(I-Q)^{-1}_{++}\,$ are the $L$-periodization
of operators $\,\CD_{2,L}\big(\frac{\CP\,\ee^{-\gamma_L\hspace{-0.03cm}\CP}}
{\CI-\ee^{-\gamma_L\hspace{-0.03cm}\CP}}\big)_{++}\,$
that converge to $\,\CD_{2}\big(\frac{\CP\,\ee^{-\gamma\CP}}{\CI-\ee^{-\gamma\CP}}
\big)_{++}$. \,Finally, $\,(Q^{-1}G^\pm)_{-+}(I-Q)^{-1}_{++}\,$ are the
$L$-periodization of $\,(\ee^{\gamma_L\hspace{-0.03cm}\CP})_{--}\CD^\pm_{2,L}\big(
\frac{\CP}{\CI-\ee^{-\gamma_L\hspace{-0.03cm}\CP}}\big)_{++}\,$ with $\,g^\pm_{s,t,L}\,$
inverted that converge to $\,(\ee^{\gamma\CP})_{--}\CD^\pm_{2}\big(
\frac{\CP}{\CI-\ee^{-\gamma\CP}}\big)_{++}\,$ with $\,g^\pm_{s,t}\,$ inverted. 
\,The convergence of $\,\Sigma^\pm_{+-}\,$ and $\,\Sigma^\pm_{--}\,$ is
obtained in a similar way.

\hspace{12cm}$\square$
\vskip 0.2cm

The Fredholm determinants $\,\det(I+\Sigma^\pm)\,$ and
$\,\det(\CI+\varSigma^\pm)\,$ are well defined, \,see Appendix \ref{app:C}.
\,We shall prove in Appendix \ref{app:E} \m the following result:
\vskip 0.3cm

\noindent{\bf Lemma 3.}
\qq
|\det(\CI+\varSigma^\pm)|\geq C>0\,.
\label{estdet}
\qqq
\vskip 0.3cm

\noindent  In virtue of Proposition C9 of Appendix \ref{app:C}, it follows
from (\ref{estdet}) that the operators $\,\CI+\varSigma^\pm\,$ are invertible
and that the operators $\,\CR^\pm=\CI-(\CI+\Sigma^\pm)^{-1}\,$ are of Schwartz
$\R_{\not=0}\times\R_{\not=0}$ type. The convergence of $\,\Sigma^\pm\,$ to
$\,\varSigma^\pm\,$ with speed $\,L^{-1}\,$ as operators of Schwartz
$\R_{\not=0}\times\R_{\not=0}$ type implies in turn by Propositions C4 and C6
of Appendix \ref{app:C} that
\qq
\det(I+\Sigma^\pm)\,=\,\det(\CI+\varSigma^\pm)\,+\,O(L^{-1})\,.
\label{detSvarS}
\qqq
As a consequence, $\,|\det(I+\Sigma^\pm)|\geq\frac{1}{2}C\,$ for
$\,L\,$ large enough, \,and the operators
$\,R^\pm=I-(I+\Sigma^\pm)^{-1}\,$ are of Schwartz
$\R_{\not=0}\times\R_{\not=0}$ type and, as such, they converges with speed
$\,L^{-1}\,$ to $\,\CR^\pm$, \,see Proposition C10 of Appendix
\ref{app:C}.

\subsection{Solution of the decoupled Fredholm equations}
\label{subsec:soldecFr}

\noindent The decoupled Fredholm equations (\ref{ZY12+-})
in $\,L^2_0(S^1_L)\,$ take the form
\qq
(I+\Sigma^\pm)Z^\pm=Z^\pm_{12}\,,
\label{newFred}
\qqq
where
\qq
Z^\pm_{12}&=&-\Big(\big(((G^\pm)^{-1}-I)Q\big)_{++}\m+Q_{++}
\m+((G^\pm)^{-1}Q)_{-+}\m-E_-\Big)Y^\pm_{12}\cr
&=&-\Big(D_{3,L}^\pm P_{++}\m+Q_{++}\m
+D^\pm_{2,L}\m(PQ)_{++}\m-E_-\Big)(g^\pm_{s,t,L}-g_0)\,.
\qqq
We shall first consider the limiting version (\ref{limvvarS}) in $\,L^2(\R)\,$
of the above equations taking the form
\qq
(\CI+\varSigma^\pm)\CZ^\pm=\CZ^{\pm}_{12}\,,
\label{limFred}
\qqq
where the functions
\qq
\CZ^\pm_{12}&=&-\Big(\big(((\CG^\pm)^{-1}-\CI)\CQ\big)_{++}+\CQ_{++}
+((\CG^\pm)^{-1}\CQ)_{-+}-\CE_-\Big)(g_{s,t}^\pm-g_0)\cr
&=&\,-\Big(\CD^\pm_3\CP_{++}+\CQ_{++}
+\CD^\pm_2(\CP\CQ)_{++}-\CE_-\Big)(g_{s,t}^\pm-g_0).
\qqq
satisfy the following estimates:
\vskip 0.3cm

\noindent{\bf Lemma 4.} \,For $\,p\not=0\,$ and $\,\ell,k=0,1,\dots$,
\qq
\big|\partial_p^\ell\m\widehat{\CZ^\pm_{12}}(p)\big|\,\leq\,
\frac{C_{\ell,k}}{(1+p^2)^k}
\qqq
for some constants $\,C_{\ell,k}$.
\vskip 0.3cm

\noindent{\bf Proof of Lemma 4.} We have
\qq
\widehat{\CZ^\pm_{12}}(p)&=&-
\frac{_1}{^{2\pi}}\int_0^\infty\hspace{-0.1cm}\Big(\widehat{\CD^\pm_3}(p,q)\,
q\,+\,\widehat{\CD^\pm_2}(p,q)\,q\,\ee^{-\gamma q}
\Big)\m\widehat{(g^\pm_{s,t}-g_0)}(q)
\,dq\cr
&&\hspace{3.1cm}
-\m\Big(\ee^{-\gamma p}\,\theta(p)-\theta(-p)\Big)\m\widehat{(g^\pm_{s,t}
-g_0)}(p)\ \qquad
\label{forasstf}
\qqq
and the assertion follows since the operators $\,\CD^\pm_{i}\,$ are of Schwartz
$\R_{\not=0}\times\R_{\not=0}$ type and $\,\widehat{(g^\pm_{s,t}-g_0)}\,$ are
Schwartz functions.

\hspace{12cm}$\square$
\vskip 0.3cm

\noindent The solutions of the Fredholm equations\,(\ref{limFred}) have
the form
\qq
\CZ^\pm=\CZ^\pm_{12}-\CR^\pm\CZ^\pm_{12}\,.
\qqq
From the fact that $\,\CR^\pm=\CI-(\CI+\varSigma^\pm)^{-1}\,$
are of $\R_{\not=0}\times\R_{\not=0}$ Schwartz
type and from Lemma 4, we infer
that for $\,p\not=0\,$ and $\,\ell,k=0,1,\dots$,
\qq
\big|\partial_p^\ell\m\widehat{\CZ^\pm}(p)\big|\,\leq\,
\frac{C_{\ell,k}}{(1+p^2)^k}
\label{estim2}
\qqq
for some constants $\,C_{\ell,k}$.
\vskip 0.2cm

Let us recall now that
the functions $\,\CY^\pm_1\,$ that satisfy (\ref{9.2}) 
are related to $\,\CZ^\pm\,$ by the first of Eqs.\,(\ref{limvvarS})
that may be solved for the derivative of $\,\CY^\pm_1\,$ by setting
\qq
\CY_1^\pm{}'=-\ii\,\CP(\CI-\CK^0)^{-1}\CZ^\pm=-\ii\,
\CP(\CI-\ee^{-\gamma|\CP|})^{-1}\CZ^\pm\,.
\label{CPCY1}
\qqq
The estimate (\ref{estim2}) implies that for
$\,p\not=0\,$ and $\,\ell,k=0,1,\dots$,
\qq
\big|\partial_p^\ell\m\widehat{\CY_1^\pm{}'}(p)\big|\,\leq\,
\frac{C_{\ell,k}}{(1+p^2)^k}
\label{CY1pm}
\qqq
for some new constants $\,C_{\ell,k}$.  \,It follows, in particular, that
the functions $\,\CY_1^\pm{}'\,$ are smooth and satisfy the uniform bounds
\qq
\big|\partial_x^l\CY^\pm_1{}'(x)|\,\leq\,C_l
\qqq
for $\,l=0,1,\dots$.
\vskip 0.2cm

Let us pass now to the finite-volume Fredholm equations (\ref{newFred}).
\vskip 0.3cm

\noindent{\bf Lemma 5.} \,There exist functions
$\,\CZ^\pm_{12,L}\,$ on $\,\R\,$ such that $\,Z^\pm_{12}\,$
are their $L$-periodization in the sense of Definition 3 and 
for $\,p\not=0$, $\,\ell,k=0,1,\dots$,
\qq
\big|\partial_p^\ell\m\widehat{\CZ^\pm_{12,L}}(p)-\partial_p^\ell\m
\widehat{\CZ^\pm_{12}}(p)\big|\,\leq\,\frac{L^{-1}C_{\ell,k}}{(1+p^2)^k}\,.
\label{estim41}
\qqq
\vskip 0.2cm

\noindent{\bf Proof of Lemma 5.} \,We shall set
\qq
\widehat{\CZ^\pm_{12,L}}(p)\,=\,-\frac{_1}{^L}\sum\limits_{n=1}^\infty
\Big(\widehat{\CD^\pm_{3,L}}(p,p_n)\,p_n\,+\,\widehat{\CD^\pm_{2,L}}(p,p_n)\,
p_n\m\ee^{-\gamma_Lp_n}\Big)\m\widehat{(g^\pm_{s,t,L}-g_0)}(p_n)\ 
\cr
-\m\Big(\ee^{-\gamma_Lp}\,\theta(p)-\theta(-p)\Big)
\m\widehat{(g^\pm_{s,t,L}-g_0)}(p)\,.
\label{forasstf2}
\qqq
That $\,Z^\pm_{12}\,$ are the $L$-periodization of $\,Z^\pm_{12,L}\,$
follows from the fact that $\,D_{i,L}\,$ are
the $L$-periodization of $\,\CD_{i,L}$. \,A comparison of (\ref{forasstf2})
and (\ref{forasstf}) shows that the estimate (\ref{estim41})
is a consequence of the convergence of $\,\CD_{i,L}\,$ to $\,\CD_{i}\,$
as Schwartz-type operators and the convergence of $\,g^\pm_{s,t,L}-g_0\,$ to
$\,g^\pm_{s,t}-g_0\,$ as Schwartz functions, and of the bound
\qq
\big|\partial_p^\ell\m\widehat{\CD^\pm_{i}}(p,p_n)-
\partial_p^\ell\m\widehat{\CD^\pm_{i}}(p,q)\big|&\leq&\pi L^{-1}
\hspace{-0.1cm}\sup\limits_{|r-p_n|\leq\pi L^{-1}}
\big(|\partial_p^{\ell}\partial_r\widehat{\CD^\pm_{i}}(p,r)\big|
\,\leq\,\frac{L^{-1}C_{\ell,k}}{(1+p^2)^k(1+q^2)^k}
\qqq
holding for $\,|q-p_n|\leq\pi L^{-1}$, \,and of a similar estimate for
$\,\big|\partial_p^\ell\m\widehat{(g^\pm_{s,t}-g_0)}(p_n)-
\partial_p^\ell\m\widehat{(g^\pm_{s,t}-g_0)}(q)\big|\,$ and, \,finally, \,of
(\ref{beta0Lbeta0}). 

\hspace{12cm}$\square$

The solutions of the Fredholm equations (\ref{newFred}) have the form
\qq
Z^\pm=Z^\pm_{12}-R^\pm Z^\pm_{12}\,.
\qqq
We shall need below a result about $\,Z^\pm\,$ analogous to
Lemma 5 about $\,Z^\pm_{12}$.
\vskip 0.3cm

\noindent{\bf Lemma 6.} \,There exist functions $\,\CZ^\pm_{L}\,$ on
$\,\R\,$ such that $\,Z^\pm\,$ are their $L$-periodization and that
for $\,p\not=0\,$ and $\,\ell,k=0,1,\dots$,
\qq
\big|\partial_p^\ell\m\widehat{\CZ^\pm_{L}}(p)-\partial_p^\ell\m
\widehat{\CZ^\pm}(p)\big|\,\leq\,\frac{L^{-1}C_{\ell,k}}{(1+p^2)^k}\,.
\label{estim4}
\qqq
\vskip 0.1cm

\noindent{\bf Proof of Lemma 6.} \,We take
\qq
\widehat{\CZ^\pm_{L}}(p)
=\widehat{\CZ^\pm_{12,L}}(p)-\frac{_1}{^L}\hspace{-0.1cm}
\sum\limits_{0\not=n\in\Z}\hspace{-0.1cm}\widehat{\CR^\pm_L}(p,p_n)\,
\widehat{\CZ^\pm_{12,L}}(p_n)\,,
\qqq
where $\,\CR^\pm_L\,$ are operators of Schwartz $\R_{\not=0}\times\R_{\not=0}$
type that converge to $\,\CR^\pm\,$ with speed $\,L^{-1}\,$ and such that
$\,R^\pm\,$ are their $L$-periodization (their existence is a consequence
of the convergence of $\,R^\pm\,$ to $\,\CR\,$ as Schwartz operators,
see Definition 4 and the end of Sec.\,\ref{subsec:momest}).
Then the functions $\,\CZ^\pm_L\,$ have the desired
properties.

\hspace{12cm}$\square$

Let us consider now the functions
\qq
Y_1^\pm{\m}'=-\ii\m P(1-K^0)^{-1}Z^\pm\,.
\label{Ypm1}
\qqq
in $\,L^2_0(S^1_L)$.
\,They are the decoupled versions of the recentered functions
$\,Y'_1$, \,where $\,Y_1\,$ is related by (\ref{9.1})
to the functions $\,X_{s,t,L;1}\,$ that appear in the finite-volume
formulae (\ref{PsitlambdaL}) and (\ref{hattaucw}) for the generating
function $\,\Psi_{t,L}(\lambda)\,$ of FCS. \,The functions $\,Y^\pm_1{}'\,$
are the $L$-periodization of the functions 
\qq
\CY^\pm_{1,L}{}\hspace{-0.12cm}'=-\ii\,\CP(\CI-\ee^{-\gamma_L|\CP|})^{-1}
\CZ^\pm_L\,.
\qqq
The estimates (\ref{estim2}) and (\ref{estim4}) imply that
\qq
\big|\partial_p^\ell\m\widehat{\CY^\pm_{1,L}\hspace{-0.14cm}{}'}\m(p)
-\partial_p^\ell\m
\widehat{\CY^\pm_1{\m}'}(p)\big|\,\leq\,\frac{L^{-1}C_{\ell,k}}{(1+p^2)^k}\,.
\label{estim5}
\qqq
for $\,\ell,k=0,1,\dots$. \,Since for $\,l=0,1,\dots$,
\qq
(\ii\partial_x)^{l}Y^\pm_1{\hspace{0.02cm}}'(x)=
\frac{_1}{^L}\hspace{-0.1cm}\sum\limits_{0\not=n\in\Z}
p_n^l\,\ee^{-\ii p_nx}\,\widehat{\CY^\pm_{1,L}{\m}'}(p_n)\,,
\qqq
we infer that
\qq
&&\Big|\partial_x^l{Y}^\pm_1{}'(x)
-\partial_x^l\CY^\pm_1{}'(x)\Big|
\,\leq\,\frac{_1}{^L}\sum_{0\not=n\in\Z}|p_n|^l
\big|\widehat{\CY^\pm_{1,L}\hspace{-0.14cm}{}'}\,(p_n)
-\widehat{\CY^\pm_1{}'}(p_n)\big|\cr
&&+\Big|\frac{_1}{^L}\hspace{-0.1cm}\sum\limits_{0\not=n\in\Z}p_n^l\,
\ee^{-\ii p_nx}\,\widehat{\CY^\pm_1{}'}(p_n)\,-\,
\frac{_1}{^{2\pi}}\int p^l\,\ee^{-\ii px}\,\widehat{\CY^\pm_1{}'}(p)\,dp
\m\Big|.
\qqq
The first term on the right is estimated directly from (\ref{estim5})
with $\,\ell=0\,$ by $\,\frac{1}{2}L^{-1}C_l\,$ whereas the second one
is bounded using (\ref{CY1pm}) for $\,\ell=0,1\,$ by
$\,\frac{1}{2}L^{-1}C_l(1+|x|)$. It follows that
\qq
\Big|\partial_x^l{Y}^\pm_1{}'(x)
-\partial_x^l\CY^\pm_1{}'(x)\Big|
\,\leq\,L^{-1}C_l(1+|x|)
\label{1stter}
\qqq
proving the uniform convergence on compacts of $\,Y^\pm_1{\m}'\,$ to
$\,\CY^\pm_1{\m}'\,$ with all derivatives.

\subsection{Corrections coupling the right- and left-movers}
\label{subsec:corrcoup}

\noindent The estimates (\ref{estim4}) and (\ref{estim2}) provide the needed control
of the solutions of the decoupled Fredholm equations (\ref{newFred}) and of
their infinite volume version (\ref{limFred}). The complete finite-volume Fredholm
equation coupling the right- and left- movers has the form (\ref{become})
in the frame centered at $\,O^+_L\,$ and its solution $\,\widetilde{Z}\,$
decomposes according to (\ref{tilZ}), where $\,\delta\widetilde Z\,$ solves
Eq.\,(\ref{coupleq}). In the present subsection, we shall estimates
$\,\delta\widetilde Z$. 
\vskip 0.2cm

The main tool that will be employed 
is the summation by parts formula
\qq
\sum\limits_{n=1}^mu_nv_n=u_ms_m-u_1s_0-\sum_{n=1}^{m-1}(u_{n+1}-u_n)s_n\qquad
\text{for}\qquad s_m=\sum\limits_{n=0}^mv_n
\label{sbpf}
\qqq
that will allow to obtain fast-decay type estimates.
As an example of its use, let us prove the following result that will be
applied below:
\vskip 0.3cm

\noindent{\bf Lemma 7.} \,If functions $\,\CX_L\,$ on $\,\R\,$ satisfy for
$\,p\not=0$, $\,\ell=0,1\,$ and $\,k=0,1,\dots\,$ the uniform in $\,L\,$
bounds
\qq
\big|\partial_p^\ell\widehat{\CX_L}(p)|\,\leq\,\frac{C_{\ell,k}}{(1+p^2)^k}
\label{CXkell}
\qqq
then 
\qq
\Big|\frac{_1}{^L}\hspace{-0.05cm}\sum\limits_{0\not=n\in\Z}
p_n^l\,\ee^{-\ii p_nx}\,\widehat{\CX_L}(p_n)
\,\ee^{\pm\ii p_nM_L}\Big|\,\leq\,L^{-1}C_l(1+|x|)
\label{n+-}
\qqq
for $\,l=0,1,\dots$.
\vskip 0.3cm

\noindent{\bf Proof of Lemma 7.} \,Clearly, the bound (\ref{n+-}) would
hold if we dropped the factor $\,L^{-1}(1+|x|)\,$ on the right-hand side.
To extract such a factor, we apply the summation by parts
formula (\ref{sbpf}) for $\,u_n=p_n^l\,\ee^{-\ii p_nx}\,\widehat{\CX_L}(p_n)\,$
and $\,v_n=\ee^{\pm\ii p_nM_L}$.
\,In that case, $\,u_m\mathop{\longrightarrow}\limits_{m\to\infty}0\,$
and
\qq
s_m=\frac{1-\ee^{\pm\ii p_{m+1}M_L}}{1-\ee^{\pm\ii p_1M_L}}
\qqq
are uniformly bounded for sufficiently large $\,L\,$ in view of
(\ref{ML/L}). \,Besides, by (\ref{CXkell}),
\qq
|u_{n+1}-u_n|\leq\frac{L^{-1}C_{k}(1+|x|)}{(1+p_n^2)^k}
\qqq
for some constants $\,C_k$. \,Hence
\qq
\Big|\frac{_1}{^L}\hspace{-0.05cm}\sum\limits_{n=1}^\infty
p_n^l\,\ee^{-\ii p_nx}\,\CX(p_n)\,\ee^{\pm\ii p_nM_L}\Big|=\Big|
L^{-1}u_1\,+\,L^{-1}\sum\limits_{n=1}^\infty(u_{n+1}-u_n)s_n\Big|
\leq\frac{_1}{^2}L^{-1}C_l(1+|x|)\,.
\qqq
The sum over the negative $\,n\,$ is estimated the same way and
the bound (\ref{n+-}) follows.

\hspace{12cm}$\square$
\vskip 0.2cm

Let us define the functions
\qq
\widetilde Y^\pm_1{\m}'\,=-\ii\m P(I-K^0)^{-1}
T_{M_L}^{\pm1}Z^\pm=-\ii\m P(I-\ee^{-\gamma_L|P|})^{-1}
T_{M_L}^{\pm1}Z^\pm\,.
\label{widetilYpm1}
\qqq
in $\,L^2_0(S^1_L)$.
\vskip 0.3cm

\noindent{\bf Corollary.} \,The functions 
$\,\widetilde Y^\pm_1{\m}'\,$ are smooth and they satisfy the bounds
\qq
\big|\partial_x^l\widetilde Y^\pm_1{\m}'(x)\big|\,\leq\,L^{-1}C_l(1+|x|)
\label{tilYpm1}\,.
\qqq
for $\,l=0,1,\dots$.
\vskip 0.3cm

\noindent{\bf Proof of Corollary.} We note that the inequalities
(\ref{CXkell}) hold for the functions
$\,\CX_L=-\ii\,\CP(I-\ee^{-\gamma_L|\CP|})^{-1}\CZ_L^\pm\,$ 
as a consequence  of (\ref{estim2}) and (\ref{estim4}) and that
\qq
\widetilde Y^\pm_1{\m}'(x)
=\frac{_1}{^L}\hspace{-0.05cm}\sum\limits_{0\not=n\in\Z}
\hspace{-0.1cm}\ee^{-\ii p_nx}\,\widehat{\CX_L}(p_n)\,\ee^{\pm\ii p_nM_L}
\qqq
so that (\ref{tilYpm1}) follows from the bound (\ref{n+-}).

\hspace{12cm}$\square$
\vskip 0.3cm

\noindent{\bf Lemma 8.} \,There exist operators $\,\CD_{i,L},\ i=7,\dots,10$,
\,on $\,L^2(\R)\,$ of fast-decay type converging to zero with speed
$\,L^{-1}\,$ whose $L$-periodizations are  
\qq
&&((G^\pm)^{-1})_{+-}\m T_{M_L}^{\mp1}(G^\mp P^{-1})_{-+}\,,
\qquad\hspace{0.6cm}((G^\pm)^{-1})_{-+}\m T_{M_L}^{\mp1}(G^\mp P^{-1})_{+-}
\hspace{1.3cm}\text{for}\ \, i=7\,\ \text{and}\ \,8\,,\qquad\\
&&((G^\pm)^{-1}-I)_{++}\m T_{M_L}^{\mp1}(G^\mp P^{-1})_{+-}\,,
\qquad((G^\pm)^{-1}-I)_{--}\m T_{M_L}^{\mp1}(G^\mp P^{-1})_{-+}\qquad
\text{for}\ \, i=9\,\ \text{and}\ \,10\,.\qquad
\qqq
\vskip 0.3cm

\noindent The proof of Lemma 8 is again based on the summation by parts
formula (\ref{sbpf}). The details may be found in Appendix \ref{app:F}.
Lemma 8 implies immediately that the operators
$\,\delta\widetilde\Sigma\,$ with the components given by the relations
(\ref{deltaSigma++})-(\ref{deltaSigma--}) are of fast-decay
type and that they converge to zero with speed $\,L^{-1}$.
\vskip 0.4cm
 
\noindent{\bf Lemma 9.} The operators $\,\Sigma^+T_{M_L}^{-1}\Sigma^-T_{M_L}\,$
are of fast-decay type and they converge to zero with speed $\,L^{-1}\,$
as such.
\vskip 0.3cm

\noindent The proof of Lemma 9 goes as for Lemma 8 in Appendix \ref{app:F}
using the Schwartz property of $\m\Sigma^\pm\,$ and the summation by parts.

\hspace{12cm}$\square$
\vskip 0.1cm

Recalling the decomposition (\ref{tilS}), \m we shall write
\qq
\widetilde\Sigma=D_L+\delta D_L
\qqq
for
\qq
D_L=\Sigma^++T_{M_L}^{-1}\Sigma^-T_{M_L}+\Sigma^+T_{M_L}^{-1}\Sigma^-T_{M_L}\,,
\qquad\delta D_L=-\Sigma^+T_{M_L}^{-1}\Sigma^-T_{M_L}+\delta\widetilde\Sigma\,.
\qqq
From the fact that $\,\Sigma^\pm\,$ are operators in $\,L^2_0(S^1_L)\,$
of Schwartz $\R_{\not=0}\times\R_{\not=0}$ type that converge as such to
$\,\varSigma^\pm\,$ with speed $\,L^{-1}$ and Lemma 9 it follows
that the operators $\,D_L\,$ are the $L$-periodization of fast-decay
operators in $\,L^2(\R)\,$ satisfying uniform in $\,L\,$ fast-decay bounds.
Note that
\qq
&&\det(I+D_L)=\det\big((I+\Sigma^+)(I+T_{M_L}^{-1}\Sigma^-T_{M_L}))
=\det(I+\Sigma^+)\,\det(I+\Sigma^-)\cr
&&=\det(\CI+\varSigma^+)\,
\det(\CI+\varSigma^-)+O(L^{-1})
\qqq
by Corollary C2 of Appendix \ref{app:C}. From Lemma 3, it follows then that
$\,\det(I+D_L)\,$ are bounded away from zero for $\,L\,$ sufficiently large.
Finally, by Lemmas 8 and 9, operators $\,\delta D_L\,$ are of fast-decay
type and converge to zero with speed $\,L^{-1}$. Hence the pair
$\,(D_L,\delta D_L)\,$ satisfies the assumptions of Proposition C8
of Appendix \ref{app:C} from which we infer that the operators
$\m\,I+\widetilde\Sigma\m\,$ are invertible for $\,L\,$ large enough and
the operators
\qq
\widetilde R=I-(I+\widetilde\Sigma)^{-1}
\qqq
are of fast-decay type being the $L$-periodization of
fast-decay type operators $\,\widetilde\CR_L\,$ on $\,L^2(\R)$
such that the bounds (\ref{fast-dec}) on the momentum-space kernels
of $\,\widetilde\CR_L\,$ are uniform in $\,L$.
\vskip 0.2cm

The Fredholm equation (\ref{coupleq}) for the corrective term
due to the coupling of right- and left-movers has the form
\qq
\big(I+\widetilde\Sigma\big)\delta\widetilde Z=\delta\widetilde Z_{12}\,,
\label{coupleq1}
\qqq
where
\qq
\delta\widetilde Z_{12}&=&-\Big(T_{M_L}^{-1}\Sigma^-T_{M_L}
+\delta\widetilde\Sigma\Big)Z^+-\Big(\Sigma^+
+\delta\widetilde\Sigma\Big)T_{M_L}^{-1}Z^-\cr
&&-\Big(T_{M_L}^{-1}\big(((G^-)^{-1}-I)Q\big)_{++}T_{M_L}
+T_{M_L}^{-1}\big((G^-)^{-1}Q\big)_{\hspace{-0.03cm}-+}T_{M_L}\Big)Y^+_{12}\cr
&&-\Big(\big(((G^+)^{-1}-I)Q\big)_{\hspace{-0.03cm}++}
+\big((G^+)^{-1}Q\big)_{-+}\Big)T_{M_L}^{-1}Y^-_{12}
\qqq
\vskip 0.3cm

\noindent{\bf Lemma 10.} \,There exist functions
$\,\delta\widetilde Z_{12,L}\,$ on $\,\R\,$ such that
$\,\delta\widetilde Z_{12}\,$ are their $L$-periodization and
for $\,k=0,1,\dots$,
\qq
\big|\widehat{\delta\widetilde Z_{12,L}}(p)\big|\,\leq\,\frac{L^{-1}C_k}
{(1+p^2)^k}\,.
\qqq
\vskip 0.2cm

\noindent{\bf Proof of Lemma 10.} \,For all terms of
$\,\delta\widetilde Z_{12,L}\,$ except for $\,-(\delta\widetilde\Sigma)Z^+\,$
and $\,-(\delta\widetilde\Sigma)T_{M_L}^{-1}Z^-$, \,this is shown as in
Proof of Lemma 8 in Appendix \ref{app:F} using the Schwartz-type estimates
for $\,\Sigma^\pm$, $\,Z^\pm$, $\,D^\pm_{3,L}$, $\,D_{2,L}$ and $\,Y^\pm_{12}\,$
and the summation by parts formula (\ref{sbpf}). For the other two terms,
\,it follows from the fast-decay type result for
$\,\delta\widetilde\Sigma\,$ established above and the bounds
(\ref{estim4}) and (\ref{estim2}) with $\,\ell=0$.

\hspace{12cm}$\square$
\vskip 0.2cm

\noindent Since the solution of (\ref{coupleq1}) takes the form
\qq
\delta\widetilde Z\,=\,\delta\widetilde Z_{12}-\widetilde R
\,\delta\widetilde Z_{12}\,,
\qqq
we infer from Lemma 10 and the result about $\,\widetilde R\,$ that
there exist
functions $\,\delta\widetilde\CZ_L\,$ on $\,\R\,$ such that
that $\,\delta\widetilde Z\,$ are their $L$-periodization and
for $\,k=0,1,\dots\,$
\qq
\Big|\widehat{\delta\widetilde\CZ_L}(p)\Big|\,\leq\,\frac{L^{-1}C_k}
{(1+p^2)^k}\,.
\qqq

Let us now consider the functions
\qq
\delta\widetilde Y_1\hspace{-0.03cm}{}'=-\ii\m P(I-K^0)^{-1}\delta\widetilde Z
\label{deltilY1}
\qqq
in $\,L^2_0(S^1_L)$. \,They are the $L$-periodization of the functions
\qq
\delta\widetilde\CY_{1,L}\hspace{-0.32cm}{}'\ =\ii\,\CP
(\CI-\ee^{-\gamma_L|\CP|})^{-1}\delta\widetilde\CZ_L\,.
\qqq
so that for $\,k=0,1,\dots$,
\qq
\Big|\widehat{\delta\widetilde\CY_{1,L}\hspace{-0.32cm}{}'\,\,}\,\m(p)\Big|
\,\leq\,\frac{L^{-1}C_k}{(1+p^2)^k}
\qqq
and, as a result,
\qq
\big|\partial_x^l\delta\widetilde Y_1\hspace{-0.02cm}{}'(x)\m\big|\,=\,
\Big|\frac{_1}{^L}\hspace{-0.1cm}\sum\limits_{0\not=n\in\Z}p_n^l\,\ee^{-\ii p_nx}
\,\widehat{\delta\widetilde\CY_{1,L}\hspace{-0.32cm}{}'\,\,}\,\m(p_n)\,
\Big|\,\leq\,L^{-1}C_l
\label{deltilY1'}
\qqq
for $\,l=0,1,\dots$.

\subsection{Infinite-volume limits of functions $\,X'_{s,t,L;1}$}

\noindent From the relations
(\ref{Y1Z+-}), (\ref{ZTZ}) and (\ref{tilZ}), we infer that
\qq
T_{O^+_L}^{-1}Y_1\hspace{-0.03cm}{}'=-\ii\m P(I-K^0)^{-1}\tilde Z
=-\ii\,P(I-K^0)^{-1}(Z^++T_{M_L}^{-1}Z^-+\delta\widetilde Z)
=Y^+_1{}'+\widetilde Y^-_1{}'+\delta\widetilde Y_1\hspace{-0.02cm}{}'
\qqq
with the last equality following from (\ref{Ypm1}), (\ref{widetilYpm1})
and (\ref{deltilY1}). The established estimates (\ref{1stter}),
(\ref{tilYpm1}) and (\ref{deltilY1'}) imply then that
\qq
\big|\partial_x^lT_{O^+_L}^{-1}Y_1\hspace{-0.02cm}{}'
-\partial_x^l\CY^+_1{}'\big|\,\leq\,L^{-1}C_l(1+|x|)\,.
\qqq
Similarly,
\qq
T_{O^-_L}^{-1}Y_1\hspace{-0.02cm}{}'=T_{M_L}T_{O^+_L}^{-1}Y_1\hspace{-0.02cm}{}'
=-\ii\,P(I-K^0)^{-1}(T_{M_L}Z^++Z^-+T_{M_L}\delta\widetilde Z)
=\widetilde Y^+_1{}'+Y^-_1{}'+T_{M_L}\delta\widetilde Y_1\hspace{-0.02cm}{}'
\qqq
and now,
\qq
\big|\partial_x^lT_{O^-_L}^{-1}Y_1\hspace{-0.02cm}{}'
-\partial_x^l\CY^-_1{}'\big|\,\leq\,L^{-1}C_l(1+|x|)\,.
\qqq
From the identity (\ref{9.1}) it follows that
\qq
T_{O^\pm}^{-1}(X'_{s,t,L;1}-1)
=T_{O^\pm_L}^{-1}(g'_{s,t,L}-1)-T_{O^\pm_L}^{-1}Y_1\hspace{-0.02cm}{}'\,.
\qqq
and from (\ref{9.2}) that
\qq
{\CX^\pm_{s,t}}\hspace{-0.07cm}{}'-1={g^\pm_{s,t}}\hspace{-0.1cm}{}'
-1-{\CY^\pm_1}'\,,
\qqq 
Since $\,T_{O^\pm_L}^{-1}(g_{s,t,L}-g_0)\,$ converges to $\,g^\pm_{s,t}-g_0\,$
uniformly on compacts with all derivatives, see (\ref{separat}) and
(\ref{unifconv}), \m we finally obtain the desired result: 
\vskip 0.4cm

\noindent{\bf Proposition 2.} \,The functions $\,X'_{s,t,L;1}\,$
considered in the frames centered at $\,O^\pm_L\,$ converge to
$\,{\CX^\pm_{s,t;1}}\hspace{-0.25cm}{}'\,\ $ uniformly on compacts with
all derivatives.
\vskip 0.4cm

\noindent{\bf Remark.} \,The speed of convergence is proportional to $\,L^{-1}$.

\nsection{Conclusions}
\label{sec:concl}

\noindent We have obtained an exact expression (\ref{Psitlambda0}) and
(\ref{Psitlambda}) for the infinite-volume generating function for Full
Counting Statistics (FCS) of energy transfers in an inhomogeneous
nonequilibrium state with a preimposed kink-like inverse temperature profile
in a broad class of unitary CFTs. The expression involves a complexified
version of the Schwarzian action
of functions $\,\CX\,$ on the line obtained from conformal
welding of the boundaries of an infinite strip in complex plane after
the twist by diffeomorphisms related to the inverse-temperature
profile. It depends only on the inverse-temperature profile and the CFT central
charge. The latter enters as an overall power, so that the generating function
could be computed from free massless fields, although such a computation
would lead to a more complicated expression involving determinants.
Our infinite-volume formula was obtained by taking the thermodynamic limit
of a finite-volume one involving conformal welding of boundaries of
complex annuli. When deriving the finite-volume formula in a way
inspired by \cite{FH2}, we have obtained, as a byproduct that may be of
independent interest, an expression for the extension of the characters of
unitary positive-energy representations of Virasoro algebra to 1-parameter
groups of circle diffeomorphisms.
The rigorous control of the thermodynamic limit of the finite-volume FCS
formula required a considerable effort in order to derive the asymptotic
behavior of solutions of a Riemann-Hilbert problem related to
conformal welding of tori from annuli. The original motivation of this work
was to prove rigorously in the context of profile states the long-time
large-deviations FCS formula of \cite{BD0}. Although the picture how such
a large-deviations regime arises from our all-times expression for
the generating function for FCS is very clear, a rigorous control of such
a regime appeared to require a refined asymptotic analysis of the solutions
of the Riemann-Hilbert problem corresponding to conformal welding of complex
cylinders from an infinite band that proved difficult and was postponed to
a future work. The functions $\,\CX\,$ involved in our infinite-volume
generating-function formula may, however, be computed numerically using
the conformal welding algorithms developed in \cite{ShM} and we wish to
perform such a computation in the future to access the finite-time corrections
to the long-time large-deviations regime of FCS of energy transfers.
The similar approach to FCS of charge transfers in CFTs with $u(1)$
current algebra, for which nonequilibrium states with chemical potential
profiles were studied in \cite{LLMM1} and \cite{GLM}, is another project
left for a future study. A relation between the approach to FCS 
based on profile states and the original approach of \cite{LL} and
\cite{MA} based on scattering amplitudes of free fields also needs a closer
examination for finite times although the correspondence between the
long-time regimes in the two approaches is transparent. 
\vskip 1cm

\appendix

\setcounter{equation}{0}
\setcounter{section}{0}

\section{}
\label{app:A}

\renewcommand{\theequation}{\thesection.\arabic{equation}}
\noindent In a conformal field theory on a circle of circumference $\,L$,
the energy-momentum components with Euclidian time dependence are
\qq
T_\pm(x\pm\ii vt)=\frac{_{2\pi}}{^{L^2}}
\sum\limits_{n=-\infty}^\infty
\ee^{\pm\frac{2\pi\ii n}{L}(x\pm\ii vt)}\big(L^\pm_n-\frac{_c}{^{24}}\delta_{n,0}\big),
\label{TbarT}
\qqq
where $\,L^\pm_n\,$ are generators of two commuting unitary Virasoro 
representation in the space of states. They satisfy the relations
\qq
T_\pm(x\pm\ii vt)\,=\,\ee^{tH_L}T_\pm(x)\,\ee^{-tH_L},
\qqq
where the Hamiltonian of the theory
\qq
H_L\,=\,v\int_0^L\big(T(x)+\bar T(-x)\big)\,dx
\,=\,\frac{_{2\pi v}}{^L}\big(L_0+\bar L_0-\frac{_c}{^{12}}\big).
\qqq
The normalized vacuum vector $\,\big|0\big\rangle\,$ 
is the unique state for which $\,L_n\big|0\big\rangle=0=\bar L_n\big|0\big\rangle\,$
for $\,n\geq0$. It follows then by a straightforward calculation that
\qq
\big\langle0\big|\,T_\pm(x\pm\ii vt)\big|0\big\rangle\,
=\,-\frac{_{\pi c}}{^{12L^2}}
\label{A1}
\qqq
with the vacuum energy $\,\langle0|H_L|0\rangle=-\frac{\pi c v}{6L}\,$
and that for $\,t_1\not=t_2$,
\qq
\big\langle0\big|\,\CT\big(T_\pm(x_1\pm\ii vt_1)
\,T_\pm(x_2\pm\ii vt_2)\big)
\big|0\big\rangle&=&\big(\frac{_{\pi c}}{^{12L^2}}\big)^{\hspace{-0.05cm}2}
+\frac{_{2\pi^2c}}{^{L^4}}\frac{_{z_1^2z_2^2}}{^{(z_1-z_2)^4}}
\label{A2}
\qqq
where $\,z_1=\ee^{\frac{2\pi\ii}{L}(x_1\pm\ii vt_1)}\,$ and $\,z_2
=\ee^{\frac{2\pi\ii}{L}(x_2\pm\ii vt_2)}$.
\vskip 0.3cm
\

\section{}
\label{app:B}

\noindent From the residue theorem,
\qq
&&\int\frac{\ee^{-\ii py}}{\sinh^4\big(\frac{\pi}{\gamma}(y\pm\ii 0)\big)}dy
-\int\frac{\ee^{-\ii p(y\mp\gamma\ii)}}{\sinh^4\big(\frac{\pi}{\gamma}(y\pm\ii 0)
\big)}dy=\mp\frac{_{\pi\ii}}{^3}\,\partial_z^3\Big|_{z=0}\,
\frac{z^4\ee^{-ipz}}{\sinh^4\big(\frac{\pi}{\gamma}z\big)}\cr
&&=\mp\frac{_{\pi\ii}}{^3}\frac{_{(\gamma)^4}}{^{\pi^4}}\,
\partial_z^3\Big|_{z=0}
\frac{e^{-\ii pz}}{(1+\frac{1}{6}\frac{\pi^2}{(\gamma)^2}z^2)^4}
=\mp\frac{_{\pi\ii}}{^3}\frac{_{(\gamma)^4}}{^{\pi^4}}\,
\partial_z^3\Big|_{z=0}\Big(
e^{-\ii pz}\big(1-\frac{_2}{^3}\frac{_{\pi^2}}{^{(\gamma)^2}}z^2\big)\Big)\cr
&&=\mp\frac{_{\pi\ii}}{^3}\frac{_{(\gamma)^4}}{^{\pi^4}}\,
\partial_z^2\Big|_{z=0}\Big(
(-\ii p)\,e^{-\ii pz}\big(1-\frac{_2}{^3}\frac{_{\pi^2}}{^{(\gamma)^2}}z^2\big)
-\frac{_4}{^3}\ee^{-\ii pz}\frac{_{\pi^2}}{^{(\gamma)^2}}z\Big)\cr
&&=\mp\frac{_{\pi\ii}}{^3}\frac{_{(\gamma)^4}}{^{\pi^4}}\,
\partial_z\Big|_{z=0}\Big(
(-\ii p)^2e^{-\ii pz}
-\frac{_{8}}{^3}\frac{_{\pi^2}}{^{(\gamma)^2}}(-\ii p)\,e^{-\ii pz}z
-\frac{_4}{^3}\ee^{-\ii pz}\frac{_{\pi^2}}{^{(\gamma)^2}}\Big)\cr
&&=\mp\frac{_{\pi\ii}}{^3}\frac{_{(\gamma)^4}}{^{\pi^4}}\,
\Big((-\ii p)^3-\frac{_{4\pi^2}}{^{(\gamma)^2}}(-\ii p)\Big)\cr
&&=\pm\frac{_{1}}{^3}\frac{_{(\gamma)^4}}{^{\pi^3}}\,p
\big(p^2+\frac{_{4\pi^2}}{^{(\gamma)^2}}\big).
\qqq
Hence
\qq
\int\frac{\ee^{-\ii py}}{\sinh^4\big(\frac{\pi}{\gamma}(y\pm\ii 0)\big)}dy=
\pm\frac{_{(\gamma)^4}}{^{3\pi^3}}\frac{_{
p\big(p^2+\frac{_{4\pi^2}}{^{(\gamma)^2}}\big)}}{1-\ee^{\mp\gamma p}}.
\label{FTcalc}
\qqq
\vskip 0.3cm
\

\section{}
\label{app:C}

\noindent We collect here few results concerning operators
of fast-decay and Schwartz type introduced in Sec.\,\ref{subsec:fdSchw}
in Definitions 1 to 4. The two cases will be covered separately
as they often differ and both are needed in the main text.
\vskip -0.7cm
\ 

\subsection{Products of operators of fast-decay and Schwartz type}

\noindent Let us start by two Propositions that are straightforward to prove. 

\vskip 0.3cm

\noindent{\bf Proposition C1.} \,If $\,\CD_1,\CD_2\,$ are operators on
$\,L^2(\R)\,$ of fast-decay type then so is their product $\,\CD_1\CD_2$.
If $\,\CD_{1,L},\CD_{2,L}\,$ are families of fast-decay type
operators converging with speed $\,L^{-1}\,$ to fast-decay type
operators $\,\CD_1,\CD_2$, \,respectively, \,then the products
$\,\CD_{1,L}\CD_{2,L}\,$ converge with speed $\,L^{-1}\,$ to the product
$\,\CD_1\CD_2\,$ as operators of fast-decay type. If $\,D_1,D_2\,$ are
operators on $\,L^2_0(S^1_L)\,$ of fast-decay type then so is their
product $\,D_{1}D_{2}$. 
\vskip 0.3cm

\noindent{\bf Proposition C2.} \,If $\,\CD_1,\CD_2\,$ are operators on
$\,L^2(\R)\,$ of Schwartz $\widehat\CJ\times\widehat\CJ'$ and
$\widehat\CJ'\times\widehat\CJ''$ type, respectively, then $\,\CD_1\CD_2\,$
is of Schwartz $\widehat\CJ\times\widehat\CJ''$ type. If
$\,\CD_{1,L},\CD_{2,L}\,$ are families of such operators converging
with speed $\,L^{-1}\,$ to $\,\CD_1,\CD_2$, \,respectively, \,then
$\,\CD_{1,L}\CD_{2,L}\,$ converges with speed $\,L^{-1}\,$ to
$\,\CD_1\CD_2\,$ as operators of Schwartz
$\widehat\CJ\times\widehat\CJ''$ type. If $\,D_1,D_2\,$ are operators
on $\,L^2_0(S^1_L)\,$ of Schwartz $\widehat\CJ\times\widehat\CJ'$ and
$\widehat\CJ'\times\widehat\CJ''$ type, respectively, then $\,D_1D_2\,$
is of Schwartz $\widehat\CJ\times\widehat\CJ''$ type.
\vskip 0.3cm

\noindent The next result is a little more subtle. 
\vskip 0.3cm

\noindent{\bf Proposition C3.} \,If $\,D_{1,L},D_{2,L}\,$ are families of
operators on $\,L^2(S^1_L)\,$ of Schwartz $\R_{\sigma}\times\R_{\sigma'}$
and $\R_{\sigma'}\times\R_{\sigma''}$ type for $\,\sigma,\sigma',\sigma''
=\pm\,$  converging with speed $\,L^{-1}\,$ to operators $\,\CD_1,\CD_2\,$
on $\,L^2_0(\R)\,$ of the same Schwartz type then the operators
$\,D_{1,L}D_{2,L}\,$ on $\,L^2_0(S^1_L)\,$ of Schwartz
$\R_{\sigma}\times\R_{\sigma''}$ type converge with speed $\,L^{-1}\,$
to $\,\CD_{1}\CD_{2}$. 
\vskip 0.3cm

\noindent{\bf Proof.} \,Let $\,\CD_{1,L}\,$ and $\,\CD_{2,L}\,$ be operators
on $\,L^2(\R)\,$ of Schwartz $\R_{\sigma}\times\R_{\sigma'}$ and
and $\R_{\sigma'}\times\R_{\sigma''}$ type, respectively, converging with
speed $\,L^{-1}\,$ to, respectively, $\,\CD_1\,$ and $\,\CD_2\,$ and such
that $\,D_{1,L}\,$ and $\,D_{2,L}\,$ are the $L$-periodization of, respectively,
$\,\CD_{1,L}\,$ and $\,\CD_{2,L}\,$ (the existence of such operators follows
from our assumptions in view of Definition 4 of Sec.\,\ref{subsec:fdSchw}).
Let $\,\CD_{3,L}\,$ be the operators on $\,L^2(\R)\,$ with the momentum-space
kernels
\qq
\widehat{\CD_{3,L}}(p,q)=\frac{1}{L}\sum\limits_{p_n\in\R_{\sigma'}}
\widehat{\CD_{1,L}}(p,p_n)\,\widehat{\CD_{2,L}}(p_n,q)\,.
\qqq
Note that the product operators $\,D_{3,L}=D_{1,L}D_{2,L}\,$ are the
$L$-periodization of $\,\CD_{3,L}$. \,Let $\,\CD_3=\CD_1\CD_2\,$ with
the momentum-space kernel
\qq
\widehat{\CD_3}(p,q)=\frac{1}{2\pi}\int_{\R_{\sigma'}}\widehat{\CD_1}(p,r)\,
\widehat{\CD_2}(r,q)\,dr\,.
\qqq
We shall prove Proposition C3 by showing that $\,\CD_{3,L}\,$ converge
with speed $\,L^{-1}\,$ to $\,\CD_3\,$ as operators of Schwartz
$\R_{\sigma}\times\R_{\sigma''}$ type. To this end, let us estimate
for $\,(p,q)\in\R_{\sigma}\times\R_{\sigma''}$
\qq
&&\Big|\partial_p^{\ell_1}\partial_q^{\ell_2}\widehat{\CD_{3,L}}(p,q)-
\partial_p^{\ell_1}\partial_q^{\ell_2}\m\widehat{\CD_3}(p,q)\Big|\cr\cr
&&\leq\,\frac{1}{L}\sum\limits_{p_n\in\R_{\sigma'}}\Big|\partial_p^{\ell_1}
\widehat{\CD_{1,L}}(p,p_n)\,\partial_q^{\ell_2}\widehat{\CD_{2,L}}(p_n,q)
-\partial_p^{\ell_1}\m
\widehat{\CD_1}(p,p_n)\,\partial_q^{\ell_2}\m\widehat{\CD_2}(p_n,q)\Big|\cr
&&+\,\Big|\frac{1}{L}\sum\limits_{p_n\in\R_{\sigma'}}
\partial_p^{\ell_1}\widehat{\CD_1}(p,p_n)\,
\partial_q^{\ell_2}\widehat{\CD_2}(p_n,q)-
\frac{1}{2\pi}\int_{\R_{\sigma'}}\partial_p^{\ell_1}\widehat{\CD_1}(p,r)\,
\partial_q^{\ell_2}\widehat{\CD_2}(r,q)\,dr\Big|.
\label{12onrhs}
\qqq
The $1^{\rm st}$ term on the right is easily bounded using the convergence of
$\,\CD_{i,L}\,$ to $\,\CD_i\,$ by
\qq
\frac{L^{-1}C_{\ell_1,\ell_2,k}}{(1+p^2)^k(1+q^2)}\,
\sum\limits_{p_n\in\R_{\sigma'}}\frac{1}{L}\,\frac{1}{(1+p_n^2)^{2k}}
\ \leq\ \frac{L^{-1}C'_{\ell_1,\ell_2,k}}{(1+p^2)^k(1+q^2)}\,.
\label{estF.4}
\qqq
The $2^{\rm nd}$ term on the right-hand side of (\ref{12onrhs}) is estimated
by
\qq
&&\frac{1}{2\pi}\sum\limits_{\widehat J_n\subset\R_{\sigma'}}\int_{\widehat J_n}
\Big|\partial_p^{\ell_1}\widehat{\CD_1}(p,p_n)\,
\partial_q^{\ell_2}\widehat{\CD_2}(p_n,q)-\partial_p^{\ell_1}\widehat{\CD_1}(p,r)\,
\partial_q^{\ell_2}\widehat{\CD_2}(r,q)\Big|\,dr\cr
&&+\,\frac{1}{2\pi}\int_{\widehat J_0\cap\R_{\sigma'}}\Big|
\partial_p^{\ell_1}\widehat{\CD_1}(p,r)\,
\partial_q^{\ell_2}\widehat{\CD_2}(r,q)\Big|\,dr
\qqq
for
\qq
\widehat J_n=\Big]\frac{_{2\pi(n-\frac{1}{2})}}{^L},
\frac{_{2\pi(n+\frac{1}{2})}}{^L}\Big].
\label{hatJn}
\qqq
so that $\,p_n\,$ is the middle-point of $\,\widehat J_n$.
\,The $1^{\rm st}$ line is estimated by (\ref{estF.4}) using the bounds of
the $\,r$-derivative of $\,\partial_p^{\ell_1}\widehat{\CD_1}(p,r)\,
\partial_q^{\ell_2}\widehat{\CD_2}(r,q)\,$ and the $2^{\rm nd}$ line using
the bounds on that function and the small length
$\,|\widehat J_0\cap\R_{\sigma'}|=\pi L^{-1}$. \,Altogether, the left-hand
side of (\ref{12onrhs}) is then bounded by
$\,L^{-1}C_{\ell_1,\ell_2,k}(1+p^2)^{-k}(1+q^2)^{-k}\,$ for
some $\,L$-independent constants $\,C_{\ell_1,\ell_2,k}$,
\,as required.

\hspace{12cm}$\square$

\subsection{Fredholm determinants}

\noindent Let $\,\CD\,$ be the operator of fast-decay type on
$\,L^2(\R)$. Then $\,\CI+\CD\,$ is a Fredholm operator and its determinant
may be defined by the series \cite{GGK}
\qq
\det(\CI+\CD)=\sum\limits_{r=0}^\infty\frac{1}{{r!(2\pi)^r}}\int_{\R^r}
{\det}_{r\times r}\big(\widehat\CD(q_i,q_j)\big)\,dq_1\cdots dq_r\,.
\label{Fredser}
\qqq
The determinant of an $\,r\times r\,$ matrix $\,M=(M_{ij})\,$ may be viewed
as an $r$-linear function $\,d_r(m_1,\dots,m_r)\,$ of the row vectors
of $\,M$, \,where $\,(m_i)_j=M_{ij}$. \,We shall frequently use below
the Hadamard inequality that states that
\qq
\big|{\det}_{r\times r}(M)\big|\,\leq\,\prod_{i=1}^r\Vert m_i\Vert\,,
\label{Hadamard}
\qqq
where $\,\Vert m\Vert\,$ stands for the Euclidian norm of the vector $\,m$.
\,In particular, we infer that
\qq
\big|{\det}_{r\times r}\big(\widehat\CD(q_i,q_j)\big)\big|\,\leq\,
\prod\limits_{i=1}^r\Big(\sqrt{r}\,\frac{C_k}{(1+q_i^2)^{k}}\Big)
\,=\,
r^{\frac{r}{2}}\m C_k^r\prod\limits_{i=1}^r\frac{1}{(1+q_i^2)^k}
\qqq
which assures the convergence of the series (\ref{Fredser}).
\vskip 0.3cm

\noindent{\bf Proposition C4.} \,Let $\,\CD_L\,$ and $\,\CD\,$ be operators
on $\,L^2(\R)\,$ of
fast-decay type such that $\,\CD_L\,$ converge to $\,\CD\,$ with
speed $\,L^{-1}$. Then
\qq
\big|\det(\CI+\CD_L)-\det(\CI+\CD)\big|\,\leq\,\,L^{-1}C
\qqq
for some $\,L$-independent constant $\,C$.
\vskip 0.3cm

\noindent{\bf Proof.} \,Viewing the determinant as the $r$-linear function
of row vectors, we may write
\qq
{\det}_{r\times r}\big(\widehat{\CD_L}(q_i,q_j)\big)-{\det}_{r\times r}
\big(\widehat{\CD}(q_i,q_j)\big)=\sum\limits_{k=1}^r d_r(m_{1,L},\dots,m_{k-1,L},
m_{k,L}-m_k,m_{k+1},\dots,m_r)\,,
\qqq
where $\,(m_{i,L})_j=\widehat{\CD_L}(q_i,q_j)\,$ and $\,(m_{i})_j
=\widehat{\CD}(q_i,q_j)$. \,Then, by the Hadamard inequality,
\qq
\big|{\det}_{r\times r}\big(\widehat{\CD_L}(q_i,q_j)\big)-{\det}_{r\times r}
\big(\widehat{\CD}(q_i,q_j)\big)\big|&\leq&\sum\limits_{k=1}^r
\Big(\prod\limits_{i=1}^{k-1}\Vert m_{i,L}\Vert\Big)\Vert m_{k,L}-m_k\Vert
\Big(\prod\limits_{i=k+1}^r\Vert m_i\Vert\Big)\cr
&\leq&
r\,L^{-1}\prod\limits_{i=1}^r\frac{\sqrt{r}\, C_k}{(1+q_i^2)^k}\,=\,
L^{-1}r^{\frac{r}{2}+1}C_k^r\prod\limits_{i=1}^r\frac{1}{(1+q_i^2)^k}\,.
\qqq
The assertion of Proposition C4 follows now from the Fredholm series 
representation (\ref{Fredser}) for $\,\det(\CI+\CD_L)\,$ and $\,\det(\CI+\CD)$.

\hspace{12cm}$\square$
\vskip 0.2cm 

If $\,D\,$ is an operator on $\,L^2_0(S^1_L)\,$ of fast-decay type in
the sense of Definition 4 of Sec.\,\ref{subsec:fdSchw} then $\,I+D\,$
is a Fredholm operator and its determinant may be defined by the series
\qq
\det(I+D)\,=\sum\limits_{r=0}^\infty\frac{1}{r!}\sum_{(n_1,\dots n_r)\in\Z_{\not=0}^{\ r}}
{\det}_{r\times r}\big(D_{n_i,n_j}\big)\,=\sum\limits_{r=0}^\infty\frac{_1}{r!\,L^r}
\sum_{(n_1,\dots n_r)\in\Z_{\not=0}^{\ r}}
{\det}_{r\times r}\big(\widehat\CD(p_{n_i},p_{n_j})\big)
\label{FredserD}
\qqq
if $\,\CD\,$ is a fast-decay operator on $\,L^2(\R)\,$ such that
$\,D\,$ is its $L$-periodization. The convergence of the series follows
from the Hadamard inequality that implies the bound
\qq
\big|{\det}_{r\times r}\big(\widehat\CD(p_{n_i},p_{n_j})\big|\,\leq\,
r^{\frac{r}{2}}\,C_k^r\prod\limits_{i=1}^r\frac{1}{(1+p_{n_i}^2)^k}
\qqq
and the uniform in $L$ convergence of the series
$\,\sum\limits_{0\not=n\in\Z}
\frac{1}{L}\,\frac{1}{(1+p_n^2)^k}\,$ for $\,k\geq 1$.
\vskip 0.3cm

\noindent{\bf Proposition C5.} Let $\,D_L\,$ be operators
on $\,L^2(S^1_L)\,$ of fast-decay type and let $\,\delta D_L\,$ be
similar operators converging with speed $\,L^{-1}\,$ to zero. Suppose
that $\,D_L\,$ are the $L$-periodization of operators $\,\CD_L\,$
on $\,L^2(\R)\,$ of fast-decay type satisfying uniform in $\,L\,$
fast-decay bounds. \,Then for $\,\widetilde D_L=D_L+\delta D_L$,
\qq
\big|\det(I+\widetilde D_L)-\det(I+D_L)\big|\,\leq\,L^{-1}C
\label{estdetdelt}
\qqq
for some $\,L$-independent constant $\,C$.
\vskip 0.3cm

\noindent{\bf Proof.} \,Let $\,\delta\CD_L\,$ be fast-decay operators on
$\,L^2(\R)\,$ converging with speed $\,L^{-1}\,$ to zero and such that
$\,\delta D_L\,$ are their $L$-periodization (their existence follows from
Definition 4 of Sec.\,\ref{subsec:momest}). Set $\,\widetilde\CD_L=\CD_L+\delta\CD_L$.
Then
\qq
&&\Big|\det(I+\widetilde D_L)-\det(I+D_L)\Big|\cr
&&\leq\,\sum\limits_{r=0}^\infty
\frac{1}{r!\,L^r}\sum\limits_{(n_1,\dots,n_r)\in\Z_{\not=0}^{\ r}}
\Big|{\det}_{r\times r}\big(\widehat{\widetilde\CD_L}(p_{n_i},p_{n_j}))\big)-
{\det}_{r\times r}\big(\widehat{\CD_L}(p_{n_i},p_{n_j})\Big|.
\qqq
Using the Hadamard inequality as in Proof of Proposition C4 above,
we obtain the bound
\qq
\big|{\det}_{r\times r}\big(\widehat{\widetilde\CD_L}(p_{n_i},p_{n_j})\big)-
{\det}_{r\times r}\big(\widehat{\CD_L}(p_{n_i},p_{n_j})\big|\,\leq\,
L^{-1}r^{\frac{r}{2}+1}\,C_k^r\prod\limits_{i=1}^r\frac{1}{(1+p_{n_i}^2)^k}
\qqq
from which (\ref{estdetdelt}) follows.

\hspace{12cm}$\square$
\vskip 0.2cm

\noindent{\bf Corollary C1.} \,Let $\,D_L\,$ be operators on $\,L^2_0(S^1_L)\,$
of fast-decay type converging with speed $\,L^{-1}\,$ to operator
$\,\CD\,$ on $\,L^2(\R)\,$ of fast-decay type and let $\,D\,$ be the
$L$-periodization of $\,\CD$. \,Then
\qq
\big|\det(I+D_L)-\det(I+D)\big|\,\leq\,L^{-1}C
\label{detbdC}
\qqq
for some $\,L$-independent constant $\,C$.
\vskip 0.3cm

\noindent{\bf Proof.} \,We set $\,D'_L=D\,$ and $\,\delta D'_L=D_L-D\,$
and apply Proposition C5 to the pair $\,(D'_L,\delta D'_L)$.

\hspace{12cm}$\square$
\vskip 0.3cm

\noindent{\bf Proposition C6.} \,If $\,\CD\,$ is an operator on $\,L^2(\R)\,$
of Schwartz $\R_{\not=0}\times\R_{\not=0}$ type and $\,D\,$ is its
$L$-periodization then
\qq
\big|\det(I+D)-\det(\CI+\CD)\big|\,\leq\,L^{-1}C
\label{topr0}
\qqq
for some $\,L$-independent constant $\,C$.
\vskip 0.3cm

\noindent{\bf Proof.} \,Let $\,\CD_L\,$ be the operators on $\,L^2(\R)\,$
with momentum space kernels
\qq
\widehat{\CD_L}(p,q)\,=\,\sum\limits_{0\not=m,n\in\Z}
\bm 1_{\widehat J_m}(p)\,\widehat\CD(p_m,p_n)\,\bm 1_{\widehat J_n}(q)\,,
\qqq
where $\,\bm 1_{\widehat J_n}\,$ is the characteristic function of the
interval $\,\widehat J_n$,\ see (\ref{hatJn}). \,We have the
identity
\qq
\det(I+D)=\det(\CI+\CD_L)
\qqq
and for $\,p\in\widehat J_m\,$ and $\,q\in\widehat J_n\,$ with $\,m,n\not=0$,
\qq
\big|\widehat{\CD_L}(p,q)-\widehat\CD(p,q)\big|\,=\,
\big|\widehat\CD(p_m,p_n)-\widehat\CD(p,q)\big|\,\leq\,
\frac{L^{-1}C_k}{(1+p^2)^k(1+q^2)^k}
\label{thebdus}
\qqq
for some $\,C_k\,$ by the Schwartz-type property of $\,\CD$.
\,This bound may fail, however, for $\,p\,$ or $\,q\,$ in
$\,\widehat J_0\,$ in which case $\,\CD_L(p,q)=0\,$ and $\,\CD(p,q)\,$
may be of order 1 with a possible discontinuity at $\,p=0\,$ and/or $\,q=0$.
\,If we define $\,\CD'_L\,$ as the operator on $\,L^2(\R)\,$ with
the momentum-space kernel
\qq
\widehat{\CD'_L}(p,q)=\bm 1_{\R\setminus\widehat J_0}(p)\,\widehat\CD(p,q)\,
\bm 1_{\R\setminus\widehat J_0}(q)
\qqq
then repeating the argument from Proof of Proposition C4, one shows using
the bound (\ref{thebdus}) that
\qq
\big|\det(I+D)-\det(\CI+\CD'_L)|\,=\,
\big|\det(\CI+\CD_L)-\det(\CI+\CD'_L)|\,\leq\,\frac{_1}{^2}L^{-1}C\,.
\label{topr1}
\qqq
On the other hand,
\qq
&&|\det(\CI+\CD'_L)-\det(\CI+\CD)\big|\,\leq\,
\sum\limits_{r=1}^\infty\frac{1}{{(r-1)!
\m(2\pi)^r}}\int_{\widehat J_0}dq_1\int_{\R^{r-1}}
\big|{\det}_{r\times r}\big(\widehat\CD(q_i,q_j)\big)\big|\,dq_2\cdots dq_r\cr
&&\leq\,\sum\limits_{r=1}^\infty\frac{r^{\frac{r}{2}}C_k^r}{{(r-1)!\m(2\pi)^r}}
\int_{\widehat J_0}dq_1\int_{\R^{r-1}}\prod\limits_{i=1}^r\frac{1}{(1+q_i^2)}\,
dq_2\cdots dq_r\,\leq\,\frac{_1}{^2}L^{-1}C\,,
\qqq
where the $\,L^{-1}\,$ factor is due to the length $\,\frac{2\pi}{L}\,$
of $\,\widehat J_0$. Together with (\ref{topr1}) this
gives (\ref{topr0}).

\hspace{12cm}$\square$
\vskip 0.2cm

\noindent{\bf Corollary C2.} \,If $\,D_L\,$ are operators on $\,L^2_0(S^1_L)\,$
of Schwartz $\R_{\not=0}\times\R_{\not=0}$ type converging with speed $\,L^{-1}\,$
to operator $\,\CD\,$ on $\,L^2(\R)\,$ of \m Schwartz
$\R_{\not=0}\times\R_{\not=0}$ type then
\qq
\big|\det(I+D_L)-\det(\CI+\CD)\big|\,\leq\,L^{-1}C
\qqq
for some $\,L$-independent constant $\,C$.
\vskip 0.3cm

\noindent{\bf Proof.} \,This follows directly from Corollary C1 and
Proposition C6 since the Schwartz type convergence implies fast-decay
type one.

\hspace{12cm}$\square$
\vskip -0.8cm
\ 

\subsection{Inverses of Fredholm operators}

\noindent{\bf Proposition C7.} \,If $\,\CD\,$ is a fast-decay type operator
on $\,L^2(\R)\,$ and $\,\det(\CI+\CD)\not=0\,$ then
the Fredholm operator $\,\CI+\CD\,$ is invertible and
$\,\CR=\CI-(\CI+\CD)^{-1}\,$ is of fast-decay type. \,If, moreover,
operators $\,\CD_L\,$ on $\,L^2(\R)\,$ of fast-decay type converge
to $\,\CD\,$ with speed $\,L^{-1}\,$ then $\,\CR_L=\CI-(\CI+\CD_L)^{-1}\,$
are well defined for $\,L\,$ large enough and are of fast-decay type
and they converge to $\,\CR\,$ with speed $\,L^{-1}$.
\vskip 0.3cm

\noindent{\bf Proof.} \,The invertibility of $\,\CI+\CD\,$ follows since
this operator has no zero eigenvalue and $\,(\CI+\CD)^{-1}\,$ is also
a Fredholm operator. The momentum-space kernel of $\,\CR\,$ 
is given by the Fredholm series \cite{GGK}
\qq
\widehat\CR(p,q)\,=\,\frac{1}{{\det(\CI+\CD)}}
\sum\limits_{r=0}^\infty\frac{1}{{r!\m(2\pi)^r}}
\int_{\R^r}{\det}_{(r+1)\times(r+1)}\big(\widehat\CD(q_i,q'_j)\big)
\,\,dq_1\cdots dq_r\,,
\label{cof1}
\qqq
where
\qq
q_0=p\,,\qquad q'_0=q\,,\qquad q_i=q_i'\quad\ \text{for}\quad\ i=1,\dots,r\,.
\label{qiqj}
\qqq
By the Hadamard inequality,
\qq
\Big|{\det}_{(r+1)\times(r+1)}\big(\widehat\CD(q_i,q'_j)\big)\Big|\,\leq\,
(r+1)^{\frac{r+1}{2}}\,C_k^{r+1}\,\frac{1}{(1+p^2)^k}
\prod\limits_{i=1}^r\frac{1}{(1+q_i^2)^k}
\qqq
and similarly, applying it to the column vectors,
\qq
\Big|{\det}_{(r+1)\times(r+1)}\big(\widehat\CD(q_i,q'_j)\big)\Big|\,\leq\,
(r+1)^{\frac{r+1}{2}}\,C_k^{r+1}\,\frac{1}{(1+q^2)^k}
\prod\limits_{i=1}^r\frac{1}{(1+q_i^2)^k}\,.
\qqq
Using the geometric mean of those estimates, we infer that
\qq
\big|\widehat\CR(p,q)\big|&\leq&\frac{1}{{|\det(\CI+\CD)|}}
\sum\limits_{r=0}^\infty\frac{{(r+1)^{\frac{r+1}{2}}\,C_{2k}^{r+1}}}{{r!\m(2\pi)^r}}
\frac{1}{(1+p^2)^k(1+q^2)^k}
\int_{\R^r}\prod\limits_{i=1}^r\frac{1}{(1+q_i^2)^{2k}}\,dq_1\cdots dq_r\cr
&\leq&\frac{C_k}{(1+p^2)^k(1+q^2)^k}
\qqq
for some new constants $\,C_k$. \,This proves that $\,\CR\,$ is
of fast-decay type.
\vskip 0.1cm

Now, if $\,\CD_L\,$ converge with speed $\,L^{-1}\,$
to $\,\CD\,$ then, by Proposition C4, $\,\det(\CI+\CD_L)\,$ converges
with speed $\,L^{-1}\,$ to $\,\det(\CI+\CD)\,$ and hence is bounded away
from zero for $\,L\,$ sufficiently large. On the other hand, using
the Hadamard inequalities as in Proof of Proposition C4, we obtain the bounds
\qq
&&\Big|{\det}_{(r+1)\times(r+1)}\big(\widehat{\CD_L}(q_i,q'_j)\big)-
{\det}_{(r+1)\times(r+1)}\big(\widehat\CD(q_i,q'_j)\big)\Big|\cr
&&\leq\,L^{-1}(r+1)^{\frac{r+1}{2}+1}\,C_{2k}^{r+1}\,\frac{1}{(1+p^2)^k(1+q^2)^k}
\prod\limits_{i=1}^r\frac{1}{(1+q_i^2)^{2k}}
\qqq
and, finally, the estimate
\qq
\big|\widehat{\CR_L}(p,q)-\widehat\CR(p,q)\big|\,\leq\,
\frac{L^{-1}C_k}{(1+p^2)^k(1+q^2)^k}\,.
\qqq
for some $\,L$-independent constants $\,C_k$. This proves that
$\,\CR_L\,$ converge to $\,\CR\,$ with speed $\,L^{-1}\,$ as operators
of fast-decay type.

\hspace{12cm}$\square$
\vskip 0.2cm

\noindent{\bf Proposition C8.} \,Let, as
in Proposition C5, $\,D_L\,$ be operators
on $\,L^2(S^1_L)\,$ of fast-decay type and let $\,\delta D_L\,$ be
similar operators converging with speed $\,L^{-1}\,$ to zero.
Suppose that $\,D_L\,$ are the $L$-periodization of operators $\,\CD_L\,$
on $\,L^2(\R)\,$ of fast-decay type satisfying uniform in $\,L\,$
fast-decay bounds. Assume additionally that there exists $\,L_0>0\,$
such that the Fredholm determinants $\,\det(I+D_L)\,$ are bounded away
from zero uniformly in $\,L\leq L_0$. \,Then for $\,L\,$ large
enough the Fredholm operators $\,I+\widetilde D_L\,$ for
$\,\widetilde D_L=D_L+\delta D_L\,$ are invertible and the operators
$\,\widetilde R_L=I-(I+\widetilde D_L)^{-1}\,$ are of fast-decay
type. Besides there exist operators $\,\widetilde\CR_L\,$ on
$\,L^2(\R)\,$ of fast-decay type satisfying uniform in $\,L\,$
fast-decay bounds and such that $\,\widetilde R_L\,$ are their
$L$-periodization. 
\vskip 0.3cm

\noindent{\bf Proof.} From Proposition C5 it follows that
$\,\det(I+\widetilde D_L)\,$ are bounded away from zero for $\,L\,$ large
enough so that $\,I+\widetilde D_L\,$ are invertible. Let
$\,\delta\CD_L\,$ be the operators on $\,L^2(\R)\,$ of fast-decay type
converging to zero with speed $\,L^{-1}\,$ and such that
$\,\delta D_L\,$ are their $L$-periodization. Set 
$\,\widetilde\CD_L=\CD_L+\delta\CD_L$. \,The matrix elements
of $\,\widetilde R_L\,$ are then given by the Fredholm series \cite{GGK} 
\qq
(\widetilde R_L)_{m,n}&=&\frac{1}{\det(I+\widetilde D_L)}\,
\sum\limits_{r=0}^\infty\,\frac{_1}{r!}\sum\limits_{(n_1,\dots,n_r)\in
\Z_{\not=0}^{\ r}}{\det}_{(r+1)\times(r+1)}\big((\widetilde D_L)_{n_i,n'_j}\big)\cr
&=&\frac{1}{\det(I+\widetilde D_L)}\,\sum\limits_{r=0}^\infty\,
\frac{1}{{r!\,L^{r+1}}}\hspace{-0.2cm}\sum\limits_{(n_1,\dots,n_r)\in
\Z_{\not=0}^{\ r}}\hspace{-0.2cm}{\det}_{(r+1)\times(r+1)}
\big(\widehat{\widetilde\CD_L}(p_{n_i},p_{n'_j})\big)\,.
\label{3333}
\qqq
where $\,n_0=m$, $\,n_0'=n$, $\,n_i=n'_i\,$ for $\,i=1,\dots,r$, \,and
\m the $2^{\rm nd}$ equality follows from the fact that $\,\widetilde D_L\,$
are the $L$-periodization of $\,\widetilde\CD_L$. \,Let us now define
an operator $\,\widetilde\CR_L\,$ on $\,L^2(\R)\,$ with the momentum-space
kernel
\qq
\widehat{\widetilde\CR_L}(p,q)\,=\,\frac{1}{\det(I+\widetilde D_L)}\,
\sum\limits_{r=0}^\infty\,\frac{1}{{r!\,L^{r}}}\hspace{-0.2cm}
\sum\limits_{(n_1,\dots,n_r)\in\Z_{\not=0}^{\ r}}\hspace{-0.2cm}
{\det}_{(r+1)\times(r+1)}\big((\widehat{\widetilde\CD_L}({\rm q}_i,{\rm q}'_j)
\big)\,,
\label{4444}
\qqq
where
\qq
{\rm q}_0=p\,,\qquad {\rm q}'_0=q\,,\qquad {\rm q}_i=p_{n_i}={\rm q}_i'\quad\
\text{for}\quad\ i=1,\dots,r\,.
\qqq
Clearly, $\,\widetilde R_L\,$ is the $L$-periodization of $\,\widetilde\CR_L$.
\,Since operators $\,\widetilde\CD_L\,$ satisfy uniform fast-decay bounds,
we get from the Hadamard inequality the uniform estimate
\qq
\Big|{\det}_{(r+1)\times(r+1)}\big(\widehat{\widetilde\CD_L}
({\rm q}_i,{\rm q}'_j)\big|\Big|
\,\leq\,(r+1)^{\frac{r+1}{2}}\m C_{2k}^{r+1}\m\frac{1}{(1+p^2)^k(1+q^2)^k}
\prod\limits_{i=1}^r\frac{1}{(1+p_{n_i}^2)^{2k}}
\qqq
leading to the uniform fast-decay bounds
\qq
\big|\widehat{\widetilde\CR_L}(p,q)\big|\,\leq\,
\frac{C_k}{(1+p^2)^k(1+q^2)^k}\,.
\qqq
\vskip -0.2cm
\hspace{12cm}$\square$
\vskip 0.3cm 

\noindent{\bf Proposition C9.} \,If $\,\CD\,$ is an operator on $\,L^2(\R)\,$
of Schwartz $\R_{\not=0}\times\R_{\not=0}$ type and
$\,\det(\CI+\CD)\not=0\,$ then $\,\CR=\CI-(\CI+\CD)^{-1}\,$ is also
of Schwartz $\R_{\not=0}\times\R_{\not=0}$ type. \,If, moreover, operators
$\,\CD_L\,$ on $\,L^2(\R)\,$ of Schwartz $\R_{\not=0}\times\R_{\not=0}$ type
converge to $\,\CD\,$ with speed $\,L^{-1}\,$ then the operators
$\,\CR_L=\CI-(\CI+\CD_L)^{-1}$,
\,well defined and of Schwartz $\R_{\not=0}\times\R_{\not=0}$ type for $\,L\,$
large enough, \,converge as such to $\,\CR\,$ with speed
$\,L^{-1}$. 

\vskip 0.3cm

\noindent{\bf Proof.} \,The momentum-space kernel of $\,\CR\,$ is given by
(\ref{cof1}). \,Since for $\,(q_i,q'_i)\,$ as in (\ref{qiqj}) with
$\,q_i\not=0\not=q'_i$,
\qq
\partial_p^{\ell_1}\partial_q^{\ell_2}\,{\det}_{(r+1)\times(r+1)}\big(\widehat\CD
(q_i,q'_j)\big)\,=\,{\det}_{(r+1)\times(r+1)}\big(\widehat
\CD^{\ell_1,\ell_2}(q_i,q'_j)\big)\,,
\qqq
where 
\qq
&&\widehat\CD^{\ell_1,\ell_2}(q_0,q'_0)=\partial_p^{\ell_1}\partial_q^{\ell_2}
\widehat\CD(p,q)\,,\cr
&&\widehat\CD^{\ell_1,\ell_2}(q_0,q'_j)=\partial_p^{\ell_1}
\widehat\CD(p,q_j)\quad\ \text{for}\quad\ j=1,\dots,r\,,\cr
&&\widehat\CD^{\ell_1,\ell_2}(q_i,q'_0)=\partial_q^{\ell_2}
\widehat\CD(q_i,q)\quad\ \,\m\text{for}\quad\ i=1,\dots,r\,,\cr
&&\widehat\CD^{\ell_1,\ell_2}(q_i,q'_j)=
\widehat\CD(q_i,q_j)\hspace{0.9cm}\text{for}\quad\ i,j=1,\dots,r\,,
\qqq
we infer from the Hadamard inequalities that
\qq
\big|{\det}_{(r+1)\times(r+1)}
\big(\widehat{\CD}^{\ell_1,\ell_2}(q_i,q'_j)\big)\big|\,\leq\,
(r+1)^{\frac{r+1}{2}}C_{\ell_1,\ell_2,2k}^{r+1}\,
\frac{1}{(1+p^2)^k(1+q^2)^k}\prod\limits_{i=1}^r\frac{1}{(1+q_i^2)^{2k}}
\qqq
for some constants $\,C_{\ell_1,\ell_2,2k}\,$ and the bounds
\qq
\big|\partial_p^{\ell_1}\partial_q^{\ell_2}\m\widehat\CR(p,q)\big|\,\leq
\frac{C_{\ell_1,\ell_2,k}}{(1+p^2)^k(1+q^2)^k}
\qqq
for some new constants $\,C_{\ell_1,\ell_2,k}\,$ follow. The statement
about the convergence of $\,\CR_L\,$ to $\,\CR\,$ with speed $\,L^{-1}\,$
is inferred similarly from the bound
\qq
&&\big|{\det}_{(r+1)\times(r+1)}
\big(\widehat{\CD_L}^{\ell_1,\ell_2}(q_i,q'_j)\big)-
{\det}_{(r+1)\times(r+1)}
\big(\widehat{\CD}^{\ell_1,\ell_2}(q_i,q'_j)\big)\big|\cr
&&\leq\,L^{-1}(r+1)^{\frac{r+1}{2}+1}C_{\ell_1,\ell_2,2k}^{r+1}\,
\frac{1}{(1+p^2)^k(1+q^2)^k}\prod\limits_{i=1}^r\frac{1}{(1+q_i^2)^{2k}}\,.
\label{tocrw}
\qqq

\hspace{12cm}$\square$
\vskip 0.2cm

\noindent{\bf Proposition C10.} \,Let $\,D_L\,$ be operators on $\,L^2_0(S^1_L)\,$
of Schwartz $\R_{\not=0}\times\R_{\not=0}$ type converging with speed
$\,L^{-1}\,$ to an operator $\,\CD\,$ on $\,L^2(\R)\,$ of Schwartz
$\R_{\not=0}\times\R_{\not=0}$ type such that $\,\det(\CI+\CD)\not=0$. \,Then
for $\,L\,$ large enough, $\,I+D_L\,$ are invertible Fredholm operators and
$\,R_L=I-(I+D_L)^{-1}\,$ are operators on $\,L^2_0(S^1_L)\,$ of Schwartz
$\R_{\not=0}\times\R_{\not=0}$ type converging with speed $\,L^{-1}\,$
to $\,\CR=\CI-(\CI+\CD)^{-1}$.
\vskip 0.3cm

\noindent{\bf Proof.} \,Let $\,\CD_L\,$ be the operators of
Schwartz $\R_{\not=0}\times\R_{\not=0}$ type on $\,L^2(\R)\,$ converging
with speed $\,L^{-1}\,$ to $\,\CD\,$
and such that $\,D_L\,$ are the $L$-periodization of $\,\CD_L$, \,see
Definitions 3 and 4 in Sec.\,\ref{subsec:momest}. From Propositions C4 and C6,
\m we infer that
\qq
\big|\det(I+D_L)-\det(\CI+\CD)\big|\,\leq\,L^{-1}C
\label{convdet}
\qqq
for some $\,C$. \,It follows then \cite{GGK} that for $\,L\,$ large
enough the Fredholm operator $\,I+D_L\,$ is invertible and
$\,R_L=I-(I+D_L)^{-1}\,$ has the matrix elements given by the Fredholm
series as in (\ref{3333}) but without tilde and is the $L$-periodization
of the operator $\,\CR_L\,$ given by (\ref{4444}), again without tilde.
Now for $\,p,q\not=0$,
\qq
\partial_p^{\ell_1}\partial_q^{\ell_2}\,{\det}_{(r+1)\times(r+1)}
\big(\widehat{\CD_L}({\rm q}_i,{\rm q}'_j)\big)\,=\,{\det}_{(r+1)\times(r+1)}
\big(\widehat{\CD_L}^{\ell_1,\ell_2}({\rm q}_i,{\rm q}'_j)\big)\,,
\qqq
where
\qq
&&\widehat{\CD_L}^{\ell_1,\ell_2}({\rm q}_0,{\rm q}'_0)=\partial_p^{\ell_1}
\partial_q^{\ell_2}\widehat{\CD_L}(p,q)\,,\cr
&&\widehat{\CD_L}^{\ell_1,\ell_2}({\rm q}_0,{\rm q}'_j)=\partial_p^{\ell_1}
\widehat{\CD_L}(p,p_{n_j})\quad\ \text{for}\quad\ j=1,\dots,r\,,\cr
&&\widehat{\CD_L}^{\ell_1,\ell_2}({\rm q}_i,{\rm q}'_0)=\partial_q^{\ell_2}
\widehat{\CD_L}(p_{n_i},q)\hspace{0.55cm}\text{for}\quad\ i=1,\dots,r\,,\cr
&&\widehat{\CD_L}^{\ell_1,\ell_2}({\rm q}_i,{\rm q}'_j)=
\widehat{\CD_L}(p_{n_i},p_{n_j})\hspace{0.72cm}\text{for}\quad\ i,j=1,\dots,r
\qqq
and
\qq
\big|{\det}_{(r+1)\times(r+1)}
\big(\widehat{\CD_L}^{\ell_1,\ell_2}({\rm q}_i,{\rm q}'_j)\big)\big|\,\leq\,
(r+1)^{\frac{r+1}{2}}C_{\ell_1,\ell_2,2k}^{r+1}\,
\frac{1}{(1+p^2)^k(1+q^2)^k}\prod\limits_{i=1}^r\frac{1}{(1+p_{n_i}^2)^{2k}}
\qqq
for some constants $\,C_{\ell_1,\ell_2,2k}\,$ implying the bounds
\qq
\big|\partial_p^{\ell_1}\partial_q^{\ell_2}\m\widehat{\CR_L}(p,q)\big|\,\leq
\frac{C_{\ell_1,\ell_2,k}}{(1+p^2)^k(1+q^2)^k}\,.
\qqq
This proves that $\,\CR_L$, \,and hence also $\,R_L$, \,are operators of
the Schwartz $\R_{\not=0}\times\R_{\not=0}$ type.
\vskip 0.1cm

It remains to prove that $\,\CR_L\,$ converge to $\,\CR\,$ with speed $\,L^{-1}\,$ as Schwartz-type operators. Because of (\ref{convdet}), it is enough
to estimate 
\qq
&&\bigg|\sum_{n_1,\dots,n_r\in\Z_{\not=0}^{\ r}}\frac{1}{L^r}\,\partial_p^{\ell_1}
\partial_q^{\ell_2}\,{\det}_{(r+1)\times(r+1)}
\big(\widehat{\CD_L}({\rm q}_i,{\rm q}'_j)\big)-\int\limits_{\R^r}
\frac{1}{(2\pi)^r}\,\partial_p^{\ell_1}\partial_q^{\ell_2}\,{\det}_{(r+1)\times(r+1)}
\big(\widehat{\CD}(q_i,q'_j)\big)\,dq_1\cdots dq_r\bigg|\cr
&&=\,\bigg|\sum_{n_1,\dots,n_r\in\Z_{\not=0}^{\ r}}\frac{1}{L^r}\,{\det}_{(r+1)\times(r+1)}
\big(\widehat{\CD_L}^{\ell_1,\ell_2}({\rm q}_i,{\rm q}'_j)\big)-
\int_{\R^r}\frac{1}{(2\pi)^r}\,{\det}_{(r+1)\times(r+1)}
\big(\widehat{\CD}^{\ell_1,\ell_2}(q_i,q'_j)\big)\,dq_1\cdots dq_r\bigg|\cr
&&\leq\,\sum_{n_1,\dots,n_r\in\Z_{\not=0}^{\ r}}\frac{1}{L^r}\,
\bigg|{\det}_{(r+1)\times(r+1)}
\big(\widehat{\CD_L}^{\ell_1,\ell_2}({\rm q}_i,{\rm q}'_j)\big)-
{\det}_{(r+1)\times(r+1)}
\big(\widehat{\CD}^{\ell_1,\ell_2}({\rm q}_i,{\rm q}'_j)\big)\bigg|\cr
&&+\,\bigg|\sum_{n_1,\dots,n_r\in\Z_{\not=0}^{\ r}}\frac{1}{L^r}\,
{\det}_{(r+1)\times(r+1)}\big(\widehat{\CD}^{\ell_1,\ell_2}({\rm q}_i,
{\rm q}'_j)\big)-\int_{\R^r}\frac{1}{(2\pi)^r}\,{\det}_{(r+1)\times(r+1)}
\big(\widehat{\CD}^{\ell_1,\ell_2}(q_i,q'_j)\big)\,dq_1\cdots dq_r\bigg|.\qquad
\label{lastin}
\qqq
The $1^{\rm st}$ sum on the right-hand side is estimated as in Proof of
Proposition C7
by
\qq
&&L^{-1}(r+1)^{\frac{r+1}{2}+1}\,C_{\ell_1,\ell_2,2k}^{r+1}\,\frac{1}{(1+p^2)^k(1+q^2)^k}
\sum_{n_1,\dots,n_r\in\Z_{\not=0}^{\ r}}\frac{1}{L^r}\,\prod\limits_{i=1}^r\frac{1}{
(1+p_{n_i})^{2k}}\cr
&&\leq\,L^{-1}(r+1)^{\frac{r+1}{2}+1}\,\frac{C_{\ell_1,\ell_2,k}^{r+1}}
{(1+p^2)^k(1+q^2)^k}\,,
\qqq
with some new $\,C_{\ell_1,\ell_2,k}$, \,compare to (\ref{tocrw}).
\,For the $2^{\rm nd}$ term on the right-hand side of (\ref{lastin}),
we use for $\,q_i\in\widehat J_{n_i}\,$ with $\,i=1,\dots,r\,$ the bound 
\qq
&&\big|{\det}_{(r+1)\times(r+1)}
\big(\widehat{\CD}^{\ell_1,\ell_2}({\rm q}_i,{\rm q}'_j)\big)
-{\det}_{(r+1)\times(r+1)}\big(\widehat{\CD}^{\ell_1,\ell_2}(q_i,q'_j)
\big)\big|\cr
&&\leq\,L^{-1}(r+1)^{\frac{r+1}{2}+1}\,C^{r+1}_{\ell_1\ell_2,2k}\,\frac{1}{(1+p^2)^k
(1+q^2)^k}\prod\limits_{i=1}^r\frac{1}{(1+q_i^2)^{2k}}
\qqq
and for at least one $\,q_i\in\widehat J_0$, \,we extract the factor
$\,L^{-1}\,$ from the length of $\,\widehat J_0$, \,similarly as in the
proof of Proposition C6. Altogether, this gives for the $2^{\rm nd}$ term
on the right hand side of (\ref{lastin}) a similar bound as that
for the $1^{\rm st}$ one and permits to conclude the proof.

\hspace{12cm}$\square$

\section{}
\label{app:D}

\noindent{\bf Proof of Lemma 2.} \,The support properties of the
momentum-space kernels $\,\widehat\CD_i(p,q)\,$ of the operators
in question are evident. Now on $\,\R_+\hspace{-0.1cm}\times\R_-$,
\qq
\widehat\CD_1(p,q)&=&q^{-1}
\int \ee^{\ii px}\Big(\ee^{-\ii qg(x)}-
\ee^{-\ii qx}\Big)\;dx\,=\,q^{-1}
\int\ee^{\ii px}\Big(\int_0^1\partial_\sigma\,\ee^{-\ii q(x+\sigma(g(x)-x))}\,d\sigma
\Big)\,dx
\quad\cr
&=&-\ii\int_0^1d\sigma\int\ee^{\ii px-\ii q(x+\sigma(g(x)-x))}\,(g(x)-x)\,dx
\qqq
so that
\qq
&&\partial_{p}^{\ell_1}\partial_{q}^{\ell_2}\widehat\CD_1(p,q)\cr
&&=\,-\ii\int_0^1d\sigma\int\ee^{\ii px-\ii q(x+\sigma(g(x)-x))}\,
\big(\ii x\big)^{\ell_1}\,\big(-\ii(x+\sigma(g(x)-x))\big)^{\ell_2}
\,(g(x)-x)\,dx\cr
&&=\,-\ii\int_0^1d\sigma\int\ee^{\ii px-\ii q(x+\sigma(g(x)-x))}\,
d_1^n\Big(\big(\ii x\big)^{\ell_1}\,\big(-\ii(x+(g(x)-x))\big)^{\ell_2}
\,(g(x)-x)\Big)\,dx
\label{C2}
\qqq
for $\,n=0,1,\dots\,$ and
\qq
(d_1\CX)(x)\,=\,(\ii\partial_x)\Big(\frac{_1}{^{p-q(1+\sigma(g'(x)-1))}}\CX(x)
\Big)\,.
\qqq
The last equality in (\ref{C2}) follows by the subsequent integration by parts
over $\,x\,$ in which all the boundary terms vanish because of the
compact support of $\,g(x)-x$. 
Since $\,|p-q(1+\sigma(g'(x)-1))|\geq|p|+\epsilon|q|\,$ if
$\,p\,$ and $\,q\,$ have different signs for some
$\,\epsilon>0\,$  independent of $\,x\,$ and $\,\sigma$, \,it follows that
\qq
\big|\partial_{p}^{\ell_1}\partial_{q}^{\ell_2}\widehat\CD_1(p,q)\big|\,\leq\,
\frac{C_{\ell_1,\ell_2,n}}{(|p|+|q|)^n}
\qqq
from which the claim of Lemma 2 follows for $\,\CD_1$. \,The claim
for $\,\CD_2\,$ follows the same way. \,For $\,\CD_3$,
\qq
\widehat\CD_3(p,q)\,=\,
-\ii\,\ee^{-\gamma_\# q}
\int_0^1d\sigma\int\ee^{\ii px-\ii q(x+\sigma(g(x)-x))}\,(g(x)-x)\,dx
\qqq
on $\,\R_+\hspace{-0.1cm}\times\R_+$, \,where $\,\gamma_\#=\gamma_L\,$ or
$\,\gamma_\#=\gamma\,$ so that
\qq
&&\partial_{p}^{\ell_1}\partial_{q}^{\ell_2}\widehat\CD_3(p,q)\cr
&&=\,-\ii\,\ee^{-\gamma_\# q}
\int_0^1d\sigma\int\ee^{\ii px-\ii q(x+\sigma(g(x)-x))}\,
\big(\ii x\big)^{\ell_1}\,
\big(-\gamma_\#-\ii(x+\sigma(g(x)-x))\big)^{\ell_2}
\,(g(x)-x)\,dx\,.\qquad
\qqq
It follows that
\qq
&&(1+p^2)^k(1+q^2)^k\,\big|\partial_{p}^{\ell_1}\partial_{q}^{\ell_2}\widehat
\CD_3(p,q)\big|\cr
&&=\,\ee^{-\gamma_\# q}\;(1+q^2)^k\,\Big|\int_0^1d\sigma\int\ee^{\ii px}
\big(1-\partial_x^2)^{k}\Big(
\big(\ii x\big)^{\ell_1}\big(-\gamma_\#-\ii(x+\sigma(g(x)-x))\big)^{\ell_2}\cr
&&\hspace{7.3cm}\times\,(g(x)-x)\,\ee^{-\ii q(x+\sigma(g(x)-x))}\Big)\m dx\Big|\cr
&&\leq\,c_{\ell_1,\ell_2,k}\;\ee^{-\gamma_\# q}(1+q^2)^{2k}\,\leq\,C_{\ell_1,\ell_2,k}
\qqq
which gives the claim of Lemma for $\,\CD_3$.
For $\,\CD_4$,
\qq
\widehat\CD_4(p,q)\,=\,
-\ii\,\ee^{\gamma_\# p}
\int_0^1d\sigma\int\ee^{\ii px-\ii q(x+\sigma(g(x)-x))}\,(g(x)-x)\,dx
\qqq
on $\,\R_-\hspace{-0.1cm}\times\R_-\,$ so that
\qq
&&\partial_{p}^{\ell_1}\partial_{q}^{\ell_2}\widehat\CD_4(p,q)\cr
&&=\,-\ii\,\ee^{\gamma_\# p}
\int_0^1d\sigma\int\ee^{\ii px-\ii q(x+\sigma(g(x)-x))}\,
\big(\gamma_\#+\ii x\big)^{\ell_1}\,
\big(-\ii(x+\sigma(g(x)-x))\big)^{\ell_2}
\,(g(x)-x)\,dx\,.\qquad
\qqq
Hence
\qq
&&(1+p^2)^k(1+q^2)^k\,\big|\partial_{p}^{\ell_1}\partial_{q}^{\ell_2}
\widehat\CD_4(p_1,p_2)\big|\cr
&&=\,\ee^{\gamma_\# p}\,(1+p^2)^k
\Big|\int_0^1d\sigma\int\ee^{-\ii q(x+\sigma(g(x)-x))}(1+d_2^2)^k\Big(\ee^{\ii px}
\big(\gamma_\#+\ii x\big)^{\ell_1}\cr
&&\hspace{7cm}\times\,
\big(-\ii(x+\sigma(g(x)-x))\big)^{\ell_2}
\,(g(x)-x)\Big)\,dx\Big|,\qquad
\qqq
where
\qq
(d_2\CX)(x)=-\ii\partial_x\Big(\frac{_1}{^{1+\sigma(g'(x)-1)}}\CX(x)\Big).
\qqq
It follows that
\qq
(1+p^2)^k(1+q^2)^k\,\big|\partial_{p}^{\ell_1}\partial_{q}^{\ell_2}
\widehat\CD_4(p,q)\big|\,\leq\,c_{\ell_1,\ell_2,k}\,\ee^{\gamma_\# p}\,(1+p^2)^{2k}
\,\leq\,C_{\ell_1,\ell_2,k}
\qqq
which proves the claim of Lemma for $\,\CD_4$.
\vskip 0.2cm

For $\,\CD_5$, 
\qq
\widehat{\CD_5}(p,q)=\frac{_1}{^{2\pi}}\int_0^\infty\widehat a_+(p,r)\,
\widehat a_-(r,q)\,r\,dr
\qqq
on $\,\R_+\times\R_-$, \,where
\qq
\widehat a_\pm(p,q)=-\ii\int_0^1d\sigma
\int\ee^{\ii rx-\ii q(y+\sigma(g^{\pm1}(x)-x))}(g^{\pm1}(x)-x)\,dx
\,.\label{apq}
\qqq
Estimating as for $\,\CD_1\,$ and $\,\CD_3$, \,we infer that
for $\,pq<0$,
\qq
\big|\partial_p^\ell\partial_q^{\ell_2}\m\widehat a_\pm(p,q)\big|
\,\leq\,\frac{c_{\ell,\ell_2,k}}{(1+p^2)^k(1+q^2)^k}\,,
\label{abest1}
\qqq
and for $\,pq>0$,
\qq
\big|\partial_p^{\ell_1}\partial_q^\ell\m\widehat a_\pm(p,q)\big|
\,\leq\,\frac{c_{\ell_1,\ell,k}(1+q^2)^k}{(1+p^2)^k}\,.\label{abest2}
\qqq
The above estimates with $\,\ell=0\,$ imply that
\qq
\big|\partial_p^{\ell_1}\partial_q^{\ell_2}
\widehat{\CD_5}(p,q)\big|\,\leq\,\frac{C_{\ell_1,\ell_2,k}}
{(1+p^2)^k(1+q^2)^k}
\qqq
as claimed. For the later use, let us observe that if we consider
operator $\,\CD_{5,L}\,$ with the momentum-space kernel
\qq
\widehat{\CD_{5,L}}(p,q)=\frac{_1}{^L}
\sum\limits_{n=1}^\infty\widehat a_+(p,p_n)\,\widehat a_-(p_n,q)\,p_n
\qqq
for $\,p_n=\frac{2\pi n}{L}\,$ then the estimates (\ref{abest1})
and (\ref{abest2}) with $\,\ell=0,1\,$ show that
\qq
\big|\partial_p^{\ell_1}\partial_q^{\ell_2}\widehat{\CD_{5,L}}(p,q)
-\partial_p^{\ell_1}\partial_q^{\ell_2}\widehat{\CD_5}(p,q)\big|\,\leq\,\frac{L^{-1}C_{\ell_1,\ell_2,k}}
{(1+p^2)^k(1+q^2)^k}\,,
\qqq
i.e. \,that $\,\CD_{5,L}\,$ converge to $\,\CD_5\,$ with speed $\,L^{-1}\,$ as operators
of Schwartz $\R_+\times\R_-$ type.
Similarly we prove that $\,\CD_6\,$ is of Schwartz $\R_-\times\R_+$ type and that
its analogous modifications $\,\CD_{6,L}\,$ converge with speed $\,L^{-1}\,$ to $\,\CD_6$. 

\hspace{12cm}$\square$
\vskip 0.3cm

If $\,g_L\,$ is a family of diffeomorphisms of $\,\R\,$ such that
$\,g_L(x)=g(x)=x\,$ outside an $\,L$-independent bounded set and if
for $\,\ell=0,1,\dots,\,$
\qq
|\partial_x^\ell g_L(x)-\partial_x^\ell g(x)|\leq L^{-1}C_\ell
\qqq
uniformly in $\,x\,$ then a small modification of the above proof
shows that the operators $\,\CD_{i,L}\,$ for $\,i=1,\dots,6\,$ obtained
from $\,\CD_i\,$ by replacement of $\,g\,$ by $\,g_L\,$ and, for $\,i=5,6$,
by cumulating this change with the one discussed above, converge as
operators of Schwartz type to $\,\CD_i\,$ with speed $\,L^{-1}$.
\vskip 0.3cm
\

\section{}
\label{app:E}

\noindent{\bf Proof of Lemma 3.} $\,\CI+\varSigma^\pm\,$ is a Fredholm operator in
$\,L^2(\R)\,$ since $\,\varSigma^\pm\,$ is Hilbert-Schmidt as it is of
Schwartz $\R_{\not=0}\times\R_{\not=0}$ type. We have then to show that
$\,\CI+\varSigma^\pm\,$ has a trivial kernel, \,i.e. that, if we assume
that for some
$\,\CZ^\pm\in L^2(\R)\,$
$\,(\CI+\varSigma^\pm)\CZ^\pm\,$ vanishes, then it follows that $\,\CZ^\pm=0$.
\,The assumption implies that for $\,\ell,k=0,1,\dots\,$
and $\,p\not=0$,
\qq
\big|\partial_p^\ell\widehat{\CZ^\pm}(p)\big|\,\leq\,\frac{C_{\ell,k}}
{(1+p^2)^k}\,.
\qqq
in virtue of the Schwartz-type property of $\,\varSigma^\pm$.
If we write $\,\CZ^\pm=(\CI-\CK^0)\CP^{-1}\ii\CY'_1\,$ then $\,\CY'_1\,$
also satisfies the last bound (with new constants).
It follows that $\,\CY_1'(x)\,$ is smooth and is bounded by
$\,C(1+|x|)^{-1}\,$ for some $\,C\,$ (as $\,\widehat{\CY_1'}(p)\,$ may have a jump at $\,p=0$).
\,Besides, from the relation
\qq
\varSigma^\pm=-(\CK-\CK^0)(\CI-\CK^0)^{-1}\,,
\qqq
where $\,\CK\,$ is given by (\ref{opCK}) and (\ref{CFCQij}) or (\ref{CK11})-(\ref{CK21}) for
$\,g=g^\pm_{s,t}$, \,it follows that
\qq
(\CI-\CK)\CP^{-1}\CY'_1=0\,.
\qqq
Let $\,\CY_1\,$ be a function whose derivative is equal to $\,\CY_1'$.
$\,\CY_1\,$ is smooth and determined up to a constant and $\,|\CY_1(x)|\leq
C\ln(2+|x|)\,$ for some $\,C$. \,Then
\qq
(\CI-\CK)\CY_1=0\,,
\label{zermod}
\qqq
where $\,\CK_{11}\CY_1\,$ and $\,(\CK_{12}+\CK_{21})\CY_1\,$ are well defined
and $\,\CK_{11}1=0,\ \,(\CK_{12}+\CK_{21})1=1\,$ so that the zero mode equation
(\ref{zermod}) is solved for all choices of $\,\CY_1$. \,Let us define a
holomorphic function $\,\CY(z)\,$ on the interior of $\,\CB_{g,\gamma}\,$ by 
\qq
\CY(z)=\frac{_1}{^{2\pi\ii}}\int\Big(\frac{g'(y)}{g(y)-\ii\gamma-z}-
\frac{1}{y-z}\Big)\,\CY_1(y)\,dy\,.
\qqq
where $\,g=g^\pm_{s,t}$. \,Note that the integral
converges since $\,g(y)=g^\pm_{s,t}(y)=y\,$ for large $\,|y|$. \,Besides, if
we add a constant to $\,\CY_1\,$ then the same constant is added to
$\,\CY$. \,A straightforward estimation shows that for $\,z=z_1+\ii z_2$,
\qq
|\CY(z)|\,\leq\,C\ln(2+|z_1|)
\label{cylest}
\qqq
for some $\,C$. \,We shall show below that the boundary values
of the function $\,\CY\,$ satisfy the relation
\qq
\CY\circ p_i=\CY_1+c
\label{tshonbd}
\qqq
for $\,p_i\,$ given by (\ref{piCB}) and for the same constant $\,c\,$ for
$\,i=1,2$. \,In the variable $\,u=\ee^{\frac{2\pi}{\gamma}z}\,$ (keeping
the same notation for the function), $\,|\CY(u)|\leq C\ln\ln(|u|+|u|^{-1})\,$
in virtue of (\ref{cylest}) and a similar estimate holds in the variable
$\,w=\CW(u)$, \,where $\,\CW\,$ is the map discussed in
Sec.\,\ref{subsec:confweldcyl} with the properties (\ref{valsat}) and
(\ref{recentinf}). Besides, $\,\CY\,$ is analytic in the
complex variable $\,w\,$ everywhere except at zero and at infinity because
of (\ref{tshonbd}). Let
\qq
\CY_+(w)=\frac{_1}{^{2\pi\ii}}\oint\limits_{|w'|=R}\frac{\CY(w')}{w'-w}\,dw'
\qqq
for any $\,R>|w|$. $\,\CY_+\,$ is an entire function on $\,\mathbb C$. \,Since
\qq
\CY'_+(w)=\frac{_1}{^{2\pi\ii}}\oint\limits_{|w'|=R}\frac{\CY(w')}{(w'-w)^2}\,dw'\,,
\qqq
taking $\,R\to\infty$, \,we infer from the {\it a priori} bound
\qq
|\CY(w)|\,\leq\,C\ln\ln(|w|+|w|^{-1})
\label{CYww-1}
\qqq
that $\,\CY_+'=0\,$ and $\,\CY_+={\rm const}$. \,Similarly, \m let
\qq
\CY_-(w)=\frac{_1}{^{2\pi\ii}}\m w^{-1}\hspace{-0.25cm}
\oint\limits_{|w'|=R}\hspace{-0.1cm}\frac{\CY(w'^{-1})}{w'(w'-w^{-1})}\,dw'
\qqq
for any $\,R>|w|^{-1}$. $\,\CY_-\,$ is holomorphic on $\,\mathbb C^\times\,$
and vanishes at infinity. Taking $\,R\to\infty$, \,we infer from the {\it a priori}
bound (\ref{CYww-1}) that $\,\CY_-=0$. \,But $\,\CY=\CY_-+\CY_+\,$ as $\,\CY_+\,$
is given by the part of the Laurent series of $\,\CY\,$ with nonnegative
powers and $\,\CY_-\,$ by the one with negative ones. Hence
$\,\CY={\rm const}.\,$
and, consequently, $\,\CY_1={\rm const}.$, $\,\CY'_1=0\,$ and $\,\CZ^\pm=0$.
\vskip 0.1cm

It remains to show (\ref{tshonbd}). To this end, let us first consider the derivative
of function $\,\CY$,
\qq
\CY'(z)\,=\,\frac{_1}{^{2\pi\ii}}
\int\Big(\frac{_1}{g(y)-\ii\gamma-z}-\frac{1}{y-z}\Big)\CY'_1(y)\,dy\,.
\qqq
Due to the decay of $\,\CY'_1$, \,one has an {\it a priori} bound
\qq
|\CY'(z)|\,\leq\,C(1+|z_1|)^{-1}
\label{bdforCY'}
\qqq
for some $\,C$. \,For the boundary values of $\,\CY'$, \,we obtain the equations
\qq
\CY'\circ p_1=\CE_-({g'}^{-1}\CY'_1)+\CK_{11}({g'}^{-1}\CY'_1)
+\CK_{12}\CY'_1\,,\qquad\CY'\circ p_2=\CE_+\CY'_1+\CK_{21}({g'}^{-1}\CY'_1)
\label{nohatbdv}
\qqq
similarly as in (\ref{inft1}), (\ref{inft2}) and (\ref{inft3}), \,except that
we do not know yet that $\,\CY'\circ p_1={g'}^{-1}\CY'_1\,$ and
$\,\CY'\circ p_2=\CY'_1\,$ and we would like to show it.
Let us now consider for $\,{\rm Im}(z)>0\,$ and $\,{\rm Im}(z)<-\gamma\,$ the
holomorphic function
\qq
\CU(z)
\,=\,{\rm sgn}({\rm Im}(z))\m\frac{_1}{^{2\pi\ii}}
\int\Big(\frac{1}{g(y)-\ii\gamma-z}
-\frac{1}{y-z}\Big)\CY'_1(y)\,dy\,,
\qqq
with the boundary values
\qq
\CU\circ p_1
=\CE_+({g'}^{-1}\CY'_1)-\CK_{11}({g'}^{-1}\CY'_1)
-\CK_{12}\CY'_1\,,\qquad
\CU\circ p_2=-\CE_-\CY'_1+\CK_{21}({g'}^{-1}\CY'_1)
\label{hatbdv}
\qqq
such that
\qq
&&\CU\circ p_1-\CU\circ p_2
\,=\,\CE_+({g'}^{-1}\CY'_1)-\CK_{11}({g'}^{-1}\CY'_1)-\CK_{12}\CY'_1
+\CE_-\CY'_1-\CK_{21}({g'}^{-1}\CY'_1)\,.
\label{difY'i}
\qqq
Differentiating (\ref{zermod}) with the use of the relations
\qq
\partial\m\CK_{11}=g'\CK_{11}{g'}^{-1}\partial
-\CE_-\partial+g'\CE_-{g'}^{-1}\partial\,,\qquad\partial\m\CK_{12}=g'
\CK_{12}\partial\,,\qquad\partial\m\CK_{21}=\CK_{21}{g'}^{-1}\partial\,,
\qqq
we obtain after some algebra the identity
\qq
g'\Big(\CE_+({g'}^{-1}\CY'_1)-\CK_{11}({g'}^{-1}\CY'_1)-\CK_{12}\CY'_1
\Big)+\CE_-\CY'_1-\CK_{21}({g'}^{-1}\CY'_1)\,=\,0
\qqq
that substituted to (\ref{difY'i}) yields the equality
\qq
\CU\circ p_1\,=\,\CU\circ p_2
\,+\,\big(1-{g'}^{-1}\big)\Big(\CE_-\CY'_1-\CK_{21}({g'}^{-1}\CY'_1)\Big)\,=\,
{g'}^{-1}\m\CU\circ p_2\m.
\label{sthatCY1CY2}
\qqq
From (\ref{hatbdv}) and (\ref{CFCQij}) it follows that
\qq
\CE_-\CG\,\m\CU\circ p_1=0\,,\qquad\CE_+\,\CU\circ p_2=0\,.
\qqq
Thus
\qq
(\CG\m{g'}^{-1})_{--}\,\CU\circ p_2=0
\label{therelCU}
\qqq
But $\,(\CG\m{g'}^{-1})_{--}$ is the hermitian adjoint of
$\,(\CG^{-1})_{--}\,$ and the operator $\,(\CG^{-1})_{--}\,$ is invertible
on $\,\CE_-L^2(\R)$. The latter fact is well known but let us digress to indicate
how it is proven. First one shows that $\,(\CG^{-1})_{--}\,$ is injective since if
$\,(\CG^{-1})_{--}\CX=0\,$ for $\,\CX\in\CE_- L^2(\R)\,$ then $\,\CX\,$ is a boundary
value of a holomorphic function on the upper half-plane that vanishes at infinity
and $\,\CX\circ g\,$ is a boundary value of a holomorphic function on the lower
half-plane that also vanishes at infinity. Such functions define a holomorphic function
vanishing at one point on the Riemann sphere welded from the two compactified half-planes 
using the diffeomorphism $\,g\,$ so that $\,\CX\,$ must vanish. Similarly, one
shows that $\,\CG_{--}\,$ is injective. But
\qq
(\CG^{-1})_{--}\CG_{--}=\CE_--(\CG^{-1})_{-+}\CG_{+-}
\qqq
and the right hand side is a Fredholm operator of index zero on
$\,\CE_-L^2(\R)\,$ because $\,(\CG^{-1})_{-+}\,$ and $\,\CG_{+-}\,$ are
Hilbert-Schmidt by Lemma 2 of Sec.\,\ref{subsec:momest} and it has no
kernel by the injectivity of the left-hand side.
Hence $\,(\CG^{-1})_{--}\CG_{--}\,$ is invertible on $\,\CE_-L^2(\R)\,$
and so are $\,(\CG^{-1})_{--}\,$ and its hermitian adjoint
$\,(\CG{g'}^{-1})_{--}$. As a consequence, the relations
(\ref{therelCU}) and (\ref{sthatCY1CY2}) imply
that $\,\CU\circ p_i=0$. \,We are almost done since the latter relations together with
(\ref{hatbdv}) and (\ref{nohatbdv}) imply that
\qq
\CY'\circ p_1={g'}^{-1}\CY'_1\,,\qquad\CY'\circ p_2=\CY'_1
\qqq
so that
\qq
\CY\circ p_1=\CY_1+c_1\,,\qquad\CY\circ p_2=\CY_1+c_2
\qqq
for some constants $\,c_i$. \,But, for $\,x\,$
sufficiently large,
\qq
c_1-c_2=\CY(x-\ii\gamma)-\CY(x)=\ii\int_{-\gamma}^0\CY'(x+\ii y)\,dy
\qqq
which is bounded by $\,O\big(\frac{1}{1+|x|}\big)\,$ in virtue of
(\ref{bdforCY'}) so that $\,c_1-c_2\,$ must vanish.
This establishes (\ref{tshonbd}) completing the proof of Lemma 3.

\hspace{12cm}$\square$
\vskip -0.3cm
\

\section{}
\label{app:F}

\noindent{\bf Proof of Lemma 8.} \,For $\,\CD^\pm_{7,L}\,$
we take the operator with the momentum-space kernel
\qq
\widehat{\CD^\pm_{7,L}}(p,q)=\frac{_1}{^L}\sum\limits_{n=1}^{\infty}
\widehat a^\pm_+(p,-p_n)\,\widehat a_-^\pm(-p_n,q)\,(-p_n)\,\ee^{\pm\ii p_nM_L}
\qqq
on $\R_+\times\R_+$, \,where $\,\widehat a^\pm_\pm(p,q)\,$ and 
are given by (\ref{apq}) for $\,g=g^\pm_{s,t,L}\,$
and they satisfy the estimates (\ref{abest1}) and (\ref{abest2})
uniformly in $\,L$. \,Note that these bounds imply that
\qq
\big|\partial_p^{\ell_1}\partial_r^\ell\m\partial_q^{\ell_2}\big(
\widehat a^\pm_+(p,r)\,\widehat a^\pm_-(r,q)\,r\big)|\,\leq\,
\frac{C_{\ell_1,\ell,\ell_2,k}}{(1+p^2)^k(1+r^2)^k(1+q^2)^k}
\label{Cklll}
\qqq
for non-zero $\,p,r,q\,$ not all of the same sign.
\,Using the summation by parts formula (\ref{sbpf})
in which we set $\,u_n=\widehat a^\pm_+(p,-p_n)\,\widehat a^\pm_-(-p_n,q)
\,(-p_n)\,$ and $\,v_n=\ee^{\pm\ii p_nM_L}$, \,we infer that
for $\,L\,$ sufficiently large,
\qq
&&\widehat{\CD^\pm_{7,L}}(p,q)=\frac{_1}{^L}\bigg(
\widehat a^\pm_+(p,-p_1)\,\widehat a^\pm_-(-p_1,q)
\,p_1\cr
&&\hspace{2cm}+\sum\limits_{n=1}^\infty\Big(\widehat a^\pm_+(p,-p_{n+1})\,
\widehat a^\pm_-(-p_{n+1},q)
\,p_{n+1}\,-\,\widehat a^\pm_+(p,-p_n)\,\widehat a^\pm_-(-p_n,q)
\,p_n\Big)\,\frac{1-\ee^{\m\pm\ii p_{n+1}M_L}}{1-\ee^{\m\pm\ii p_1M_L}}\bigg)\qquad
\label{CD5bp}
\qqq
where we used the facts that by (\ref{Cklll}), 
$\,u_m\mathop{\longrightarrow}\limits_{m\to\infty}0$, \,and that
$\,s_m\,$ are bounded uniformly in $\,L\,$ sufficiently large,
see (\ref{ML/L}). The bound (\ref{Cklll}) also implies that
\qq
\Big|\Big(\widehat a^\pm_+(p,-p_{n+1})\,
\widehat a^\pm_-(-p_{n+1},q)
\,p_{n+1}\,-\,\widehat a^\pm_+(p,-p_n)\,\widehat a^\pm_-(-p_n,q)
\,p_n\Big)\Big|\,\leq\,\frac{L^{-1}C_k}{(1+p^2)(1+p_n^2)(1+q^2)^k}
\qqq
which used on the right hand side of (\ref{CD5bp}) gives the estimate
\qq
\big|\widehat{\CD^\pm_{7,L}}(p,q)\big|\,\leq\,
\frac{L^{-1}C_k}{(1+p^2)^k(1+q^2)^k}
\qqq
showing that $\,\CD^\pm_{7,L}\,$ converge to zero with speed $\,L^{-1}\,$
as operators of fast-decay type.
The case of operators $\,\widehat{\CD^\pm_{i,L}}\,$ with
$\,i=8,\cdots,10\,\m$ is treated same way using again the summation by parts
formula (\ref{sbpf}) and the estimates (\ref{Cklll}).

\hspace{12cm}$\square$

\eject

\end{document}